\documentclass[british]{scrartcl}
\usepackage{float}
\usepackage{bm}
\usepackage{amsthm}
\usepackage{amsmath}
\usepackage{amssymb}
\usepackage{graphicx}
\usepackage{esint}
\usepackage[authoryear]{natbib}
\usepackage{slashbox}
\usepackage{url}
\usepackage{afterpage}
\makeatletter

\floatstyle{ruled}
\newfloat{algorithm}{tbp}{loa}
\providecommand{\algorithmname}{Algorithm}
\floatname{algorithm}{\protect\algorithmname}

\theoremstyle{plain}
\newtheorem{thm}{\protect\theoremname}
  \theoremstyle{plain}
  \newtheorem{lem}[thm]{\protect\lemmaname}

\usepackage{dsfont} 

\usepackage{algpseudocode}
\usepackage{mathtools}

\newcommand{\R}{\mathbb{R}}
\newcommand{\Rp}{\mathbb{R}^+}

\newcommand{\contb}{\mathcal{C}_b}

\newcommand{\setX}{\mathcal{X}}

\newcommand{\ui}{[0,1)} 
\newcommand{\uid}{\ui^d}

\newcommand{\E}{\mathbb{E}}
\newcommand{\var}{\mathrm{Var}}

\newcommand{\Q}{\mathbb{Q}} 
\newcommand{\Qh}{\widehat{\mathbb{Q}}} 
\newcommand{\Qb}{\overline{\mathbb{Q}}} 

\newcommand{\pilti}{\mathcal{M}} 
\newcommand{\Poi}{\mathcal{P}} 
\newcommand{\Unif}{\mathcal{U}} 

\newcommand{\bx}{\mathbf{x}}
\newcommand{\btx}{\tilde{\mathbf{x}}} 
\newcommand{\bu}{\mathbf{u}}
\newcommand{\bv}{\mathbf{v}}
\newcommand{\by}{\mathbf{y}}
\newcommand{\bz}{\mathbf{z}}

\newcommand{\ind}{\mathds{1}} 
\newcommand{\dd}{\mathrm{d}}
\newcommand{\dx}{\dd \mathbf{x}}
\newcommand{\du}{\dd \mathbf{u}}
\newcommand{\eqdef}{:=} 
\newcommand{\bigO}{\mathcal{O}} 
\newcommand{\smallo}{{\scriptscriptstyle\mathcal{O}}} 
\newcommand{\cvz}{\rightarrow 0} 
\renewcommand{\emptyset}{\varnothing} 
\newcommand{\opA}{\alpha_N}
\newcommand{\Dst}{D^\star} 
\newcommand{\stn}{\|_\mathrm{E}} 
\newcommand{\onetoN}{1{:}N} 

\newcommand{\FK}{Feynman-Kac \,}
\newcommand{\RN}{Radon-Nikodym\,}

\newcommand{\Lip}{Lipschitz\,}

\newcommand{\comment}[1]{ \ifthenelse{ \equal{\showcomment}{true} }{ {\bf #1} }{} }
\newcommand{\showcomment}{true}
\newcommand{\Sop}{\mathcal{S}} 

\newcommand{\HSFC}{H}  
\newcommand{\IHSFC}{h} 

\makeatother

\usepackage{babel}
  \providecommand{\lemmaname}{Lemma}
\providecommand{\theoremname}{Theorem}

\begin{document}

\title{Sequential Quasi-Monte Carlo}
\author{Mathieu Gerber\thanks{Universit\'e de Lausanne, and CREST} 
\and Nicolas Chopin\thanks{CREST-ENSAE (corresponding author, nicolas.chopin@ensae.fr)}}
\date{}
\maketitle

\begin{abstract}
We derive and study SQMC (Sequential Quasi-Monte Carlo), a class of algorithms
obtained by introducing QMC point sets in particle filtering. 
SQMC is related to, and may be seen as an extension of, the array-RQMC algorithm of \cite{LEcuyer2006}. 
The complexity of SQMC is $\bigO(N\log N)$, where $N$ is
the number of simulations at each iteration, and its error rate is
smaller than the Monte Carlo rate $\bigO_P(N^{-1/2})$. The only requirement
to implement SQMC is the ability to write the simulation of particle
$\bx_{t}^{n}$ given $\bx_{t-1}^{n}$ as a deterministic function
of $\bx_{t-1}^{n}$ and a fixed number of uniform variates. 
We show that SQMC is amenable
to the same extensions as standard SMC, such as forward smoothing,
backward smoothing, unbiased likelihood evaluation, and so on. In
particular, SQMC may replace SMC within a PMCMC (particle Markov chain
Monte Carlo) algorithm. We establish several convergence results.
We provide numerical evidence that 
SQMC may significantly outperform SMC in practical scenarios.

Key-words: Array-RQMC; Low discrepancy; Particle filtering; Quasi-Monte Carlo;
Randomized Quasi-Monte Carlo; Sequential Monte Carlo
\end{abstract}

\section{Introduction}

Sequential Monte Carlo (SMC, also known as particle filtering) is
a class of algorithms for computing recursively Monte Carlo approximations
of a sequence of distributions $\pi_{t}(\dx_{t})$, $t\in 0{:}T$, $0{:}T=\{0,\ldots,T\}$. The
initial motivation of SMC was the filtering of state-space models
(also known as hidden Markov models); that is, given a latent Markov
process $(\bx_{t})$, observed imperfectly as e.g. $\by_{t}=f(\bx_{t})+\mathbf{\epsilon}_{t}$,
recover at every time $t$ the distribution of $\bx_{t}$ given the
data $\by_{0:t}=(\by_{0},\ldots,\by_{t})$. SMC's popularity stems
from the fact it is the only realistic approach for filtering and related
problems outside very specific cases (such as the linear Gaussian
model). Recent research has further increased interest in SMC, especially
in Statistics, in at least two directions. First, several papers \citep{Neal:AIS,Chopin:IBIS,DelDouJas:SMC}
have extended SMC to non-sequential problems; that is, to sample from
distribution $\pi$, one applies SMC to some artificial sequence $\pi_{t}$
that ends up at $\pi_{T}=\pi$. In certain cases, such an approach
outperforms MCMC (Markov chain Monte Carlo) significantly. Second,
the seminal paper of \citet{PMCMC} established that SMC may be used
as a proposal mechanism within MCMC, leading to so called PMCMC (particle
MCMC) algorithms. While not restricted to such problems, PMCMC is
the only possible approach for inference in state-space models such
that the transition kernel of $(\bx_{t})$ may be sampled from, but
does not admit a tractable density. Excitement about PMCMC is evidenced
by the 30 papers or so that have appeared in the last two years on
possible applications and extensions.

Informally, the error rate of SMC at iteration $t$ is $C_{t}N^{-1/2}$,
where $C_{t}$ is some function of $t$. There has been a lot of work
on SMC error rates \citep[e.g.][]{DelMoral:book}, but it seems fair
to say that most of it has focussed on the first factor $C_{t}$;
that is, whether to establish that the error rate is bounded uniformly
in time, $C_{t}\leq C$, or to reduce $C_{t}$ through more efficient
algorithmic designs, such as better proposal kernels or resampling
schemes.

In this work, we focus on the second factor $N^{-1/2}$, i.e. we want
the error rate to converge quicker relative to $N$ than the standard
Monte Carlo rate $N^{-1/2}$. To do so, we adapt to the SMC context
ideas borrowed from QMC (Quasi-Monte Carlo); that is, the idea of replacing
random numbers by low discrepancy point sets.

The following subsections contain very brief introductions to SMC and 
QMC, with an exclusive focus on the concepts
that are essential to follow this work. For a more extensive presentation
of SMC, the reader is referred to the books of \citet{DouFreiGor},
\citet{DelMoral:book} and \citet{CapMouRyd}, while for QMC and RQMC, see Chapter 5 of \citet{glasserman2003monte},
Chapters 5 and 6 of 
\citet{Lemieux:MCandQMCSampling},
and \citet{dick2010digital}.

\subsection{Introduction to SMC\label{sub:Introduction-to-SMC}}

As already mentioned, the initial motivation of SMC is the sequential analysis
of state-space models; that, is models for a Markov chain $(\bx_t)$ in $\setX\subseteq\mathbb{R}^d$, 
$$ \bx_0 \sim f^X_0(\bx_0),\quad \bx_t|\bx_{t-1} \sim f^X(\bx_t|\bx_{t-1}),$$
which is observed only indirectly through some $\by_t$, with density $\by_t|\bx_t\sim 
f^Y(\by_t|\bx_t)$. 

This kind of model arises in many areas of science: in tracking for instance, $\bx_t$ may be the position of a ship (in two dimensions) or a plane (in three dimensions), and $\by_t$ may be a noisy angular observation (radar). 
In Ecology, $\bx_t$ would be the size of a population of bats in a cave, and $\by_t$ would be $\bx_t$ plus noise. And so on. 

The most standard inferential task for such models is that of \emph{filtering};
that is, to recover iteratively in time $t$, $p(\bx_t|\by_{0:t})$, the distribution of $\bx_t$, given  the data collected up time $t$, $\by_{0:t}=(\by_0,\ldots,\by_t)$. One may also be interested in smoothing, $p(\bx_{0:t}|\by_{0:t})$, or likelihood evaluation, 
$p(\by_{0:t})$, notably  when the model depends on a fixed parameter $\theta$ which should be learnt from the data. 

A simple Monte Carlo approach to filtering is sequential 
importance sampling: choose an initial distribution $m_0(\dx_0)$, a sequence
of Markov kernels $m_t(\bx_{t-1},\dx_t)$, $t\geq 1$, 
then simulate $N$ times iteratively from
these $m_t$'s, $\bx_0^n\sim m_0(\dx_0)$, $\bx_t^n|\bx_{t-1}^n \sim m_t(\bx_{t-1}^n,\dx_t)$, 
and reweight `particle' (simulation) $\bx_t^n$ as follows: $w_0^n= G_0(\bx_0^n)$,
$w_t^n = w_{t-1}^n \times G_t(\bx_{t-1}^n,\bx_t^n)$, where the weight functions $G_t$ are defined as 
\begin{equation}\label{eq:defGt}
G_0(\bx_0)=\frac{f^Y(\by_0|\bx_0)f^X_0(\bx_0)}{m_0(\bx_0)},
\quad
G_t(\bx_{t-1},\bx_t) = \frac{f^Y(\by_t|\bx_t)f^X(\bx_t|\bx_{t-1})}{m_t(\bx_t|\bx_{t-1})},
\end{equation}
and $m_t(\bx_t|\bx_{t-1})$ in the denominator denotes the conditional probability density associated to kernel $m_t(\bx_{t-1},\dx_t)$.
Then it is easy to check that the weighted average $\sum_{n=1}^N w_t^n \varphi(\bx_t^n)
/ \sum_{n=1}^N w_t^n$ is a consistent estimate of the filtering expectation 
$\E[\varphi(\bx_t)|\by_{0:t}]$, as $N\rightarrow +\infty$. 
However, it is well known that, even for carefully chosen proposal densities 
$m_t$, sequential importance sampling quickly degenerates: as time progresses, 
more and more particles get a negligible weight. 

Surprisingly, there is a simple solution to this degeneracy problem: one may \emph{resample} the particles; that is, draw $N$ times with replacement from
the set of particles, with probabilities proportional to the weights $w_t^n$. 
In this way, particles with low weight gets quickly discarded, while particles
with large weight may get many children at the following iteration. Empirically, 
the impact of resampling is dramatic: the variance of filtering estimates typically remains stable over time, while without resampling it diverges exponentially
fast. 

The idea of using resampling may be traced back to \cite{Gordon}, and has initiated 
the whole field of particle filtering. See Algorithm \ref{alg:Generic-SMC-algorithm}
for a summary of a basic PF (particle filter). The price to pay for introducing resampling is that it creates non-trivial dependencies between the particles,
which complicates the formal study of such algorithms. In particular, establishing convergence (as $N\rightarrow +\infty$) is
non-trivial, although the error rate of SMC  is known to be $\bigO_P(N^{-1/2})$; see e.g. the central
limit theorems of \citet{DelGui}, \citet{Chopin:CLT} and \citet{Kunsch:CLT}. We shall see that
it is also the resampling step that makes the introduction of Quasi-Monte Carlo into 
SMC non-trivial.

\begin{algorithm}[htp]

At time $t=0$, 
\begin{description}
\item [{(a)}] Generate $\bx_{0}^{n}\sim m_0(\dx_0)$ for  all $n\in 1{:}N$.  
\item [{(b)}] Compute 
$w_0^n = G_{0}(\bx_{0}^{n})$ and 
$W_{0}^{n}=w_0^n/\sum_{m=1}^N w_0^m$
for all $n\in 1{:}N$.
\end{description}
From time $t=1$ to time $T$, 
\begin{description}
\item [{(a)}] Generate $a_{t-1}^{n}\sim\pilti(W_{t-1}^{1:N})$
for all $n\in 1{:}N$, 
the multinomial
distribution that produces outcome $m$ with probability $W_{t-1}^m$. See Algorithm \ref{alg:inverse_method}.
\item [{(b)}] Generate 
$\bx_{t}^{n}\sim m_t(\bx_{t-1}^{a_{t-1}^{n}},\dx_{t})$
for all $n\in 1{:}N$.
\item [{(c)}] 
Compute 
$w_t^n = G_{t}(\bx_{t-1}^{a_{t-1}^{n}},\bx_{t}^{n})$, and
$W_{t}^{n}=w_t^n/\sum_{m=1}^N w_t^m$
for all $n\in 1{:}N$. 
\end{description}
\caption{\label{alg:Generic-SMC-algorithm}Basic particle filter}
\end{algorithm}

The complexity of SMC is $\bigO(N)$. In particular, to implement the resampling
step in $\bigO(N)$ time (Step (a) at times $t\geq 1$ in Algorithm \ref{alg:Generic-SMC-algorithm}), one proceeds as follows: (a) generate $u^{1:N}=\mathrm{sort}(v^{1:N})$,
where the $v^{n}$ are independent uniform variates \citep[see p.214 of][for a well-known algorithm to generate $u^{1:N}$
directly in $\bigO(N)$ time, without any sorting]{Devroye:book}; and (b) use the
inverse transform method for discrete distributions, recalled in Algorithm
\ref{alg:inverse_method}. We will re-use Algorithm \ref{alg:inverse_method}
in SQMC. 

\begin{algorithm}[htp]
\begin{algorithmic}

\Require $u^{1:N}$ (such that $0\leq u^{1}\leq\ldots\leq u^{N}\leq1$),
$W^{1:N}$ (normalised weights)

\Ensure $a^{1:N}$ (labels in $1:N$)

\State $s\gets W^1$, $m\gets1$

\For{$n=1\to N$}

\While{$s<u^{n}$} 

\State $m\gets m+1$ 

\State $s\gets s+W^{m}$

\EndWhile

\State $a^{n}\gets m$

\EndFor 

\end{algorithmic}

\caption{\label{alg:inverse_method}Resampling Algorithm (inverse transform
method)}
\end{algorithm}

\subsection{Introduction to QMC\label{sub:Introduction-to-QMC}}

QMC (Quasi-Monte Carlo) is generally presented as a way to perform
integration with respect to the (semi-closed) hypercube of dimension $d$:
\[
\frac{1}{N}\sum_{n=1}^{N}\varphi(\bu^{n})\approx\int_{\ui^{d}}\varphi(\bu)\,\du
\]
where the $N$ vectors $\bu^{n}\in\ui^{d}$ must be chosen so as
to have ``low discrepancy'', that is, informally, to be spread evenly
over $\ui^{d}$. 
(We respect the standard convention in the QMC literature to work 
with space $\uid$, rather than $[0,1]^d$, as it turns out to be 
technically more convenient.) 

Formally, the general notion of discrepancy is
defined as 
\[
D(\bu^{1:N};\mathcal{A})=\sup_{A\in\mathcal{A}}\left|\frac{1}{N}\sum_{n=1}^{N}\ind\left(\bu^{n}\in A\right)-\lambda_{d}(A)\right|
\]
where $\lambda_{d}(A)$ is the volume (Lebesgue measure on $\mathbb{R}^{d}$)
of $A$, and $\mathcal{A}$ is a set of measurable sets. Two discrepancies
are particularly useful in this work: the extreme discrepancy, 
\[
D(\bu^{1:N})=\sup_{[\bm{a},\bm{b}]}\left|\frac{1}{N}\sum_{n=1}^{N}\ind\left(\bu^{n}\in[\bm{a},\bm{b}]\right)-\prod_{i=1}^{d}(b_{i}-a_{i})\right|
\]
which is the discrepancy relative to the set $\mathcal{A}$ of $d-$dimensional
intervals $[\bm{a},\bm{b}]\eqdef\prod_{i=1}^{d}[a_{i},b_{i}]$, $0\leq a_{i}< b_{i}<1$;
and the star discrepancy: 
\[
\Dst(\bu^{1:N})=\sup_{[\bm{0},\bm{b}]}\left|\frac{1}{N}\sum_{n=1}^{N}\ind\left(\bu^{n}\in[\bm{0},\bm{b}]\right)-\prod_{i=1}^{d}b_{i}\right|,
\]
where again $[\bm{0},\bm{b}]=\prod_{i=1}^{d}[0,b_{i}]$, $0< b_{i}<1$.
When $d=1$, the star discrepancy is the Kolmogorov-Smirnov statistic
for an uniformity test of the points $\bu^{n}$.

These two discrepancies are related as follows \citep[Proposition 2.4]{Niederreiter1992}:
\[
\Dst(\bu^{1:N})\leq D(\bu^{1:N})\leq2^{d}\Dst(\bu^{1:N}).
\]
The importance of the concept of discrepancy, and in particular of
the star discrepancy, is highlighted by the Koksma\textendash{}Hlawka
inequality \citep[see e.g.][Theorem 5.1]{Kuipers1974}:

\[
\left|\frac{1}{N}\sum_{n=1}^{N}\varphi(\bu^{n})-\int_{[0,1)^{d}}\varphi(\bu)\,\du\right|\leq V(\varphi)\Dst(\bu^{1:N})
\]
which conveniently separates the effect of the smoothness of $\varphi$
(as measured by $V(\varphi)$, the total variation in the sense
of Hardy and Krause, see Chapter 2 of \citealp{Niederreiter1992} for
a definition), and the effect of the discrepancy of the points $\bu^{1:N}$.
The quantity $V(\varphi)$ is generally too difficult to compute in
practice, and the Koksma\textendash{}Hlawka inequality is used mainly
to determine the asymptotic error rate (as $N\rightarrow+\infty$),
through the quantity $\Dst(\bu^{1:N})$.

There are several methods to construct $\bu^{1:N}$ so that $\Dst(\bu^{1:N})=\bigO(N^{-1+\epsilon})$
for any $\epsilon>0$; which is of course better than the Monte Carlo
rate $\bigO_P(N^{-1/2})$. The best known rates are $\bigO(N^{-1}(\log N)^{d-1})$
for QMC point sets $\bu^{N,1:N}$ that are allowed to depend on
$N$ (i.e. $\bu^{N,1:N}$ are not necessarily the $N$ first elements
of $\bu^{N+1,1:N+1}$) and $\bigO(N^{-1}(\log N)^{d})$ for QMC
sequences (that is $\bu^{1:N}$ are the $N$ first elements of a sequence
$(\bu^{n})$ which may be generated iteratively). For simplicity,
we will not distinguish further QMC point sets and QMC sequences,
and will use the same notation $\bu^{1:N}$ in both cases (although
our results will  apply to both types of construction). 

These asymptotic rates seem to indicate that the comparative performance of QMC
over Monte Carlo should deteriorate with $d$: for $d=10$, 
$N^{-1}(\log N)^d \leq  N^{-1/2}$ only for $N\geq 1.3\times 10^{39}$. 
But since these rates correspond to an upper bound for the error size, it is hard to determine
beforehand if and when  QMC ``breaks'' with the dimension. For instance, 
\citet[p.327]{glasserman2003monte} exhibits a  
a numerical example where QMC remains competitive relative to Monte Carlo
for $d\geq 150$ and $N\leq 10^5$. 
 
\begin{figure}
\begin{centering}
\includegraphics[scale=0.35]{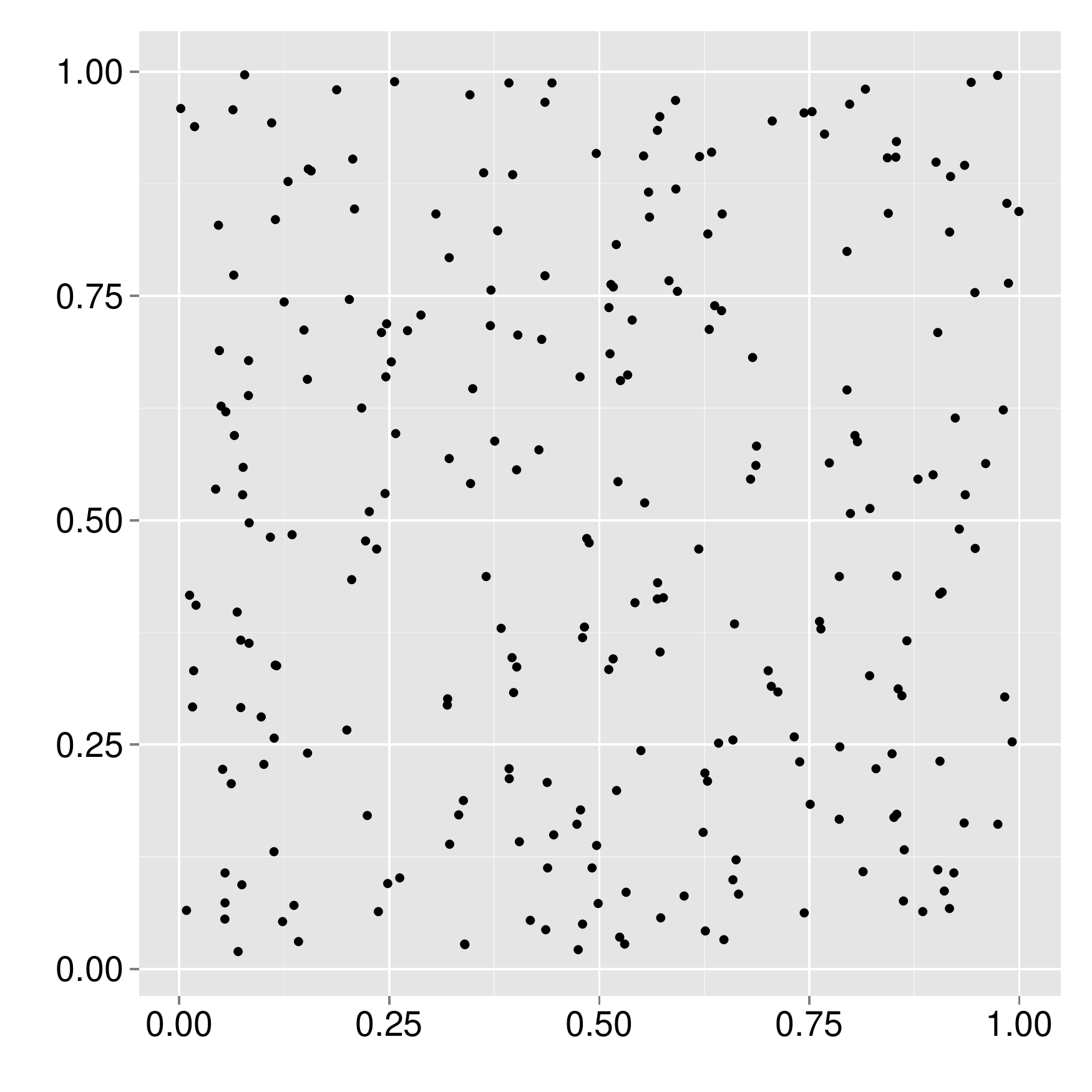}\includegraphics[scale=0.35]{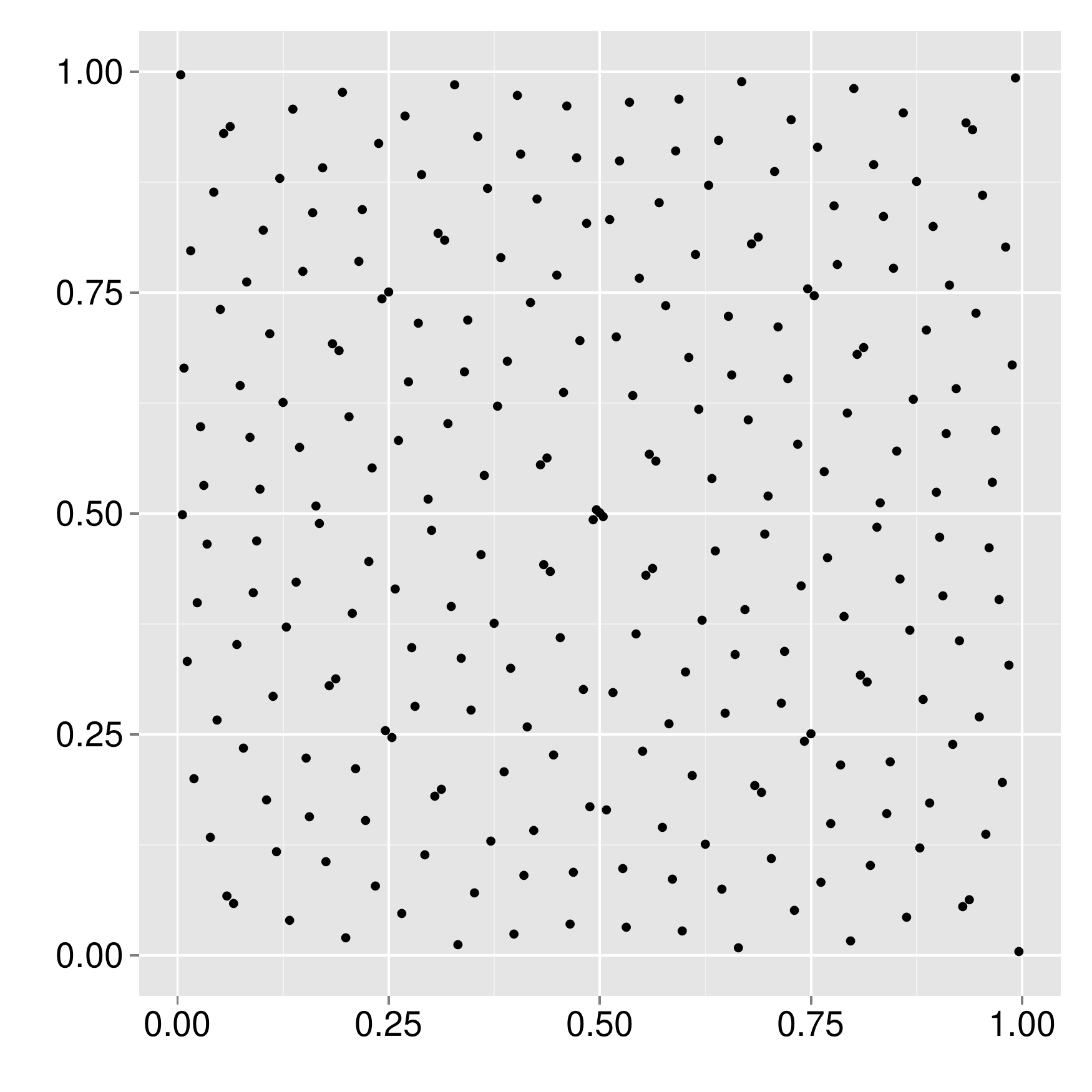} 
\par\end{centering}

\caption{QMC versus Monte Carlo: $N=256$ points sampled independently and
uniformly in $\ui^{2}$ (left); QMC sequence (Sobol') in $\ui^{2}$
of the same length (right).\label{fig:QMC-versus-randomness:} }
\end{figure}

Describing the different strategies to construct low-discrepancy point
sets is beyond the scope of this paper; see again the aforementioned
books on QMC. 
Figure \ref{fig:QMC-versus-randomness:}
illustrates the greater regularity of a QMC point set over a set of
random points.

\subsection{Introduction to RQMC}

RQMC (randomized QMC) amounts to randomize the points $\bu^{1:N}$
in such a way that (a) they still have low discrepancy (with probability one); and (b) 
each $\bu^{n}\sim\Unif(\ui^d)$ marginally.
The simplest construction of such RQMC point sets is the random shift method proposed by  \citet{Cranley1976} in which  we take $\bu^{n}=\bv^{n}+\mathbf{w}\pmod{1}$,
where $\mathbf{w}\sim\Unif(\ui^{d})$ and $\bv^{1:N}$ is a low-discrepancy
point set.

RQMC has two advantages over QMC. First, one then obtains an unbiased
estimator of the integral of interest: 
\[
\E\left\{ \frac{1}{N}\sum_{n=1}^{N}\varphi(\bu^{n})\right\} =\int_{\ui^{d}}\varphi(\bu)\,\du,
\]
which makes it possible to evaluate the approximation error through
independent replications. We will see that, in our context, this unbiasedness
property will also be very convenient for another reason: namely to
provide an unbiased estimate of the likelihood of the considered state-space model. 

Second, \citet{Owen1997a, Owen1997b,Owen1998} established that randomization may lead
to better rates, in the following sense: under appropriate conditions,
and for a certain type of randomization scheme known as nested scrambling, 
the mean square error of a RQMC estimator is $\bigO(N^{-3+\epsilon})$.
The intuition behind this rather striking result
is that randomization may lead to cancellation of certain error terms.

\subsection{A note on array-RQMC}

Consider the following problem: we have a Markov chain in $\setX$, whose evolution
may be formulated as
$$\bx_t = \Gamma_t(\bx_{t-1},\bu_t),\quad \bu_t \sim \Unif\left( \ui^d\right),
\quad t\geq 1,\quad \bx_0\mbox{ is fixed},$$
and  we wish to compute the expectation of 
$ \sum_{t=1}^T \varphi_t(\bx_t)$, for certain functions $\varphi_t$. 

From the two previous sections, we see that a simple approach to this problem would be  to generate a QMC (or RQMC)
point set $\bu^{1:N}$ in $\ui^{dT}$, $\bu^n = (\bu_1^n,\ldots,\bu_T^n)$, 
to transform   $\bu_t^n$ into $\bx_t^n=\Gamma_t(\bx_{t-1}^n,\bu_t^n)$, and finally 
to return the corresponding empirical average, $N^{-1} \sum_{t=1}^T \varphi_t(\bx_t^n)$.
The problem with this direct approach is that the dimension $dT$ of $\bu^{1:N}$
may be very large, and, as  we have seen, 
equidistribution properties of  QMC point sets (as measured by the star discrepancy) deteriorate with the dimension. 

An elegant alternative to this approach is the array-RQMC algorithm of 
\cite{LEcuyer2006}, see also 
\cite{lecot2002quasirandom}, 
\cite{lecot2004quasi}, and 
\cite{LEcuyer2009}. 
The main idea of this method is to replace the QMC point set in $\ui^{dT}$ by  $T$ QMC points sets $\bu_t^{1:N}$ in $\ui^{d}$. Then, $\bx_t^n$ is obtained as $\bx_t^n=\Gamma_t(\bx_{t-1}^{a_{t-1}^n},\bu_t^n)$, 
where the ancestor $\bx_{t-1}^{a_{t-1}^n}$ of $\bx_t^n$ is chosen so as
to be the $n$-th ``smallest'' point  among the $\bx_{t-1}^n$'s. Note that array-RQMC therefore requires to specify a total order for the state space $\setX$; for instance
one may define a certain $\omega:\setX\rightarrow \R$ so that 
$\omega(\bx)\leq \omega(\bx')$ means that $\bx$ is ``smaller'' than $\bx'$.


Array-RQMC is shown to have excellent empirical performance in the aforementioned
papers. On the other hand, it is currently lacking in terms of supporting theory 
\cite[see however][for $d=1$]{l2008randomized}; 
in particular, it is not clear how to choose the order $\omega$, beside
the obvious case where $\setX\subset \R$. 
The SQMC algorithm we develop in this paper may be seen as an extension of array-RQMC to particle filtering. In particular, it re-uses the essential idea to generate one QMC point set at each step of the simulation process.
 As an added benefit, the convergence
results we obtain for SQMC also apply to array-RQMC, provided the state space
is ordered through the Hilbert curve, as explained later.

\subsection{Background, plan and notations}

QMC is already very popular in Finance for e.g. derivative pricing
\citep{glasserman2003monte}, and one may wonder why it has not received
more attention in Statistics so far. The main reason seems to be the
perceived difficulty to adapt QMC to non-independent simulation such
as MCMC (Markov chain Monte Carlo); see however \citet{Chen2011}
and references therein, in particular \citet{TribblePhd}, for exciting
numerical and theoretical results in this direction which ought to
change this perception.

Regarding SMC, we are aware of two previous attempts to develop QMC
versions of these algorithms: \citet{Lemieux2001} and \citet{Fearnhead2005};
see also \cite{guo2006quasi} who essentially proposed the same algorithm 
as \citet{Fearnhead2005}.
The first paper casts SMC as a Monte Carlo algorithm in $d(T+1)$ dimensions,
where $d=\mathrm{dim}(\setX)$, and therefore requires to generate a low-discrepancy point set
in $\ui^{d(T+1)}$. But, as we have already explained, such an approach 
may not work well when $d(T+1)$ is too large. 

Our approach is closer to, and partly inspired by,
the RPF (regularized particle filter) of \citet{Fearnhead2005}, who, in the same spirit as array-RQMC, 
casts SMC as a sequence of $T+1$ successive importance sampling steps of dimension $d$. 
(The paper focus on the $d=1$ case.) The main limitation of the RPF is that
it has complexity $\bigO(N^2)$. This is because the importance sampling steps 
are defined with respect to a target  
which is a mixture of $N$ components,
hence the evaluation of a single importance weight costs $\bigO(N)$.

The SQMC algorithm we develop in this paper has complexity $\bigO(N\log(N))$ per time step. 
It is also based on a sequence of $T+1$ importance sampling steps, but of dimension $d+1$; 
the first component is used to determine which ancestor $\bx_{t-1}^m$ should be
assigned to particle $\bx_t^n$. For $d>1$, this requires us to ``project''
the set of ancestors  $\bx_{t-1}^{1:N}\in\setX^N$ into $\ui^N$, by means of 
a space-filling curve known as the Hilbert curve. 
The choice of this particular space-filling curve is not only
for computational convenience, but also because of its nice properties
regarding conversion of discrepancy, as we will explain in the paper.
(One referee pointed out to us that the use of Hilbert curve in the context
of array-RQMC has been suggested by \cite{wachter2008efficient}, but not implemented.)

%

The paper is organised as follows. Section \ref{sec:SQMC} derives
the general SQMC algorithm, first for $d=1$, then for any $d$ through
the use of the Hilbert curve. Section \ref{sec:Convergence-results}
presents several convergence results; proofs of these results are
in the Appendix. Section \ref{sec:Extensions} shows how several standard
extensions of SMC, such as forward smoothing, backward smoothing,
and PMCMC, may be adapted to SQMC. Section \ref{sec:Numerical-study}
compares numerically SQMC with SMC. Section \ref{sec:Conclusion}
concludes.

Most random variables  in this work will be vectors in
$\R^{d}$, and will be denoted in bold face, $\bu$ or $\bx$. In
particular, $\setX$ will be an open set of $\R^{d}$. The Lebesgue
measure in dimension $d$ is denoted by $\lambda_{d}$. Let $\mathcal{P}(\setX)$
 be the set of probability measures defined on $\setX$ dominated
by $\lambda_{d}$ (restricted to $\setX$), and $\pi(\varphi)$ be the expectation of function $\varphi$ relative to 
$\pi\in\mathcal{P}(\setX)$. Let $a:b$ be the set of
integers $\{a,\ldots,b\}$ for $a\leq b$. We also use this notation
for collections of random variables, e.g. $ \bx_{t}^{1:N}=(\bx_{t}^{1},\ldots,\bx_{t}^{N})$,
$\bx_{0:t}=(\bx_{0},\ldots,\bx_{t})$ and so on.

\section{SQMC\label{sec:SQMC}}

The objective of this section is to construct the SQMC algorithm.
To this aim, we discuss how to rewrite SMC as a deterministic function
of independent uniform variates $\bu_{t}^{1:N}$,   $t\in 0{:}T$, which then may be
replaced by  low-discrepancy point sets. 


\subsection{SMC formalisation}

A closer inspection of our basic particle filter, Algorithm \ref{alg:Generic-SMC-algorithm}, reveals that this algorithm is entirely determined by (a)
the sequence of proposal kernels $(m_t)_{t\geq 0}$ (which determine how particles are simulated) 
and (b) the sequence of weight functions $(G_t)_{t\geq 0}$ (which determine how particles are weighted). Our introduction to particle filtering focussed on 
the specific expression \eqref{eq:defGt} for $G_t$, but useful
SMC algorithms may be obtained by considering other weight functions; see e.g. the auxiliary particle filter of \cite{PittShep}, 
as explained in \cite{johansen2008note}, or the SMC algorithms for non-sequential
problems mentioned in the introduction. 

The exact expression and meaning of $m_t$ and $G_t$ will not play a particular
role in the rest of the paper, so it is best to think of SMC from now on 
as a generic algorithm, again based on 
a certain sequence $(m_t)$, $m_0(\dx_0)$ being an initial distribution, and 
$m_t(\bx_{t-1},\dx_t)$ being a Markov kernel for $t\geq 1$, and a certain 
sequence of functions, $G_0:\setX\rightarrow\Rp$, $G_t:\setX\times\setX\rightarrow \Rp$, which produces the following consistent (as $N\rightarrow +\infty$) 
estimators: 
$$
\frac{1}{N}\sum_{n=1}^{N}\varphi(\bx_{t}^{n})  \rightarrow  \Qb_{t}(\varphi),
\qquad
\sum_{n=1}^{N}W_{t}^{n}\varphi(\bx_{t}^{n}) \rightarrow  \Q_{t}(\varphi),
$$
where $\varphi:\setX\rightarrow\R$, and  $\Qb_{t}$ and
$\Q_{t}$ are defined as follows:
\begin{eqnarray}
Z_{t} & = & \E\left[G_{0}(\bx_{0})\prod_{s=1}^{t}G_{s}(\bx_{s-1},\bx_{s})\right],\label{eq:def_Zt}\\
\Qb_{t}(\varphi) & = & \frac{1}{Z_{t-1}}\E\left[\varphi(\bx_{t})G_{0}(\bx_{0})\prod_{s=1}^{t-1}G_{s}(\bx_{s-1},\bx_{s})\right],\label{eq:def_Qbt}\\
\Q_{t}(\varphi) & = & \frac{1}{Z_{t}}\E\left[\varphi(\bx_{t})G_{0}(\bx_{0})\prod_{s=1}^{t}G_{s}(\bx_{s-1},\bx_{s})\right],\label{eq:def_Qt}
\end{eqnarray}
with expectations taken with respect to the law of the non-homogeneous Markov chain
$(\bx_{t})$, e.g. 
\[
Z_{t}=\int_{\setX^{t+1}}\left\{ G_{0}(\bx_{0})\prod_{s=1}^{t}G_{s}(\bx_{s-1},\bx_{s})\right\} \, m_{0}(\dx_{0})\prod_{s=1}^{t}m_{s}(\bx_{s-1},\dx_{s}),
\]
and with the conventions that $Z_{-1}=1$ and empty products equal
one; e.g. $\Qb_{0}(\varphi)=m_{0}(\varphi)$.

For instance, for the standard filtering problem covered in our introduction, 
where $G_t$ is set to \eqref{eq:defGt},  $\Q_t(\varphi)$ is the filtering expectation
of $\varphi$, i.e. $\E[\varphi(\bx_t)|\by_{0:t}]$, and $\Qb_t(\varphi)$ is 
the predictive distribution of $\varphi$, i.e. $\E[\varphi(\bx_t)|\by_{0:t-1}]$.


\subsection{Towards SQMC: SMC as a sequence of importance sampling steps}

QMC requires to write any simulation as an explicit function
of uniform variates. We therefore make the following assumption for our
generic SMC sampler: 
to generate $\bx_0^n\sim m_0(\dx_0)$, one computes  $\bx_{0}^{n}=\Gamma_{0}(\bu_{0}^{n})$, and to generate $\bx_t^n|\bx_{t-1}^n\sim m_t(\bx_{t-1},\dx_t)$,
one computes $\bx_{t}^{n}=\Gamma_{t}(\bx_{t-1}^{n},\bu_{t}^{n})$,
where $\bu_t^n\sim \Unif(\uid)$, and the functions $\Gamma_t$ are easy to evaluate.


Iteration $0$ of Algorithm \ref{alg:Generic-SMC-algorithm} amounts
to an importance sampling step, from $m_{0}(\dx_{0})$ to $\Q_{0}(\dx_{0})=m_{0}(\dx_{0})G_{0}(\bx_{0})/Z_{0}$,
which produces the following estimator 
\[
\sum_{n=1}^{N}W_{0}^{n}\varphi(\bx_{0}^{n})=\frac{\sum_{n=1}^{N}G_{0}(\bx_{0}^{n})\varphi(\bx_{0}^{n})}{\sum_{m=1}^{N}G_{0}(\bx_{0}^{m})}
\]
of $\Q_{0}(\varphi$). To introduce QMC at this stage, we take $\bx_{0}^{n}=\Gamma_{0}(\bu_{0}^{n})$
where $\bu_{0}^{1:N}$ is a low-discrepancy point set in $\ui^{d}$.


The key remark that underpins SQMC is that iteration $t\geq1$ of
Algorithm \ref{alg:Generic-SMC-algorithm} also amounts to an importance
sampling step, but this time from 
\begin{equation}
\Qb_{t}^{N}(\dd(\widetilde{\bx}_{t-1},\bx_{t}))=\sum_{n=1}^{N}W_{t-1}^{n}\delta_{\bx_{t-1}^{n}}(\dd\widetilde{\bx}_{t-1})m_{t}(\bx_{t-1}^{n},\dx_{t})\label{eq:Q_proposal}
\end{equation}
to 
\[
\Q_{t}^{N}(\dd(\widetilde{\bx}_{t-1},\bx_{t}))=\frac{1}{\Qb_{t}^{N}(G_{t})}\Qb_{t}^{N}(\dd(\widetilde{\bx}_{t-1},\bx_{t}))G_{t}(\widetilde{\bx}_{t-1},\bx_{t})
\]
where $\Q_{t}^{N}$ and $\Qb_{t}^{N}$ are
two \emph{random} probability measures defined over $\setX\times\setX$,
a set of dimension $2d$. 
In particular, the generation of random
variables $a_{t-1}^{1:N}$ and $\bx_{t}^{1:N}$ in Steps (a) and (b)
of Algorithm \ref{alg:Generic-SMC-algorithm} is equivalent to sampling
$N$ times independently random variables $(\tilde{\bx}_{t-1}^{n},\bx_{t}^{n})$
from $\Qb_{t}^{N}(\dd(\widetilde{\bx}_{t-1},\bx_{t}))$: i.e. $\tilde{\bx}_{t-1}^{n}=\bx_{t-1}^{a_{t-1}^{n}}$
(not to be mistaken with $\bx_{t-1}^{n}$), and $\bx_{t}^{n}\sim m_{t}(\tilde{\bx}_{t-1}^{n},\dx_{t})$.

Based on these remarks, the general idea behind SQMC is to replace
at iteration $t$ the $N$ IID random numbers sampled from $\Qb_{t}^{N}(\dd(\widetilde{\bx}_{t-1},\bx_{t}))$
by a low-discrepancy point set relative to the same distribution.

When $d=1$, this idea may be implemented as follows: generate a low-discrepancy
point set $\bu_{t}^{1:N}$ in $\ui^{2}$, let $\bu_{t}^{n}=(u_{t}^{n},v_{t}^{n})$,
then set $\tilde{\bx}_{t-1}^{n}=\hat{F}_{N}^{-1}(u_{t}^{n})$, $\bx_{t}^{n}=\Gamma_{t}(\tilde{\bx}_{t-1}^{n},v_{t}^{n})$,
where $\hat{F}_{N}^{-1}$ is the generalised inverse of the empirical
CDF 
\[
\hat{F}_{N}(x)=\sum_{n=1}^{N}W_{t-1}^{n}\ind\left\{ \bx_{t-1}^{n}\leq x\right\} ,\quad x\in\setX\subset\R.
\]

It is easy to see that the most efficient way to compute $\tilde{\bx}_{t-1}^{n}=\hat{F}_{N}^{-1}(u_{t}^{n})$
for all $n\in 1{:}N$ is (a) to sort the $\bx_{t-1}^{n}$'s, i.e. to
find permutation $\sigma$ such that $\bx_{t-1}^{\sigma(1)}\leq\ldots\leq \bx_{t-1}^{\sigma(N)}$,
(b) to sort the $u_{t}^{1:N}$, call $\mathrm{sort}(u_{t}^{1:N})$
the corresponding result; (c) to obtain $a_{t-1}^{1:N}$ as the output
of Algorithm \ref{alg:inverse_method}, with inputs $\mathrm{sort}(u_{t}^{1:N})$
and $W_{t}^{\sigma(1:N)}$; and finally (d) set $\tilde{\bx}_{t-1}^{n}=\bx_{t-1}^{a_{t-1}^{n}}$.

Algorithm \ref{alg:SQMC} gives a pseudo-code version of SQMC for
any $d$, but note how Step (b) at times $t\geq1$ simplifies to what
we have just described for $d=1$. 

When $d>1$, the inverse transform method cannot be used to sample
from the marginal distribution of $\widetilde{\bx}_{t-1}$ relative to $\Qb_{t}^{N}(\dd(\widetilde{\bx}_{t-1},\bx_{t}))$,
at least unless the $\bx_{t-1}^{n}$ are ``projected'' to the real
line in some sense. This is the point of the Hilbert curve presented
in the next section.

\subsection{The Hilbert space-filling curve\label{sub:The-Hilbert-space-filling}}

\begin{figure}
\begin{centering}
\includegraphics[scale=0.1]{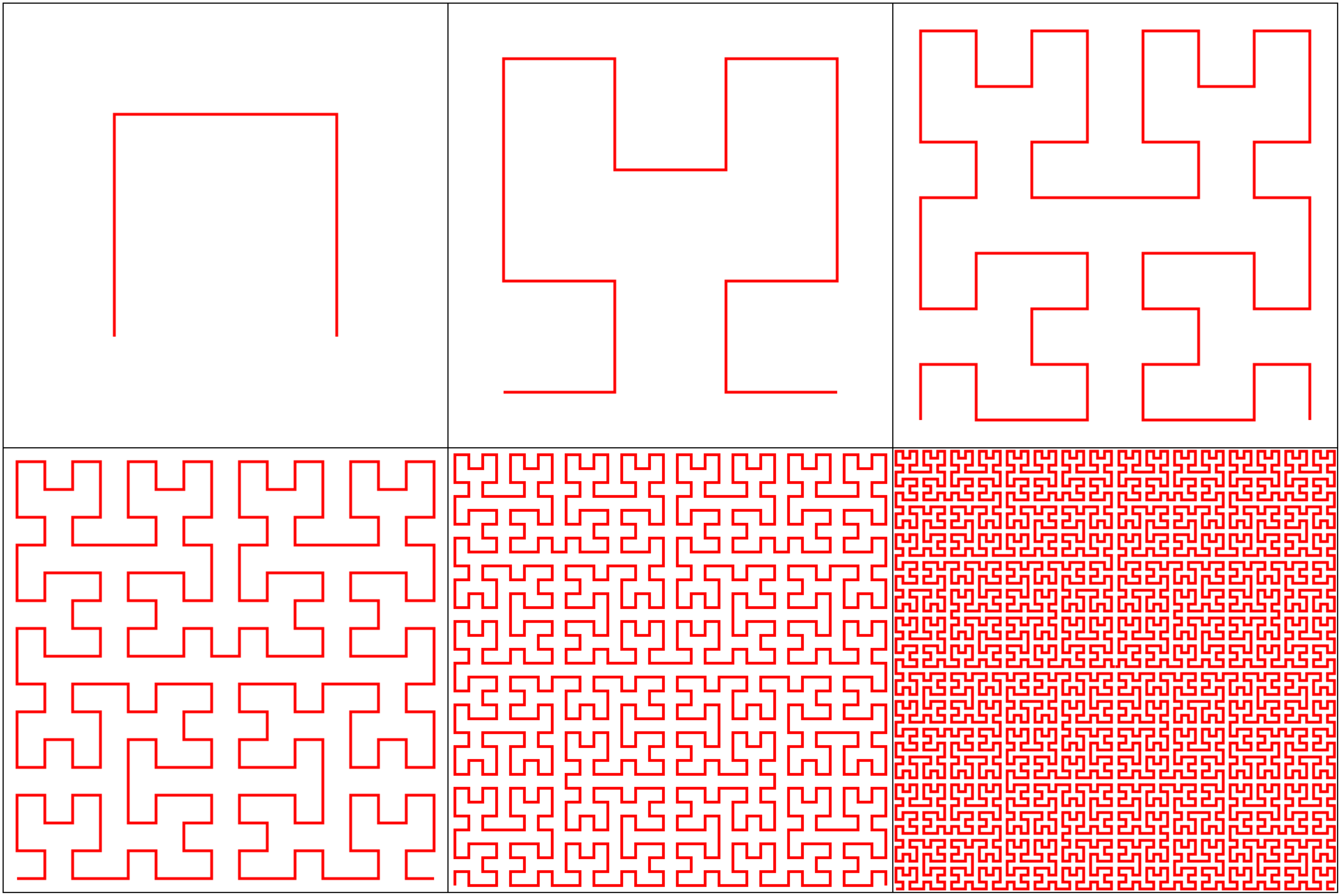} 
\par\end{centering}

\caption{\label{fig:Hilbertcurve}$\HSFC_{m}$ curve for $d=2$ and $m=1$
to $m=6$ (Source: Wikipedia)}
\end{figure}

The Hilbert curve is a continuous fractal map $\HSFC:[0,1]\rightarrow[0,1]^{d},$
which ``fills'' entirely $[0,1]^{d}$.
$H$ is obtained as the limit of a sequence $(\HSFC_{m})$, $m\rightarrow+\infty$,
the first terms of which are depicted in Figure \ref{fig:Hilbertcurve}.

The function $H$ admits a pseudo-inverse $h:[0,1]^d \rightarrow [0,1]$, i.e. 
$H\circ h(\bx)=\bx$  for all $\bx\in [0,1]^d$. $H$ is not a bijection because
certain points $\bx\in [0,1]^d$ have more than one pre-image
through $H$; however the set of such points is of Lebesgue measure 0. 


Informally, $H$ transforms $[0,1]$ into $[0,1]^d$, while preserving ``locality'': if $x$, $x'\in[0,1]$
are close, then $H(x)$ and $H(x')$ are close
as well. We will establish that $h$ also preserves discrepancy: a
low-discrepancy point set in $\uid$ remains a low-discrepancy
point set in $\ui$ when transformed through $h$. It is these
properties that give to the Hilbert curve its appeal in the SQMC context
(as opposed to other space filling curves, such as Z-ordering). We refer to \citet{Sagan1994}, 
\citet{Butz1969} and \citet{Hamilton2008}
for how to compute  $h$ in practice for any $d\geq 2$. 
For $d=1$, we simply set $H(x)=h(x)=x$ for $x\in[0,1]$. 

The following technical properties of $\HSFC$ and $\HSFC_{m}$ will
be useful later (but may be skipped on first reading). For $m\geq0$, let $\mathcal{I}_{m}^{d}=\left\{ I_{m}^{d}(k)\right\} _{k=0}^{2^{md}-1}$
be the collection of consecutive closed intervals in $[0,1]$ 
of equal size $2^{-md}$ and such that $\cup \mathcal{I}_{m}^{d}=[0,1]$. For $k\geq0$, $S_{m}^{d}(k)=\HSFC_{m}(I_{m}^{d}(k))$
belongs to $\mathcal{S}_m^d$, the set of the $2^{md}$ closed hypercubes of volume $2^{-md}$ that covers $[0,1]^d$, $\cup \mathcal{S}_m^d=[0,1]^d$; $S_{m}^{d}(k)$ and $S_{m}^{d}(k+1)$ are adjacent,
i.e. have at least one edge in common (\emph{adjacency property}).
If we split $I_{m}^{d}(k)$ into the $2^{d}$ successive closed intervals
$I_{m+1}^{d}(k_{i})$, $k_{i}=2^{d}k+i$ and $i\in 0:2^{d}-1$, then
the $S_{m+1}^{d}(k_{i})$'s are simply the splitting of $S_{m}^{d}(k)$
into $2^{d}$ closed hypercubes of volume $2^{-d(m+1)}$ (\emph{nesting property}). Finally, the limit $\HSFC$ of $\HSFC_{m}$ has the \emph{bi-measure
property}: $\lambda_{1}(A)=\lambda_{d}(\HSFC(A))$ for any measurable
set $A\subset[0,1]$, and satisfies the \textit{H\"{o}lder condition} $\|H(x_1)-H(x_2)\|_{\infty}\leq C_H |x_1-x_2|^{1/d}$ for any $x_1,\,x_2\in[0,1]$.


\subsection{SQMC for $d\geq2$}

Assume now $d\geq2$, and consider the following change of variables
at iteration $t\geq1$: 
\[
h_{t-1}^{n}=h\circ\psi(\bx_{t-1}^{n})\in [0,1]
\]
where $h:[0,1]^{d}\rightarrow[0,1]$ is the inverse of the Hilbert
curve defined in the previous section, and $\psi:\setX\rightarrow[0,1]^{d}$
is some user-chosen bijection between $\setX$ and $\psi(\setX)\subset[0,1]^{d}$. To preserve the low discrepancy property of  $\bx_{t-1}^{1:N}$ it is important to choose for $\psi$ a mapping which is discrepancy preserving.
This requires to select $\psi$ such that $\psi(\bx)=\left(\psi_1(x_1),...,\psi_d(x_d)\right)$ where the $\psi_i$'s are continuous and strictly monotone. But choosing such a $\psi$ is trivial in most applications; e.g. apply the logistic transformation component-wise when $\setX=\R^{d}$ (see Section \ref{sec:Numerical-study} for more details). 

With this change of variables, we obtain particles $h_{t-1}^{1:N}$
that lie in $[0,1]$, and \eqref{eq:Q_proposal} becomes 
\[
\Qb_{t,h}^N\left(\dd(\tilde{h}_{t-1},\bx_{t})\right)=\sum_{n=1}^{N}W_{t-1}^{n}\delta_{h_{t-1}^{n}}(\dd\tilde{h}_{t-1})m_{t}(\bx_{t-1}^{n},\dx_{t}).
\]

Sampling a low-discrepancy sequence from $\Qb_{t,h}^N(\dd \tilde{h}_{t-1},\dx_{t})$
may then proceed exactly as for $d=1$; that is: use the inverse transform
method to sample $N$ points $\tilde{h}_{t-1}^{1:N}$ from the
marginal distribution $\Qb_{t,h}^N(\dd \tilde{h}_{t-1})$,
then sample $\bx_{t}^{1:N}$ conditionally on $\tilde{\bx}_{t-1}^{1:N}$,
with $\tilde{\bx}^n_{t-1}=\psi^{-1}\circ H(\tilde{h}_{t-1}^{n})$. The
exact details of the corresponding operations are the same as for
$d=1$. We therefore obtain the general SQMC algorithm as described
in Algorithm \ref{alg:SQMC}.

\begin{algorithm}[t]
\medskip
At time $t=0$, 
\begin{description}
\item [{(a)}] Generate a QMC or a RQMC point set $\bu_{0}^{1:N}$ in $\ui^{d}$,
and compute $\bx_{0}^{n}=\Gamma_{0}(\bu_{0}^{n})$ for each $n\in \onetoN$.
\item [{(b)}] Compute 
$w_{0}^{n}=G_{0}(\bx_{0}^{n})$ 
and $W_0^n = w_0^n /\sum_{m=1}^N w_0^m$ 
 for each $n\in \onetoN$. 

\end{description}
Iteratively, from time $t=1$ to time $t=T$, 
\begin{description}
\item [{(a)}] Generate a QMC or a RQMC point set $\bu_{t}^{1:N}$ in $\ui^{d+1}$;
let $\bu_{t}^{n}=(u_{t}^{n},\bv_t^n)\in\ui\times\uid$. 
\item [{(b)}] Hilbert sort: find permutation $\sigma_{t-1}$ such that $\IHSFC\circ\psi(\bx_{t-1}^{\sigma_{t-1}(1)})
\leq\ldots\leq\IHSFC\circ\psi(\bx_{t-1}^{\sigma_{t-1}(N)})$
if $d\geq2$, or $\bx_{t-1}^{\sigma_{t-1}(1)}\leq\ldots\leq\bx_{t-1}^{\sigma_{t-1}(N)}$
if $d=1$. 
\item [{(c)}] Find permutation $\tau$ such that $u_t^{\tau(1)}\leq ...\leq u_t^{\tau(N)}$, generate $a_{t-1}^{1:N}$ using Algorithm \ref{alg:inverse_method},
with inputs $u_{t}^{\tau(1:N)}$ and $W_{t-1}^{\sigma_{t-1}(1:N)}$,
and compute  $\bx_{t}^{n}=\Gamma_{t}(\bx_{t-1}^{\sigma_{t-1}(a_{t-1}^{n})},\bv_{t}^{\tau(n)})$ for each $n\in \onetoN$. 
\item [{(e)}] Compute
$w_t^n=G_{t}(\bx_{t-1}^{\sigma_{t-1}(a_{t-1}^{n})},\bx_{t}^{n})$,
and $W_t^n = w_t^n/\sum_{m=1}^N w_t^m$ 
for each $n\in \onetoN$.
\end{description}
\caption{\label{alg:SQMC} SQMC algorithm}
\end{algorithm}

To fully define SQMC, one must choose a particular method to generate
point sets $\bu_t^{1:N}$ at each iteration. If QMC point sets are generated,
one obtains a deterministic algorithm, while if RQMC point sets are generated,
one obtains a stochastic algorithm.

\subsection{Complexity of SQMC}

The complexity of both Steps (b) and (c) (for $t\geq1$) of the SQMC
algorithm is $\bigO(N\log N)$, because they include a sort operation.
The complexity of Step (a) depends on the chosen method for generating
the  point sets $\bu_{t}^{1:N}$. For instance, \citet{Hong2003}  propose a
$\bigO(N\log N)$ method that applies to most constructions of $(t,s)$-sequences (such as the Faure, the Sobol', the Niederreiter or the Niederreiter-Xing sequences). 
The cost to randomize a QMC point set  is only $\bigO(N)$ if one chooses the simple random shift approach,  while nested scrambling methods for $(t,s)$-sequences, which are such that all the results below hold, may be implemented at cost $\bigO(N\log N)$ \citep{Owen1995, Hong2003}.


To summarise, the overall complexity of SQMC is $\bigO(N\log N)$, provided the method
to generate the point sets $\bu_t^{1:N}$ is chosen appropriately.

\section{Convergence study}
\label{sec:Convergence-results}

We concentrate on two types of asymptotic results (as $N\rightarrow+\infty$):
consistency, and stochastic bounds, that is bounds on the mean square
error for the randomized SQMC algorithm (i.e. SQMC based on randomized
QMC point sets). We leave deterministic bounds of the error 
(for when deterministic QMC point sets are used)
to future work. We find stochastic bounds more interesting, because 
(a) results from \citep{Owen1997a, Owen1997b, Owen1998} suggest
one might obtain better convergence rates than 
for deterministic bounds; and (b) 
the randomized version of SQMC has more applications, as discussed in
Section \ref{sec:Extensions}.

These results are specialised to the case  where 
the simulation of $\bx_{t}^{n}$ at time $t$ is based on the
inverse transform method, as explained in Section \ref{sub:Inverse-transform-method}.
Certain of our results require $\setX$ to be bounded, so for simplicity
we take $\setX=\uid$, and $\psi$ is set to the identity function.
(Recall that, 
to deal with certain QMC technicalities, we follow the standard practice
of taking $\setX=\uid$ rather than $\setX=[0,1]^{d}$.) The fact
that $\setX$ is bounded may not be such a strong restriction, as
our results allow for unbounded test functions $\varphi$; thus, one
may accommodate for an unbounded state space (and expectations with
respect to that space) through appropriate variable transforms. 


We introduce the following extreme norm. For any signed
measure $\mu$ over $\setX=\uid$, 
\[
\|\mu\stn=\sup_{B\in\mathcal{B}_{\uid}}\left|\mu(B)\right|,\quad\mathcal{B}_{\uid}=\left\{ B:\: B=\prod_{i=1}^{d}[a_i,b_{i}]\subset\ui^d,\: a_i<b_{i}\right\} 
\]
which generalises the extreme discrepancy in the following sense: 
\[
\|\Sop(\bx^{1:N})-\lambda_{d}\stn=D(\bx^{1:N})
\]
for any point set $\bx^{1:N}$ in $\setX$, where $\Sop$ is the operator
that associates to $\bx^{1:N}$ its empirical distribution: 
\[
\bx^{1:N}\in\setX^{N}\mapsto\Sop(\bx^{1:N})=\frac{1}{N}\sum_{n=1}^{N}\delta_{\bx^{n}}.
\]
Our consistency results will be stated with this norm. Note that $\|\pi^N-\pi\stn\rightarrow0$
implies $\left|\pi^N(\varphi)-\pi(\varphi)\right|\rightarrow0$ for
any continuous, bounded function $\varphi$, 
by portmanteau lemma \citep[][Lemma 2.2]{VanderVaart2007}.

The next subsection explains how the inverse method may be used to generate
$\bx_t^n$ given $\bx_{t-1}^n$.
The two following subsections state preliminary results that should provide
insights on the main ideas that underpin the proofs of our convergence results.  Readers interested mostly in the 
main results may skip these subsections and go directly to 
Section \ref{sub:consistency} (consistency) and Section \ref{sub:stobounds} (stochastic bounds).

This section will use the following standard notations: $\|\varphi\|_{\infty}$
for the supremum norm for functions $\varphi$, $L_{2}(\setX,\mu)$
for the set of square integrable functions $\varphi:\setX\rightarrow\R$ and
$\contb(\setX)$
for the set of continuous, bounded functions $\varphi:\setX\rightarrow\R$.

\subsection{Inverse transform method\label{sub:Inverse-transform-method}}

We discuss here how to write
the simulation of $\bx_{t}^{n}$ as $\bx_{t}^{n}=\Gamma_{t}(\bx_{t-1}^{n},\bu_{t}^{n})$,
using the inverse transform method. Our convergence results are specialised to this particular $\Gamma_t$.

For a generic distribution $\pi\in\mathcal{P}(\setX)$, $\setX\subset\R^{d}$,
let $F_{\pi}$ be the Rosenblatt transformation \citep{Rosenblatt1952} of $\pi$  defined through the following
chain rule decomposition: 
\[
F_{\pi}(\bx)=\left(u_{1},\ldots,u_{d}\right)^T,\quad\bx=(x_{1},\ldots,x_{d})^T\in\setX,
\]
where, recursively, $u_{1}=F_{\pi,1}(x_{1}),$ $F_{\pi,1}$
being the CDF of the marginal distribution of the first component
(relative to $\pi$), and for $i\geq2$, $u_{i}=F_{\pi,i}(x_{i}|x_{1:i-1})$,
$F_{\pi,i}(\cdot|x_{1:i-1})$ being the CDF of component $x_{i}$,
conditional on $(x_{1},\ldots,x_{i-1}$), again relative to $\pi$. Similarly, we define the multivariate
GICDF (generalised inverse CDF) $F_{\pi}^{-1}$ through the following
chain rule decomposition: 
\[
F_{\pi}^{-1}(\bu)=\left(x_{1},\ldots,x_{d}\right)^T,\quad\bu=(u_{1},\ldots,u_{d})^T\in\ui^{d},
\]
where, recursively, $x_{1}=F_{\pi,1}^{-1}(u_{1}),$ $F_{\pi,1}^{-1}$
being the GICDF of the marginal distribution of the first component
(relative to $\pi$), and for $i\geq2$, $x_{i}=F_{\pi,i}^{-1}(u_{i}|x_{1:i-1})$,
$F_{\pi,i}^{-1}(\cdot|x_{1:i-1})$ being the GICDF of component $x_{i}$,
conditional on $(x_{1},\ldots,x_{i-1}$), again relative to $\pi$.
Note that this function depends on the particular order of the components
of $\pi$. For some probability kernel $K:\setX\rightarrow\mathcal{P}(\setX)$,
define similarly $F_K(\bx,\cdot)$ and $F_{K}^{-1}(\bx,\cdot)$ as, respectively, the Rosenblatt transformation and the multivariate GICDF
of distribution $K(\bx,\dx')$ for a fixed $\bx$. 

It is well known that taking $\Gamma_{0}=F_{m_{0}}^{-1}$, and $\Gamma_{t}=F_{m_{t}}^{-1}$
lead to valid simulations algorithms, i.e. if $\bx_{0}^{n}=F_{m_{0}}^{-1}(\bu_{0}^{n})$,
resp. $\bx_{t}^{n}=F_{m_{t}}^{-1}(\bx_{t-1},\bu_{t}^{n})$, then $\bx_{0}^{n}\sim m_{0}(\dx_{0})$,
resp. $\bx_{t}^{n}|\bx_{t-1}^{n}\sim m_{t}(\bx_{t-1}^{n},\dx_{t})$.

\subsection{Preliminary results: importance sampling\label{sub:Preliminary-results:-importance}}

Since SQMC is based on importance sampling (e.g. Iteration
$0$ of Algorithm \ref{alg:SQMC}), we need to establish the validity
of importance sampling based on low-discrepancy point sets;
see  \citet{Gotz2002,Aistleitner2014} for other results on QMC-based importance sampling.


\begin{thm}
\label{thm:IS_1} Let $\pi$ and $q$ be two probability
measures on $\ui^d$ such that the \RN derivative $w(\bx)=\pi(\dx)/q(\dx)$
is continuous and bounded. Let $(\bx^{1:N})$ be a  sequence of point sets
in $\ui^d$ such that $\|\Sop(\bx^{1:N})-q\stn\rightarrow0$
as $N\rightarrow+\infty$, and define 
\[
\pi^N=\sum_{n=1}^{N}W^{n}\delta_{\bx^{n}},\quad W^{n}=\frac{w(\bx^{n})}{\sum_{m=1}^{N}w(\bx^{m})}.
\]
Then, $\|\pi^N-\pi\stn\rightarrow0$ as $N\rightarrow+\infty$.

\end{thm}
See Section \ref{sub:Proof-IS1} of the Appendix for a proof. 

Recall
that in our notations we drop the dependence of point sets on $N$,
i.e. we write $(\bx^{1:N})$ rather than $(\bx^{N,1:N})$, although
in full generality $\bx^{1:N}$ may not necessarily be the $N$ first
points of a fixed sequence.

The next theorem gives the stochastic error rate when a RQMC point
set is used.
\begin{thm}
\label{thm:IS_2} Consider the set-up of Theorem \ref{thm:IS_1}.
Let $(\bu^{1:N})$ be a  sequence of random point sets in
$[0,1)^{d}$ such that $\bu^{n}\sim\Unif(\ui^{d})$ marginally and,
$\forall\varphi\in L_{2}(\uid,\lambda_{d})$, 
\[
\var\left(\frac{1}{N}\sum_{n=1}^{N}\varphi(\bu^{n})\right)=\smallo\Big(r(N)\Big),
\]
where $r(N)\rightarrow 0$ as $N\rightarrow+\infty$. Let $\bx^{1:N}=F_{q}^{-1}(\bu^{1:N})$  and assume that
either one of the following two conditions is verified: 
\begin{enumerate}
\item \label{H:thmIS_2:1} $F_{q}^{-1}$ is continuous and, for any $\epsilon>0$,
there exists a $N_{\epsilon}\in\mathbb{N}$ such that, almost surely,
$\Dst(\bu^{1:N})\leq\epsilon$, $\forall N\geq N_{\epsilon}$; 
\item \label{H:thmIS_2:2} for any $\epsilon>0$ there exists a $N_{\epsilon}\in\mathbb{N}$
such that, almost surely, 
\[
\|\Sop(\bx^{1:N})-q\stn\leq\epsilon,\quad\forall N\geq N_{\epsilon}.
\]

\end{enumerate}
Then, for all $\varphi\in L_{2}(\setX,\pi)$, 
\[
\E\left|\pi^N(\varphi)-\pi(\varphi)\right|=\smallo\Big(r(N)^{1/2}\Big),\quad\var\left(\pi^N(\varphi)\right)=\smallo\Big(r(N)\Big).
\]

\end{thm}
See Section \ref{sub:Proof-IS1} of the Appendix for a proof.

To fix ideas, note that several RQMC strategies reach the Monte Carlo error rate and therefore fulfil the assumptions
above with $r(N)=N^{-1+\epsilon}$ for any $\epsilon>0$ \citep[see e.g.][]{Owen1997a,Owen1998}. In addition,  nested scrambling methods for  $(t,s)$-sequences in base $b$ \citep{Owen1995, Matousek1998,Hong2003}   are such that  $r(N)=N^{-1}$. This result is established  for $N=\lambda b^{m}$ in \citet{Owen1997a,Owen1998} and extended for an arbitrary  $N$ in \citet{Gerber:QMCarbitrarysize}.

\subsection{Preliminary results: Hilbert curve and discrepancy}

We motivated the use of the Hilbert curve as a way to
transform back and forth between $[0,1]^{d}$ and $[0,1]$ while preserving
low discrepancy in some sense. This section formalises this idea.

For a probability measure $\pi$ on $\uid$, we write $\pi_{\IHSFC}$
the image by $\IHSFC$ of $\pi$.  For a kernel $K:\uid\rightarrow\mathcal{P}(\setX)$,
we write $\pi_{h}\otimes K_h\left(\dd(h_1,\bx_{2})\right)$ the image
of $\pi\otimes K$ by the mapping $(\bx_{1},\bx_{2})\in\uid\times\setX\mapsto(h(\bx_{1}),\bx_{2})$,
where $\pi\otimes K$ denotes the joint probability measure $\pi(\dx_{1})K(\bx_{1},\dx_{2})$. 

The  following  theorem is a technical result on the conversion
of discrepancy through $h$.

\begin{thm}
\label{thm:Hilbert} Let $(\pi^N)$ be a sequence of probability
measure on $[0,1)^{d}$ such that, $\|\pi^N-\pi\stn\cvz$, where
$\pi(\dx)=\pi(\bx)\lambda_{d}(\dx)$ admits a bounded probability density $\pi(\bx)$.
Then 
\[
\|\pi_h^N-\pi_h\stn\rightarrow0,\quad\mbox{as }N\rightarrow+\infty.
\]

\end{thm}
See Section \ref{sub:Proof-hilbert1} of the Appendix for a proof.

%


The following theorem is an extension of \citet[][``Satz 2'']{Hlawka1972}, which establishes the validity, in the context of QMC, of the multivariate GICDF approach described in Section \ref{sub:Inverse-transform-method}. More precisely, for a probability measure $\pi$  on $\ui^d$, \citet[][``Satz 2'']{Hlawka1972} show that  $\|\Sop\left(F^{-1}_{\pi}(\bu^{1:N})\right)-\pi\stn\leq c \Dst(\bu^{1:N})^{1/d}$ (under some conditions on $F_{\pi}$, see below). 

\begin{thm}
\label{thm:LD} Let  $K:\ui^{d_{1}}\rightarrow\mathcal{P}\left(\ui^{d_{2}}\right)$ be a Markov
kernel and assume that: 
\begin{enumerate}
\item\label{H:thmLD:1} For a fixed $\bx_1\in\ui^{d_1}$, the $i$-th coordinate of $F_{K}\left(\bx_{1},\bx_{2}\right)$ is strictly increasing in $x_{2i}\in\ui$, $i\in 1:d_{2}$, and,
viewed as a function of $\bx_{1}$ and $\bx_{2}$,
$F_{K}\left(\bx_{1},\bx_{2}\right)$ is Lipschitz;
\item \label{H:thmLD:2} $\pi^N(\dx)=\sum_{n=1}^{N}W_{N}^{n}\delta_{\bx_1^n}(\dx)$, $\bx_1^n\neq \bx_1^m$ $\forall n\neq m\in 1:N$, and 
$\max_{n\in 1:N}W_N^n\cvz$.
\item \label{H:thmLD:3} The sequence $(\pi^N)$ is such that $\|\pi^N-\pi\stn\cvz$
as $N\rightarrow+\infty$, where $\pi(\dx)=\pi(\bx)\lambda_{d_1}(\dx)$ admits a strictly positive
bounded density $\pi$. 
\end{enumerate}
Let $(\bu^{1:N})$, $\bu^{n}=(u^{n},\bv^{n})\in\ui^{1+d_2}$, be a sequence of point sets in $[0,1)^{1+d_{2}}$
such that $\Dst(\bu^{1:N})\rightarrow0$ as $N\rightarrow+\infty$, and define $P_h^{N}=\left(h^{1:N},\bx_{2}^{1:N}\right)$ where
\[
h^{n}=F_{\pi_h^N}^{-1}(u^{n}),\quad\btx_{1}^{n}=\HSFC(h^{n}),\quad\bx_{2}^{n}=F_{K}^{-1}\left(\btx_{1}^{n},\bv^{n}\right).
\]
Then 
\[
\|\Sop(P_h^{N})-\pi_h^N\otimes K_h\stn\cvz,\quad\mbox{as }N\rightarrow+\infty.
\]
\end{thm}
See Section \ref{sub:Proof-hilbert3}  of the Appendix for a proof.

Assumption \ref{H:thmLD:1} regarding the regularity of the vector-valued function $F_K$ is the main assumption of the above theorem and comes from \citet[][``Satz 2'']{Hlawka1972}.
It is verified as soon as kernel $K$ admits a density
that is continuously differentiable on $[0,1)^{d}$ \citep[][p.232]{Hlawka1972}. 
Assumption \ref{H:thmLD:2} is a technical condition, which will always hold under the assumptions
of our main results.

\subsection{Consistency}\label{sub:consistency}

We are now able to establish the consistency of SQMC; see Appendix
\ref{sub:app_consistency} for a proof of the following theorem. 
 For convenience, let $F_{m_t}(\bx_{t-1},\bx_t)=F_{m_0}(\bx_0)$  when $t=0$.

\begin{thm}
\label{thm:consistency} Consider the set-up of Algorithm \ref{alg:SQMC}
where, for all $t\in0:T$, $(\bu_{t}^{1:N})$ is a (non random) sequence of point sets in
$\ui^{d_{t}}$, with $d_{0}=d$ and $d_{t}=d+1$ for $t>0$,
such that $\Dst(\bu_t^{1:N})\rightarrow 0$ as $N\rightarrow +\infty$. 
Assume the following holds for all $t\in 0{:}T$:
\begin{enumerate}
\item \label{H:thmPF1:1} The components of $\bx_t^{1:N}$ are pairwise distinct, $\bx_t^n\neq \bx_t^m$ for $n\neq m$. 
\item \label{H:thmPF1:2} $G_{t}$ is continuous and bounded; 
\item \label{H:thmPF1:3} $F_{m_{t}}(\bx_{t-1},\bx_{t})$ verifies  Assumption \ref{H:thmLD:1} of Theorem \ref{thm:LD} ;  
\item \label{H:thmPF1:4} $\Q_{t}(\dx_{t})=p_{t}(\bx_{t})\lambda_d(\dx_{t})$ where
$p_{t}(\bx_{t})$ is a strictly positive bounded density. 
\end{enumerate}
Let $\Qh_{t}^{N}(\dx_t)=\sum_{n=1}^{N}W_{t}^{n}\delta_{\bx_{t}^{n}}(\dx_t)$.
Then, under Assumptions \ref{H:thmPF1:1}-\ref{H:thmPF1:4}, as $N\rightarrow+\infty$,
\[
\|\Qh_{t}^{N}-\Q_{t}\stn\cvz,\quad \forall t\in 0{:}T.
\]

\end{thm}

Assumption \ref{H:thmPF1:1} is stronger than necessary because for the result to hold it is enough that the number of identical particles  does not grow too quickly as $N\rightarrow +\infty$. Note that this is  a very weak restriction since Assumption \ref{H:thmPF1:1} holds almost surely  when RQMC point sets are used,
since then the particles are generated from a continuous GICDF. The assumption that the weight functions $(G_t)$ are bounded is standard in SMC literature \citep[see e.g.][]{DelMoral:book}.

\subsection{Stochastic bounds}\label{sub:stobounds}

Our second main result concerns stochastic bounds for the randomized
version of SQMC, i.e. SQMC based on randomized point sets $(\bu_{t}^{n})$. 
See Section \ref{sec:proof_stobounds} of the Appendix for a proof of
the next result. 
\begin{thm}
\label{thm:PF2} Consider the set-up of Algorithm \ref{alg:SQMC} 
where  $(\bu_{t}^{1:N})$, $t\in0:T$, are independent sequences of random point sets in
$\ui^{d_{t}}$, with $d_{0}=d$ and $d_{t}=d+1$ for $t>0$, such that, for all $t\in0:T$, $\bu_t^{n}\sim\Unif(\ui^{d_t})$ marginally and
\begin{enumerate}
\item \label{H:thmPF2:1} For any $\epsilon>0$, there exists a $N_{\epsilon,t}>0$
such that, almost surely, $\Dst(\bu_{t}^{1:N})\leq\epsilon$, $\forall N\geq N_{\epsilon,t}$. 
\item \label{H:thmPF2:2} For any function $\varphi\in L_{2}\left(\ui^{d_{t}},\lambda_{d_{t}}\right)$,
$
\var\left(\frac{1}{N}\sum_{n=1}^{N}\varphi(\bu_{t}^{n})\right)\leq C^*\,\sigma_{\varphi}^{2}r(N)
$
where  $\sigma_{\varphi}^{2}=\int\left\{ \varphi(\bu)-\int\varphi(\bv)\dd \bv\right\} ^{2}\du$,
and where both $C^*$ and $r(N)$ do not depend on $\varphi$. 
\end{enumerate}
In addition, assume that the  Assumptions of Theorem \ref{thm:consistency} are verified and that  $F_{m_0}^{-1}$ is continuous. Let $\varphi\in L_{2}(\ui^d,\Q_{t})$ for all $t\in 0{:}T$.
Then, $\forall t\in 0{:}T$, 
\[
\E\left|\Qh_{t}^{N}(\varphi)-\Q_{t}(\varphi)\right|=\bigO\Big(r(N)^{1/2}\Big),\quad 
\var\Big(\Qh_{t}^{N}(\varphi)\Big)=\bigO\Big(r(N)\Big).
\]

\end{thm}

Note that the implicit constants in the line above may depend on $\varphi$. 
Assumptions \ref{H:thmPF2:1} and \ref{H:thmPF2:2} are verified for $r(N)=N^{-1}$ if $\bu_t^{1:N}$ is the first $N$ points of a nested scrambled $(t,s)$-sequences in base $b\geq 2$. This result is established for $N=\lambda b^m$ in \citet{Owen1997a,Owen1998} and can be extended to any pattern of $N$ using \citet[][Lemma 1]{Hickernell2001}. Consequently, for this construction of RQMC point sets, Theorem \ref{thm:PF2}  shows that the approximation error of SQMC goes to zero at least as fast as for SMC. However, contrary to the $\bigO(N^{-1})$ convergence rate of SMC, this rate for SQMC  based on nested scrambled $(t,s)$-sequences is not exact but results from a worst case analysis. We can therefore expect to reach faster convergence on a smaller class of functions. The following result shows that it is indeed the case on the class on continuous and bounded functions; see Section  \ref{sec:proof_smallo} of the Appendix for a proof. 

\begin{thm}
\label{thm:PF2Bis} Consider the set-up of Algorithm \ref{alg:SQMC}
where  $(\bu_{t}^{1:N})$,  $t\in 0{:}T$,  are $(t,d_t)$-sequences in base $b\geq 2$, with $d_{0}=d$ and $d_{t}=d+1$ for $t>0$, independently scrambled such that results in \citet{Owen1997a,Owen1998} hold. Let $N=\lambda b^{m}$, $1\leq \lambda<b$, and assume the following holds:
\begin{enumerate}
\item\label{H:thmPF3:1} Assumptions of Theorem \eqref{thm:PF2} are verified;
\item\label{H:thmPF3:3} For $t\in 1{:}T$, $F^{-1}_{m_t}(\bx_{t-1},\bx_t)$ is a continuous function of $\bx_{t-1}$.
\end{enumerate}
Let $\varphi\in\mathcal{C}_{b}(\setX)$. Then, $\forall t\in 0{:}T$,
\[
\E|\Qh_{t}^{N}(\varphi)-\Q_{t}(\varphi)|=\smallo(N^{-1/2}),\quad\var(\Qh_{t}^{N}(\varphi))=\smallo(N^{-1}).
\]
\end{thm}

Thus, for SQMC based on the first $N=\lambda b^m$ points of nested scrambled $(t,s)$-sequences in base $b$, one obtains that the stochastic
error of (the random version of) SQMC converges faster than for SMC. Note that we can relax the constraint on $N$ in Theorem \ref{thm:PF2Bis} using  \citet[][Corollary 2]{Gerber:QMCarbitrarysize}.


\section{Extensions\label{sec:Extensions}}

\subsection{Unbiased estimation of evidence, PMCMC}\label{sub:evidence}

Like SMC, the randomized version of SQMC (that is SQMC based on RQMC
point sets) provides an unbiased estimator of the
normalising constant $Z_{t}$ of the Feynman-Kac model, see \eqref{eq:def_Zt}. 
\begin{lem}
Provided that $\bu_{t}^{1:N}$ is a RQMC point set in $\ui^{d_t}$ for
$t\in0:T$ (i.e. $\bu_t^n\sim\Unif(\ui^{d_t})$ marginally), with $d_0=d$ and $d_t=d+1$ for $t>0$, the following quantity 
\[
Z_t^N=\left\{\frac{1}{N} \sum_{n=1}^N G_0(\bx_0^n) \right\}
 \prod_{s=1}^t \left\{\frac{1}{N} 
\sum_{n=1}^N G_s(\bx_{s-1}^{a_{s-1}^n},\bx_s^n)\right\}
\]
is an unbiased estimator of $Z_t$, $\mathbb{E}[Z_t^N]=Z_t$. 
\end{lem}
We omit the proof, as it follows the same steps as for SMC \citep{DelMoral1996unbiased}.

In a state-space model parametrised by $\theta\in\Theta$, $Z_{t}=Z_t(\theta)$ is the marginal likelihood of the data up to time $t$. 
One may want to implement a Metropolis-Hastings sampler with respect
to posterior density $\pi_T(\theta)\propto p(\theta)Z_{T}(\theta)$
for the full dataset and for a prior distribution $p(\theta)$, but $Z_{T}(\theta)$
is typically intractable. 

\citet{PMCMC} established that,
by substituting $Z_{T}(\theta)$ with an unbiased estimate of $Z_{T}(\theta)$
in a Metropolis sampler, one obtains an exact MCMC (Markov chain Monte
Carlo) algorithm, in the sense that the corresponding MCMC kernel
leaves invariant $\pi_{T}(\theta)$. The so obtained algorithm is
called PMMH (Particle marginal Metropolis-Hastings). \citet{PMCMC}
use SMC to obtain an unbiased estimate of $Z_{T}(\theta$), that is,
at each iteration a SMC sampler is run to obtain that estimate.
We will call PMMH-SQMC the same algorithm, but with SQMC replacing SMC
for the evaluation of an unbiased estimate of the likelihood.

The acceptance rate of PMMH depends directly on the variability of
the estimates of $Z_{T}(\theta)$. Since the point of (randomized)
SQMC is to provide estimates with a lower variance than SMC (for a
given $N$), one may expect that PMMH-SQMC may require a smaller number
of particles than standard PMMH for satisfactory acceptance rates;
see Section \ref{sec:Numerical-study} for a numerical illustration
of this.

%
%
%
%

\subsection{Smoothing}

Smoothing amounts to compute expectations $\Q_t(\varphi)$ of functions
$\varphi$ of the complete trajectory $\bx_{0:t}$; e.g. $\Q_t(\varphi)$
is the expectation of $\varphi(\bx_{0:t})$ conditional on data $\by_{0:t}$
for a state-space model with Markov process $\left(\bx_{t}\right)$
and observed process $\left(\by_{t}\right)$. See \citet{briers2010smoothing}
for a general overview on SMC smoothing algorithms. This
section discusses how to adapt certain of these algorithms to
SQMC.

\subsubsection{Forward smoothing}\label{sect:Forward}

Forward smoothing amounts to carry forward the complete trajectories
of the particles, rather than simply keeping the last component $\bx_{t}^{n}$
(as in Algorithm \ref{alg:Generic-SMC-algorithm}). A simple way to
formalise forward smoothing is to introduce a path \FK model, corresponding
to the inhomogeneous Markov process $\bz_{t}=\bx_{0:t}$, and weight
function (abusing notations) $G_{t}(\bz_{t})=G_{t}(\bx_{t})$. Then
forward smoothing amounts to Algorithm \ref{alg:Generic-SMC-algorithm}
applied to this path \FK model (substituting $\bx_{t}$ with $\bz_{t}=\bx_{0:t}$).

One may use the same remark to define a SQMC version of forward smoothing:
i.e. simply apply SQMC to the same path \FK model. The only required
modification is that the Hilbert sort of Step (b) at times $t\geq1$
must now operate on some transformation of the vectors $\bz_{t}^{n}$,
of dimension $(t+1)d$, rather than vectors $\bx_{t}^{n}$ of dimension
$d$ as in the original version. 

Forward smoothing is sometimes used to approximate the smoothing expectation
of additive functions, $\varphi(\bx_{0:t})=\sum_{s=0}^{t}\tilde{\varphi}(\bx_{s})$,
such as the score function of certain models \citep[e.g. ][]{Poyiadjis2011}.
In that case, one may instead apply SQMC to the \FK model corresponding
to the inhomogeneous Markov process $\bz_{t}=(\sum_{s=0}^{t-1}\tilde{\varphi}(\bx_{s}),\bx_{t})$.
This means that in practice, one may implement the Hilbert sort on
a space of much lower dimension (i.e. the dimension of this new $\bz_{t}$),
which is computationally more convenient.

\subsubsection{Backward smoothing}

Backward smoothing consists of two steps: (a) a forward pass, where
SMC is run from time $0$ to time $T$; and (b) a backward pass, where
one constructs a trajectory $\tilde{\bx}_{0:T}$ recursively backwards
in time, by selecting randomly each component $\tilde{\bx}_{t}$ out
of the $N$ particle values $\bx_{t}^{n}$ generated during the forward
pass. An advantage of backward smoothing is that it is less prone
to degenerate than forward smoothing. A drawback of backward smoothing
is that generating a single trajectory costs $\bigO(N)$, hence obtaining
$N$ of them costs $\bigO(N^{2}$).

Backward smoothing for SQMC may be implemented in a similar way to SMC: see Algorithm \ref{alg:backward-step-of} for the backward
pass that generates $N_B$  trajectories $\tilde{\bx}^{1:N_B}_{0:T}$ from 
the output of the SQMC algorithm. Note that backward
smoothing requires that the Markov kernel $m_{t}(\bx_{t-1},\dx_{t})$
admits a closed-form density $m_{t}(\bx_{t}|\bx_{t-1})$ with respect
to an appropriate dominating measure. Then one may compute empirical
averages over the so obtained $N_B$ trajectories to obtain smoothing estimates
in the usual way.

\begin{algorithm}
\caption{\label{alg:backward-step-of}Backward step of SQMC backward smoothing}

\begin{algorithmic}

\Require $\bx_{0:T}^{\sigma_t(1:N)}$, $W_{0:T}^{\sigma_t(1:N)}$
(output of SQMC obtained after the Hilbert sort step, i.e for all $t\in 0{:}T$,  
$h\circ\psi(\bx_t^{\sigma_t(n)})\leq h\circ\psi(\bx_t^{\sigma_t(m)})$, $n\leq m$) and
$\tilde{\bu}^{1:N_B}$ a point set in $\ui^{T+1}$; let $\tilde{\bu}^n=(\tilde{u}_0^n,\dots,\tilde{u}_T^n)$.

\Ensure $\tilde{\bx}^{1:N_B}_{0:T}$ ($N_B$ trajectories in $\setX^{T+1}$)

\State Find permutation $\tau$ such that $\tilde{u}_0^{\tau(1)}\leq...\leq \tilde{u}_0^{\tau(N_B)}$,  generate $\tilde{a}_{T}^{1:N_B}$ using Algorithm \ref{alg:inverse_method}, with inputs $\tilde{u}_0^{\tau(1:N_B)}$ and $W_T^{\sigma_T(1:N)}$, and set $\tilde{\bx}_T^{n}=\bx_T^{\tilde{a}_T^{n}}$
for all $n\in \onetoN_B$. 

\For{$t=T-1\rightarrow 0$}

\State For  $n\in \onetoN_B$, set $\tilde{\bx}_t^{n}=\bx_t^{\tilde{a}_t^n}$ where
$
\tilde{a}_t^n=F_{\pi^n_{t}}^{-1}(\tilde{u}_{T-t}^{\tau(n)})
$, $\pi_t^n=\sum_{m=1}^N \widetilde{W}^m_{t}(\tilde{\bx}^n_{t+1})\delta_{m}$ and, for $m\in \onetoN$,
$$ 
\widetilde{W}_{t}^{m}(\bx_{t+1})=W_{t}^{\sigma_t(m)}m_{t}(\bx_{t+1}|\bx_{t}^{\sigma_t(m)})/\left\{ \sum_{n=1}^{N}W_{t}^{n}m_{t}(\bx_{t+1}|\bx_{t}^{n})\right\}.
$$ 
\EndFor
\end{algorithmic} 
\end{algorithm}

\section{Numerical study}\label{sec:Numerical-study}

The objective of this section is to compare the performance of SMC
and SQMC. Our comparisons are either for the same number of particles $N$, or for the same amount of CPU time to take into account the fact that SQMC has greater complexity than SMC. These comparisons will often summarised through gain factors, which we define as ratios of mean square errors (for a certain quantity) between SMC and SQMC.

In SQMC, we generate $\bu_t^{1:N}$ as a \citet{Owen1995} nested scrambled Sobol' sequence  using the C++ library of T. Kollig and A. Keller (\url{http://www.uni-kl.de/AG-Heinrich/SamplePack.html}). Note  that both the generation and the randomization of $(t,s)$-sequences in base 2 (such as the Sobol' sequence) are very fast since logical operations can be used. In order to sort the particles according to their Hilbert index we use the C++ library of Chris Hamilton (\url{http://web.cs.dal.ca/~chamilto/hilbert/index.html}) to evaluate $H_m^{-1}(\psi(\bx))$, $m\in\mathbb{N}$. Again, Hilbert computations are very fast as they are based on logical operations  \citep[see][for more details]{Hamilton2008}.  In addition,  thanks to the nesting property of the Hilbert curve (see Section \ref{sub:The-Hilbert-space-filling})  we only need to take $m$ large enough such that  different particles are mapped into different points of $\ui$. Function $\Gamma_t$ is set to the inverse transform described in Section \ref{sub:Inverse-transform-method}, and function $\psi$ to a component-wise (rescaled) logistic transform; that is,
$
\psi(\bx)=(\psi_1(x_1),...,\psi_d(x_d))
$
with
$$
\psi_i(x_i)=\left[1+\exp\left(-\frac{x_i-\underline{x}_i}{\bar{x}_i-\underline{x}_i}\right)\right]^{-1},\quad i\in 1:d
$$
and where the constants $\bar{x}_i$ and $\underline{x}_i$ are used to solve numerical problems due to high values of $|x_i|$. For instance, when $(\bx_t)$ is a stationary process we chose  $\bar{x}_i=\mu_i+2\sigma_i$ and $\underline{x}_i=\mu_i-2\sigma_i$ where $\mu_i$ and $\sigma^2$ are respectively the mean and the standard deviation of the stationary distribution of $(\bx_t)$.

SMC is implemented using systematic resampling \citep{CarClifFearn} and all the random variables are generated using standard methods (i.e. not using the multivariate GICDF). The C/C++ code implementing both SMC and SQMC is available on-line at
\url{https://bitbucket.org/mgerber/sqmc}.


Even if Theorems \ref{thm:PF2} and \ref{thm:PF2Bis} are valid for any pattern of $N$, choosing for $N$ powers of 2 (with 2 the  base of the Sobol' sequence) is both natural and optimal  for QMC methods based on (scrambled) $(t,s)$-sequences (see e.g. \citealp{Owen1997b}; \citealp{Hickernell2001}; and Chapter 5 of \citealp{dick2010digital}). Comparing the performance of SQMC for different patterns of $N$ is beyond the scope of this paper \citep[see][for a discussion of this point]{Gerber:QMCarbitrarysize} and therefore we follow in this numerical study the standard approach in the QMC literature by considering  values of $N$ that are  powers of 2. We nevertheless do one exception to this rule for the PMMH estimation on real data (Section \ref{subsec:PMMH}) because  doubling the number of particles to reduce the variance of the likelihood estimate used in the Metropolis-Hastings ratio may be very inefficient from a computational point of view. As we will see, allowing $N$ to differ from  powers of the Sobol' base  does not seem to alter the  performance of SQMC.




One may expect the two following situations to be challenging for SQMC:
(a) small $N$ (because our results are asymptotic); and (b) large $d$ (because of the usual deterioration of QMC with respect of the dimension, and also because of the Hilbert sort step). Thus we consider examples of varying dimensions (from 1 to 10),
and we will also make $N$ vary within a large range (between $2^4$ and $2^{17}$).

\subsection{Example 1: A non linear and non stationary univariate model}

We consider the following popular toy example \citep{Gordon,Kitagawa:Smoother}:
\begin{equation}\label{simu:eq:modelUniv}
\begin{cases}
y_{t}=\frac{x_{t}^{2}}{a}+\epsilon_{t},\quad &\epsilon_t\sim \mathcal{N}_1(0,1),\quad t\geq 0\\
x_{t}=b_{1}x_{t-1}+b_{2}\frac{x_{t-1}}{1+x_{t-1}^{2}}+b_{3}\cos(b_{4}t)+\sigma\nu_{t},
\quad &\nu_t\sim \mathcal{N}_1(0,1),\quad t>0
\end{cases}
\end{equation}
and $x_0\sim \mathcal{N}_1(0,2)$, where $\mathcal{N}_d(\bm{\mu},\Sigma)$ denotes the $d$-dimensional Gaussian distribution with mean $\bm{\mu}$ and covariance matrix $\Sigma$.
We generate observations from 100 time steps of the model, with the parameters set as in \citet{Gordon}:
 $a=20$, $\bm{b}=(0.5,25,8,1.2)$,
$\sigma^{2}=10$, $x_0=0.1$. Note that inference in this model is non trivial because the observation $y_t$ does not allow to identify the sign of $x_t$, and because the weight function $G_t(x_t)$ is bimodal if $y_t>0$ (with  modes at $\pm (20y_t)^{1/2}$). In addition, we expect this model to be challenging for SQMC due to the high non linearity of the Markov transition $m_t(\bx_{t-1},\dx_t)$.

All the results presented below are based on 500 independent runs of SMC and SQMC.
Figure \ref{fig:UnivModel:Lik} presents results concerning the estimation of the log-likelihood functions evaluated at the true value of the parameters. The two top graphs show that, compared to SMC, SQMC yields  faster convergence of both the mean and the variance of the estimates.  

\afterpage{
\begin{figure}[t!]
\begin{centering}
\includegraphics[scale=0.35]{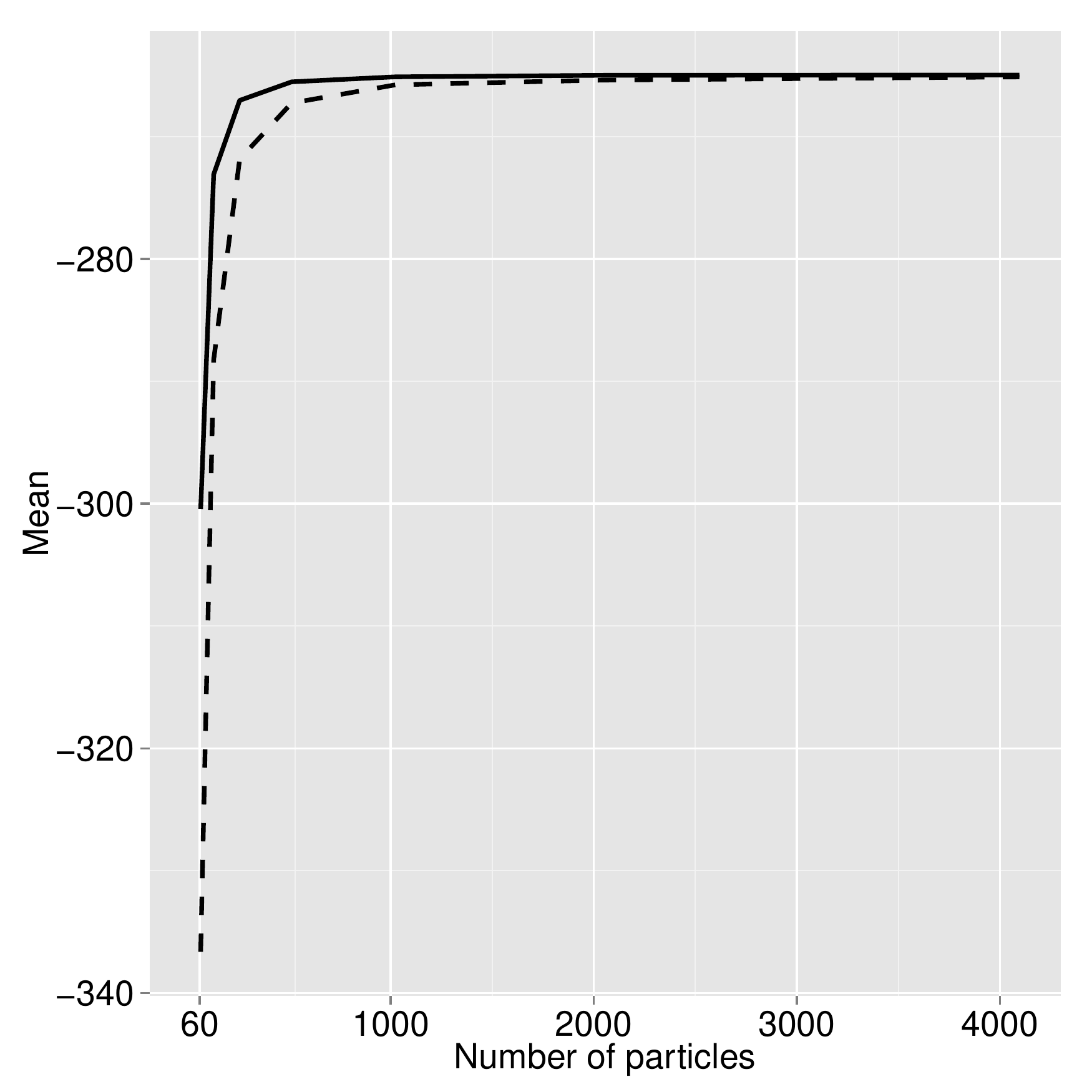}\includegraphics[scale=0.35]{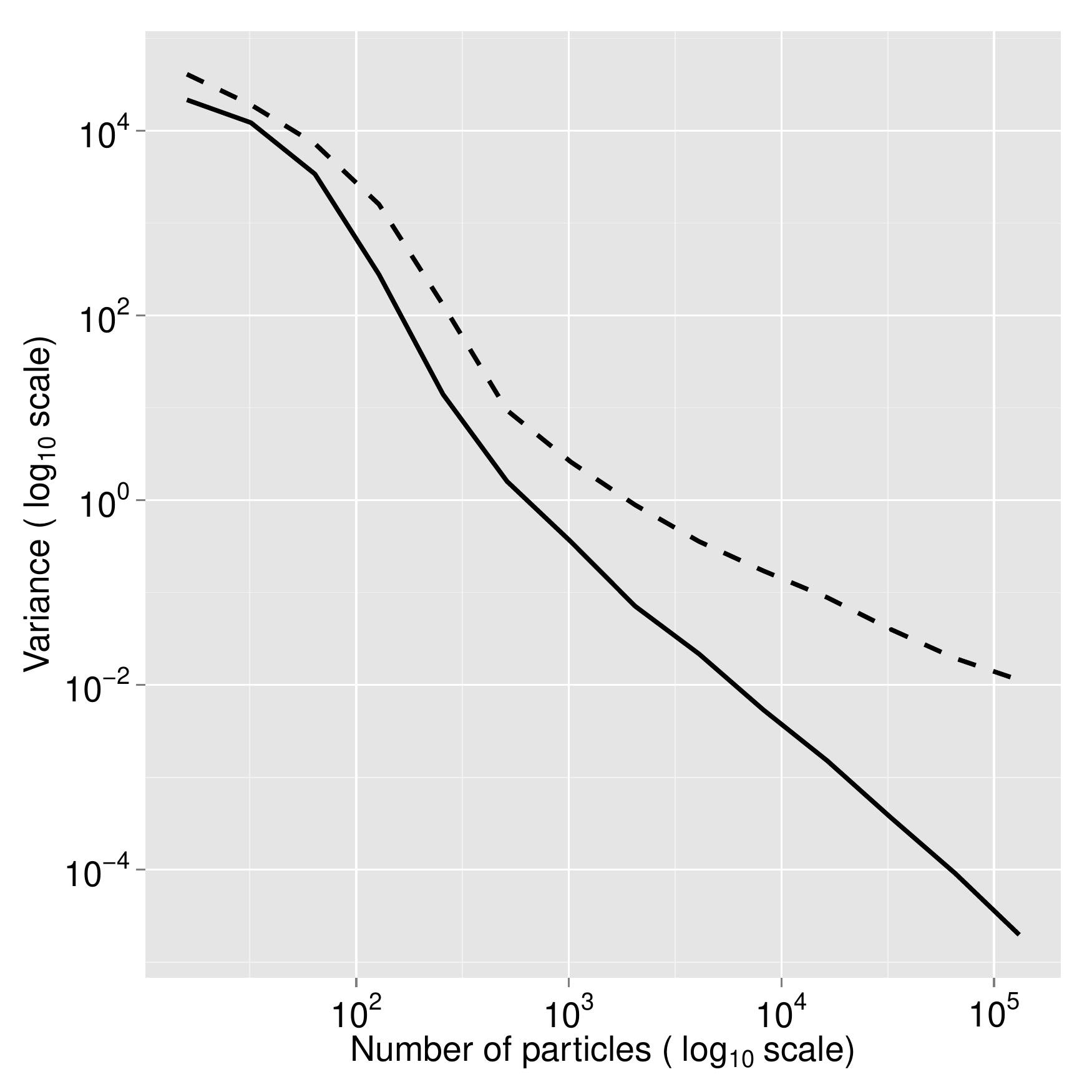} 

\includegraphics[scale=0.35]{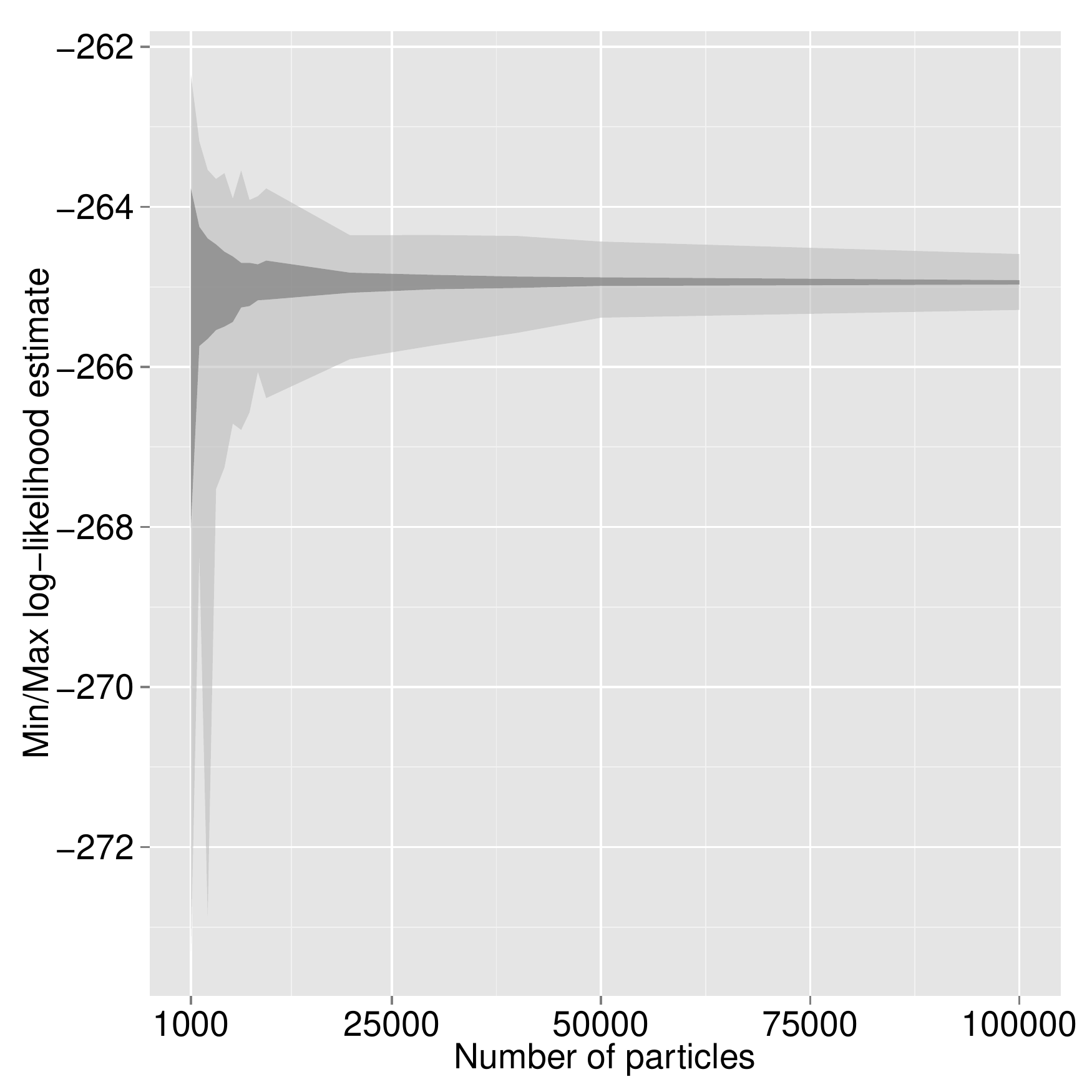}\includegraphics[scale=0.35]{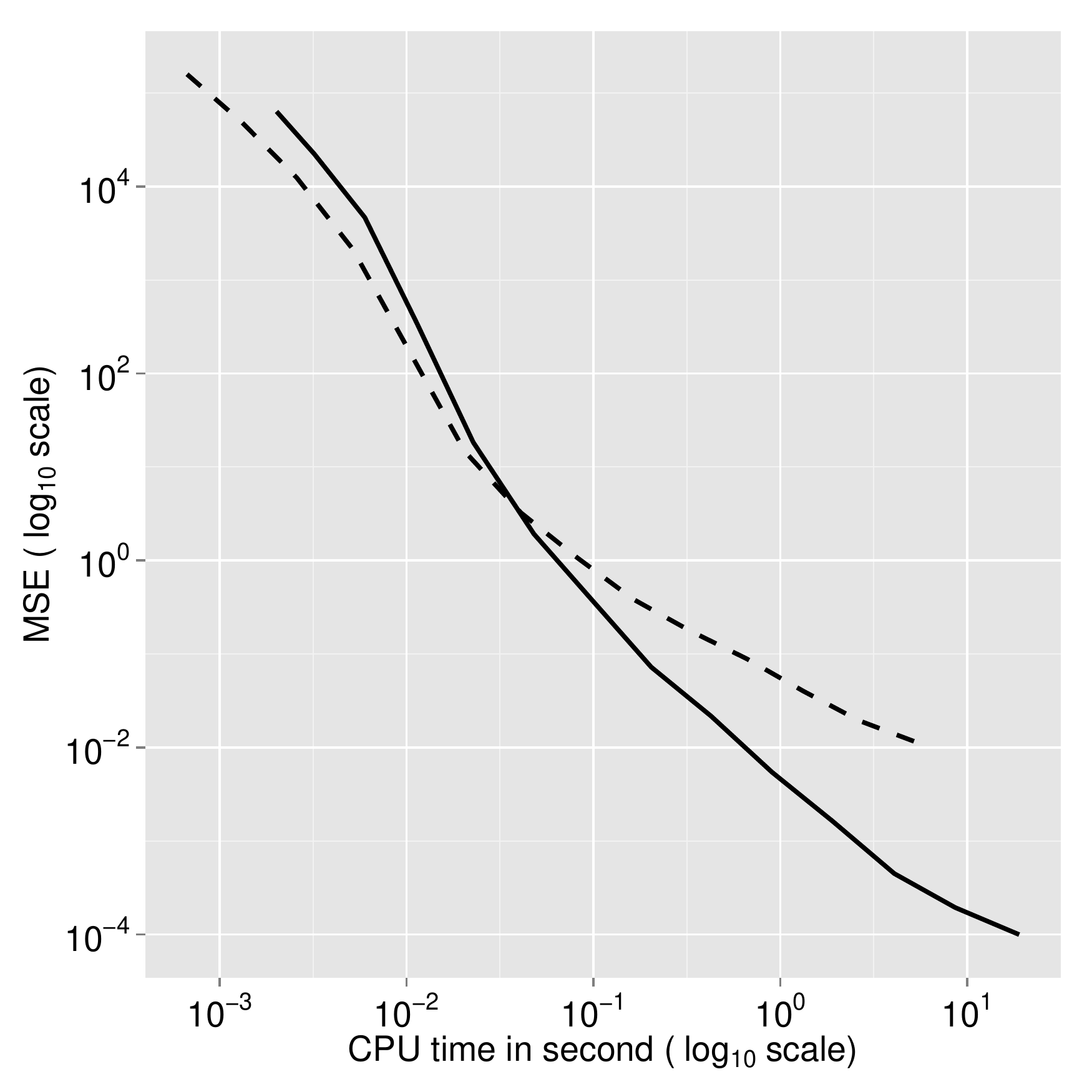} 
\par\end{centering}

\caption{Log-likelihood estimation for the toy example \eqref{simu:eq:modelUniv}. The solid lines are for SQMC while the dashed lines are for SMC. In the bottom-left graph, the dark (light) area shows the range in which lies the SQMC (SMC) estimates of the log-likelihood function. All the results are obtained from 500 independent runs of SQMC and SMC.\label{fig:UnivModel:Lik} }
\end{figure}
}

These better consistency properties of SQMC are also illustrated on the bottom left graph of Figure \ref{fig:UnivModel:Lik} where we have reported for each $N$ the range in which lies the 500 estimates of the log-likelihood. From this plot we see that  quickly the SQMC estimates stay in a very tiny interval while, on the contrary, the SMC estimates are much more dispersed, even for large values of $N$.

The bottom right panel of Figure \ref{fig:UnivModel:Lik}  shows the MSE of SQMC and SMC as a function of CPU time. One sees that the gain of SQMC over SMC does not only increase with $N$, as predicted by the theory, but also with the CPU time which is of more practical interest. On the other hand, in this particular case (log-likelihood evaluation for this univariate model), when $N$ is small the reduction in MSE brought by SQMC does not  compensate its greater running time.  Nevertheless,  we observe that  SQMC outperforms SMC very quickly, that is,   as soon as the CPU time is larger or equal to $10^{-1.5}\approx 0.03$ seconds.



\begin{figure}[t!]
\begin{centering}
\includegraphics[scale=0.35]{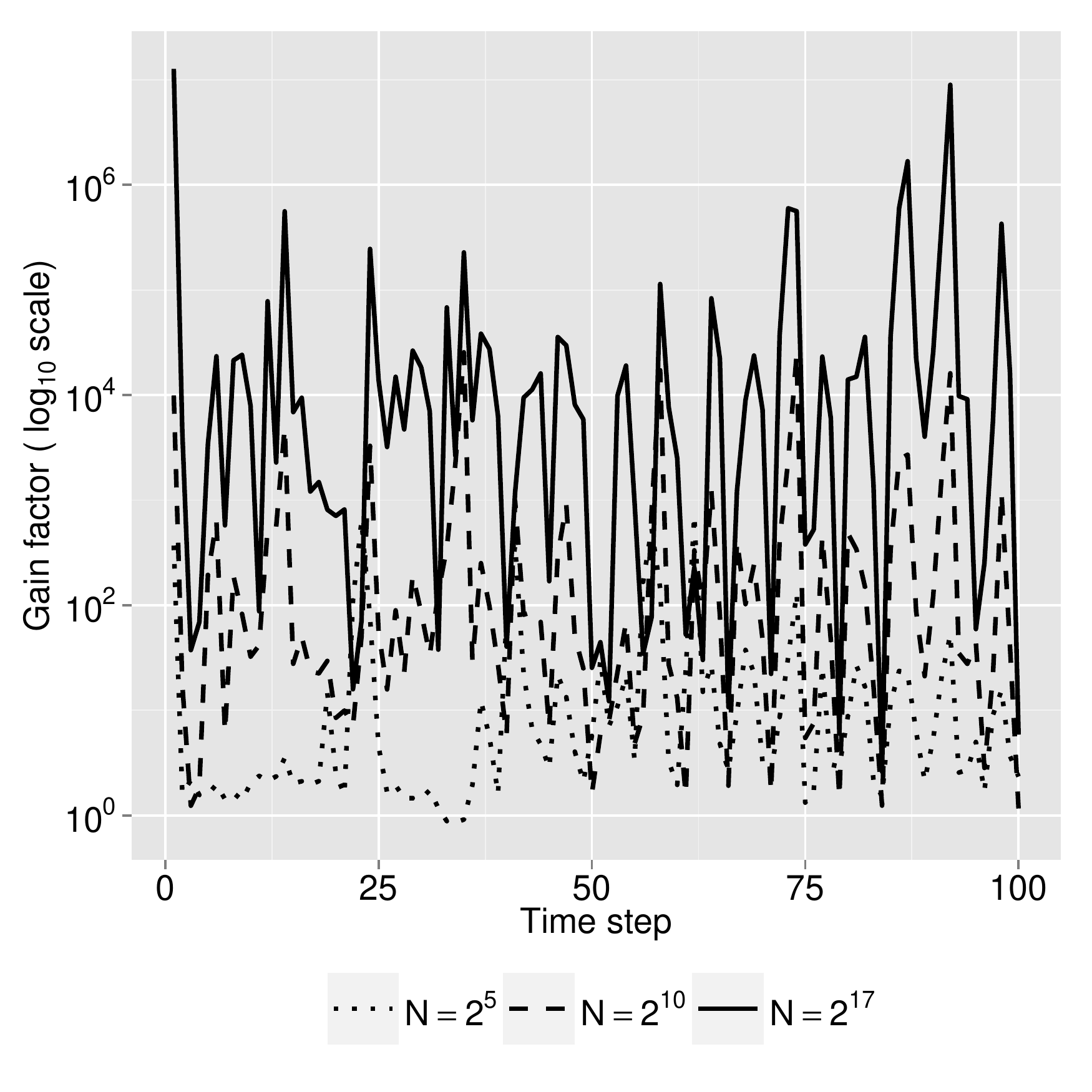}\includegraphics[scale=0.35]{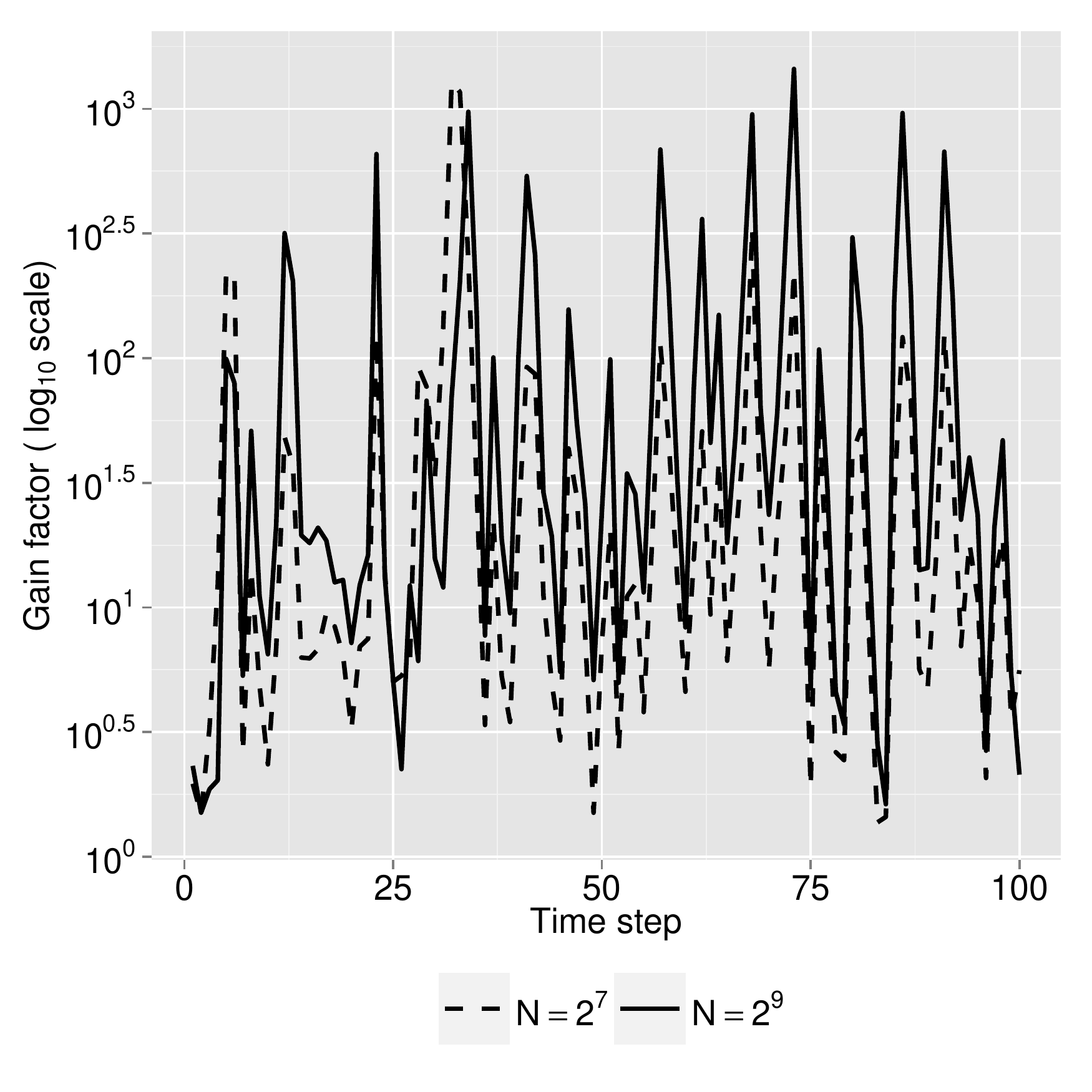}
\par\end{centering}
\caption{Filtering (left graph) and backward smoothing (right graph) for the toy example \eqref{simu:eq:modelUniv}: gain factor as a function of $t$ for the estimation of $\E[x_t|y_{0:t}]$ and for the estimation of $\E[x_t|y_{0:T}]$, obtained from  500 independent runs of SQMC and SMC.\label{fig:UnivModel:Back} }
\end{figure}

In the left graph of Figure \ref{fig:UnivModel:Back} we have reported the gain factor for the estimation of $\E[x_t|y_{0:t}]$ as a function of $t$ and for different values of $N$. From this plot we observe both significant and increasing gain of SQMC over SMC. 

The right panel of Figure \ref{fig:UnivModel:Back} compares SQMC and SMC backward smoothing for the estimation of $\E[x_t|y_{0:T}]$ as a function of $t$ and for $N\in\{2^7,2^9\}$. As for the filtering problem, SQMC significantly outperforms SMC with  gain factors that increase with the number of particles. 

\subsection{Example 2: Multivariate stochastic volatility model}\label{num:SV}

We consider the following multivariate stochastic volatility model (SV) proposed by \citet{Chan2006}:
\begin{equation}\label{simu:eq:modelSV}
\begin{cases}
\by_{t}=S_t^{1/2}\boldsymbol{\epsilon}_{t},\quad &t\geq 0\\
\bx_{t}=\boldsymbol{\mu}+\Phi(\bx_{t-1}-\boldsymbol{\mu})+
\Psi^{\frac{1}{2}}\boldsymbol{\nu}_t,\quad &t> 0
\end{cases}
\end{equation}
where $S_t=\text{diag}(\exp(x_{t1}),...,\exp(x_{td}))$,  $\Phi$ and $\Psi$ are diagonal matrices and $(\boldsymbol{\epsilon}_{t},\boldsymbol{\nu}_{t})\sim\mathcal{N}_{2d}(\bm{0}_{2d}, C)$, with $C$  a correlation matrix and $\bm{0}_{2d}=(0,\dots,0)\in\mathbb{R}^{2d}$.

In order to study the relative performance of SQMC over SMC as the dimension $d$ of the hidden process increases we perform simulations for $d\in\{1,2,4,10\}$. The parameters we use for the simulations are the same as in \citet{Chan2006}:     $\phi_{ii}=0.9$, $\mu_i=-9$, $\psi_{ii}^2=0.1$ for all $i=1,...,d$ and
$$
C=
\begin{pmatrix}
0.6\mathbf{1}_d+0.4\mathcal{I}_d&-0.1\mathbf{1}_d-0.2\mathcal{I}_d\\
-0.1\mathbf{1}_d-0.2\mathcal{I}_d&0.8\mathbf{1}_d+0.2\mathcal{I}_d
\end{pmatrix}
$$
where $\mathcal{I}_d$ is the $d$-dimensional identity matrix,
and  $\mathbf{1}_d$ is the $d\times d$ matrix having one in all its entries.
Note that the errors terms $\boldsymbol{\epsilon}_{t}$ and $\boldsymbol{\nu}_{t}$ are correlated so that the weight function $G_t$ depends now both on $\bx_{t-1}$ and on $\bx_t$. The prior distribution for $\bx_0$ is the stationary distribution of the process $(\bx_t)$ and we take $T=399$.

\begin{figure}
\begin{centering}
\begin{tabular}{cc}
$d=1$&$d=2$\\
\includegraphics[scale=0.35]{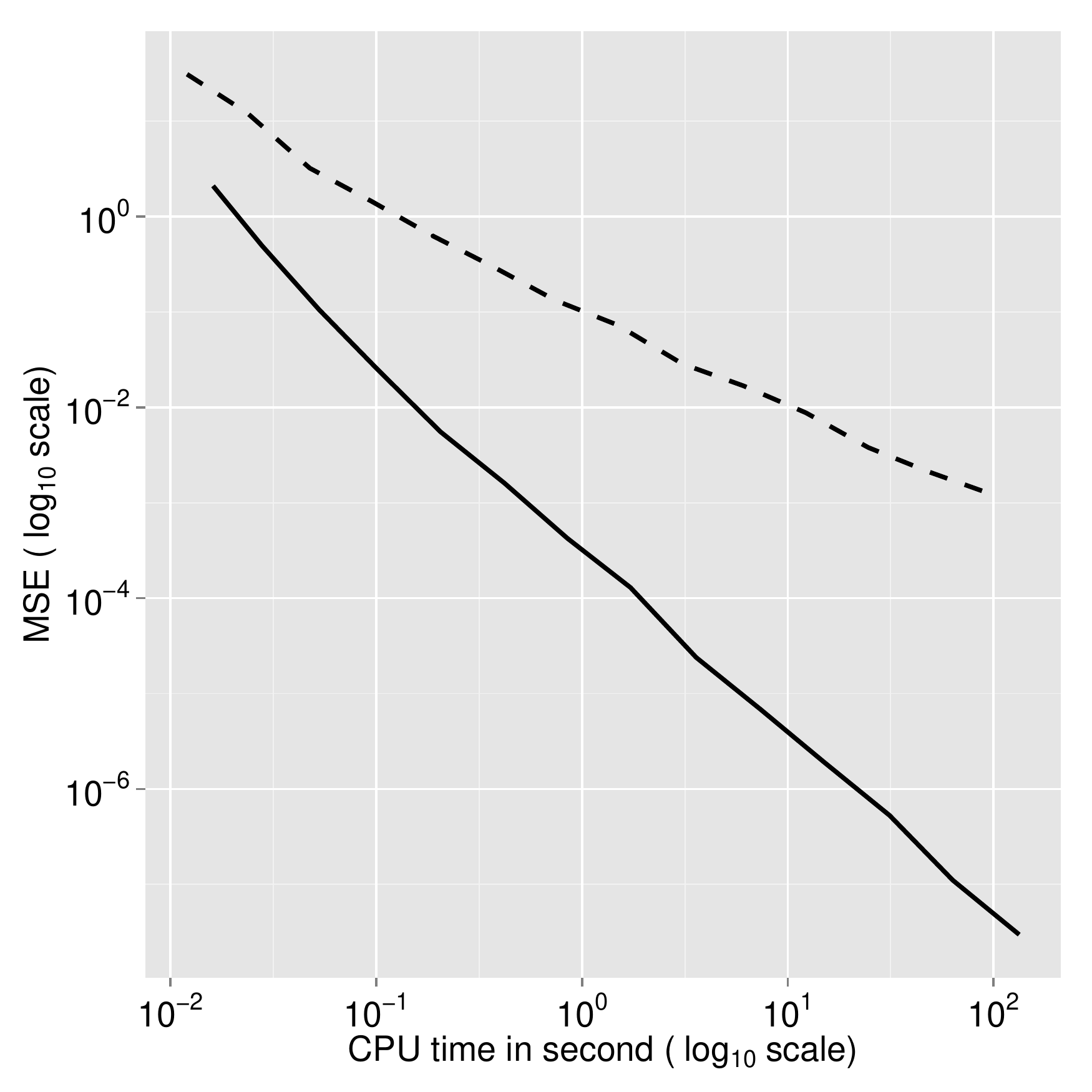}&\includegraphics[scale=0.35]{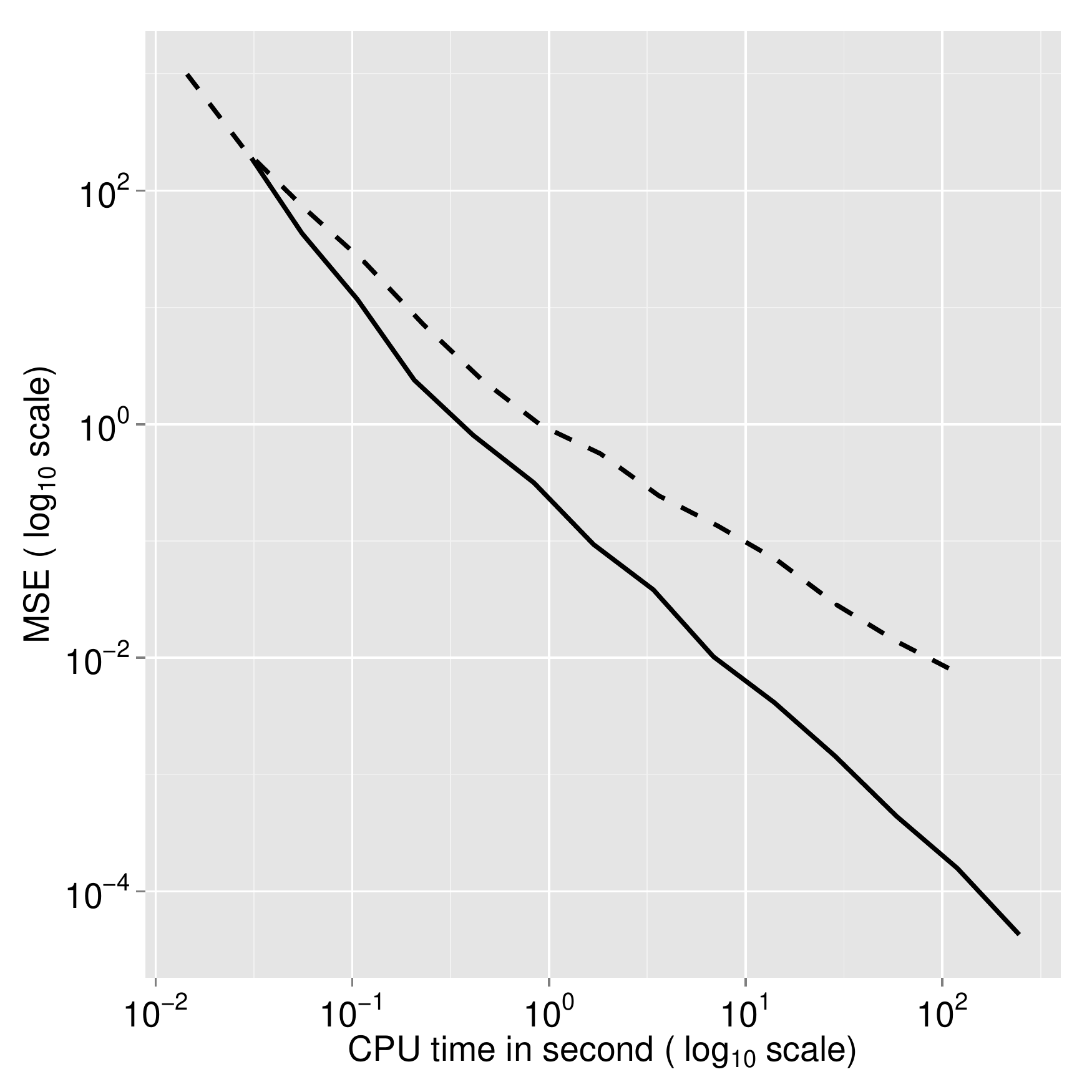}\\
$d=4$&\\
\includegraphics[scale=0.35]{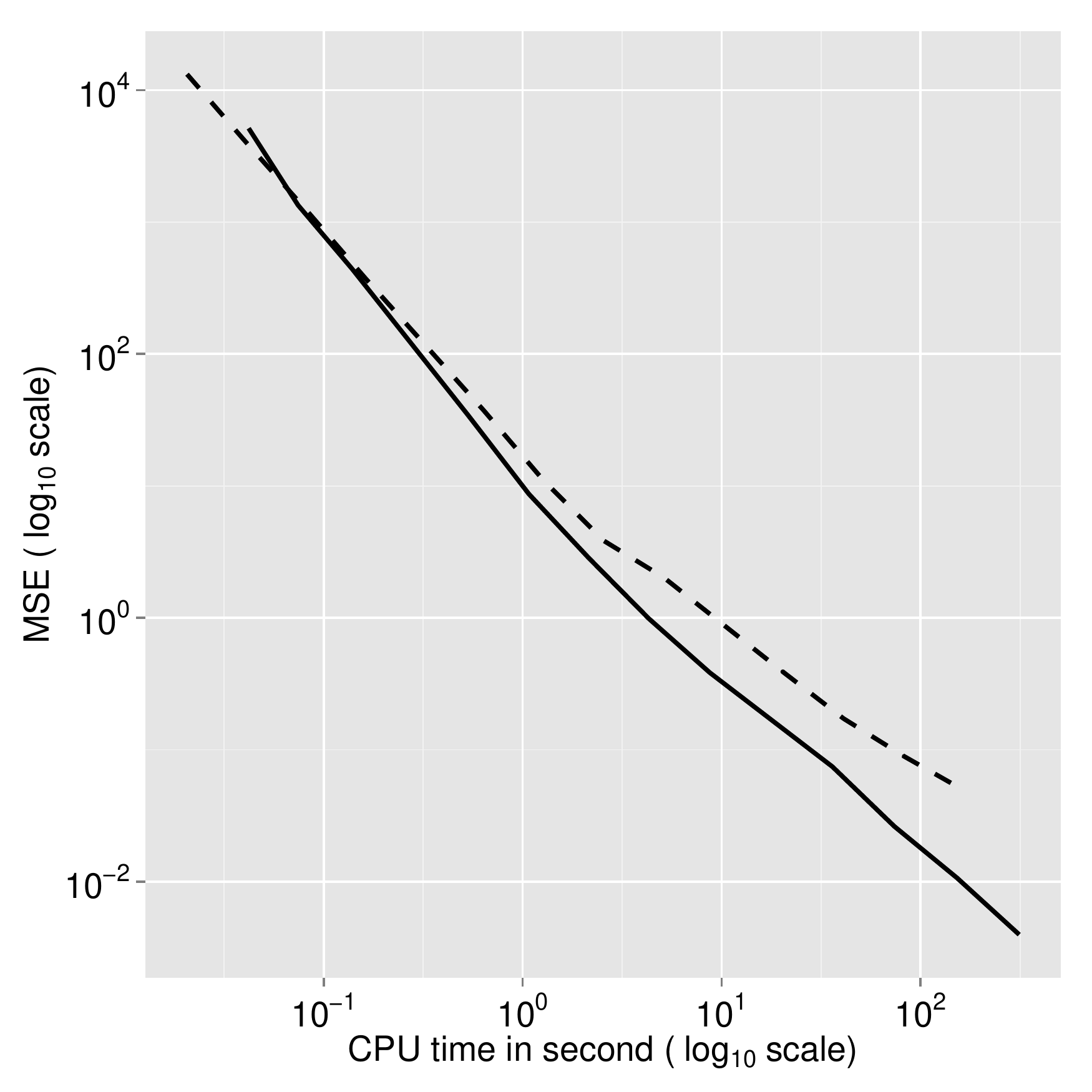} 
&\includegraphics[scale=0.35]{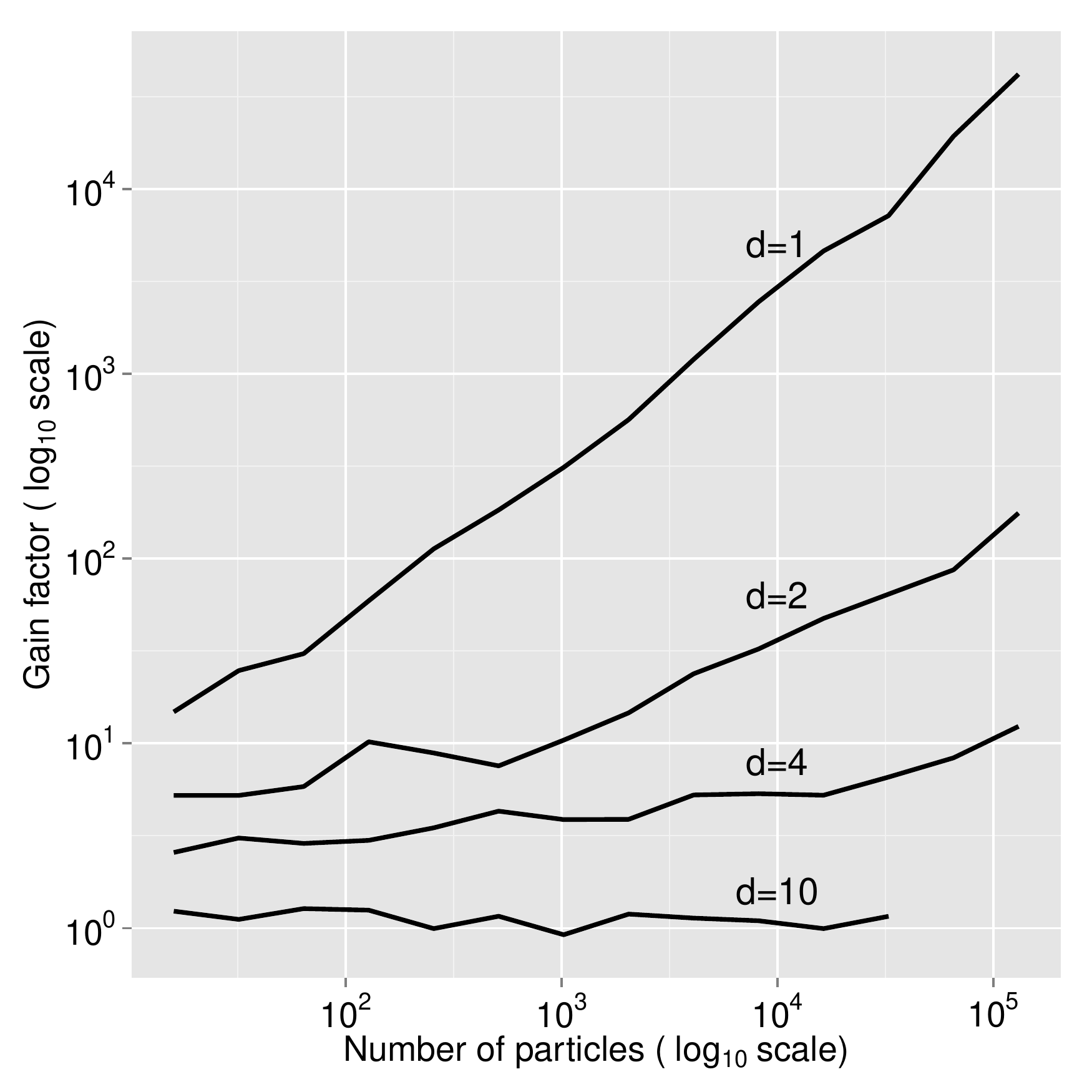}\\
\end{tabular}

\par\end{centering}

\caption{Log-likelihood estimation of SV model \eqref{simu:eq:modelSV}: MSE as a function of CPU time, 
for $d=1$, 2, 4; gain factor as a function of $N$ 
for $d=1$, 2, 4 and 10. The solid lines are for SQMC while the dashed lines are for SMC.
The graphs  are obtained from 200 independent runs of SQMC and SMC.\label{fig:SVModel:Lik} }
\end{figure}

The three first panels of Figure \ref{fig:SVModel:Lik} present results for the estimation of the log-likelihood (evaluated at the true value of the parameters and for the complete dataset $y_{0:T}$),
for $d\in\{1,2,4\}$. One sees that the gain factor increases quickly with $N$, and, more importantly, 
the MSE of SQMC converges faster than SMC even as a function of CPU time. In fact,
except for a very small interval for the univariate model, SQMC always outperforms SMC
in terms of MSE for the same CPU effort. 
We note the particularly impressive values of the gain factor we obtain for $d=1$ when $N$ is large: around $4.2\times 10^4$ for $N=2^{17}$. 
The last panel of Figure \ref{fig:SVModel:Lik} plots the gain factors as a function of 
$N$, for same values of $d$, plus $d=10$. The improvement brought by SQMC decreases
with the dimension, and in fact, for $d=10$, the gain factor
is essentially one for the considered values of $N$; yet for $d=4$ we still observe
some notable improvement; e.g. a gain factor of 10 for $N\approx 10^5$.  We now
focus on $d=1$, 2 and 4.

\begin{figure}
\begin{centering}
\begin{tabular}{cc}
\multicolumn{2}{c}{Univariate SV model}\\

\includegraphics[scale=0.35]{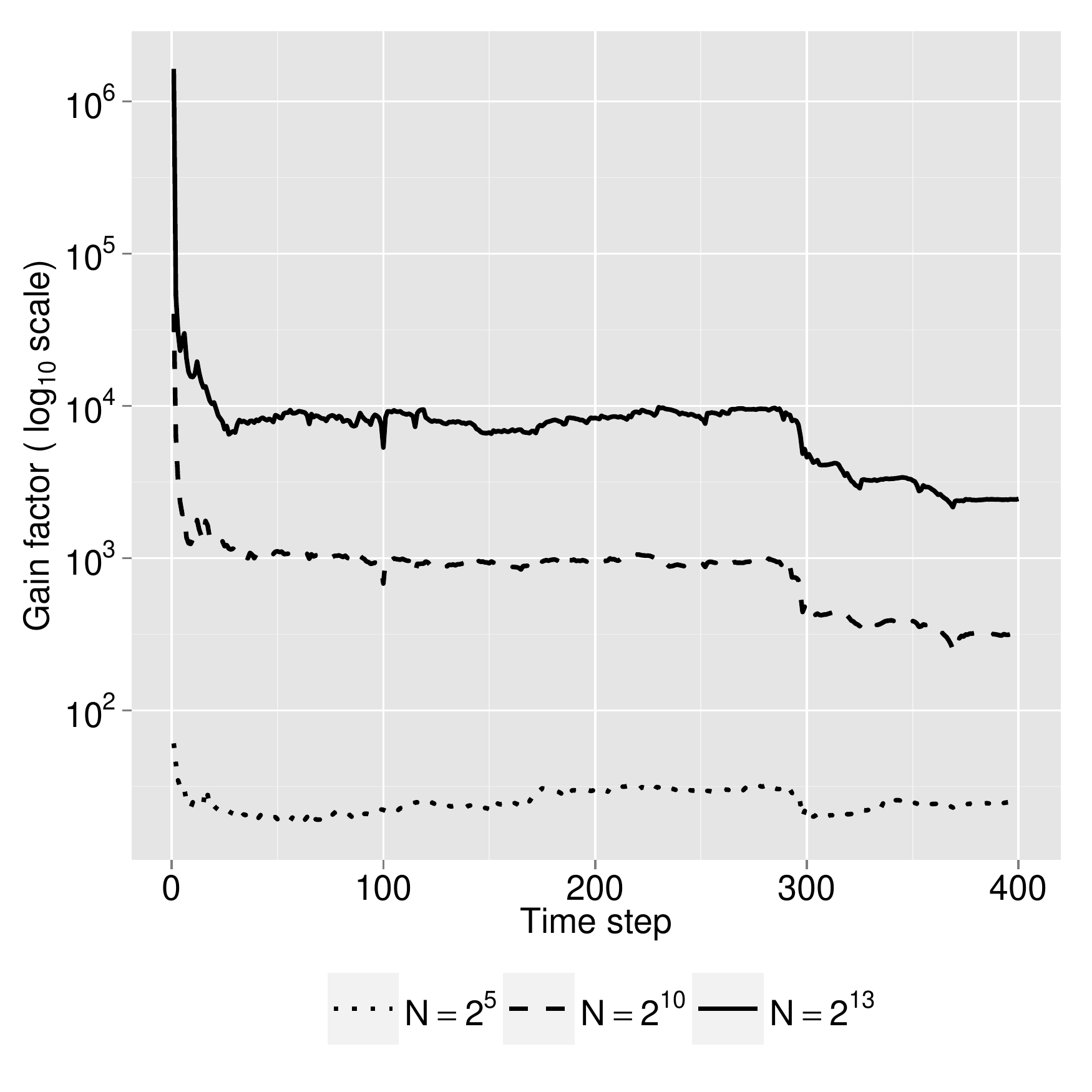}&
\includegraphics[scale=0.35]{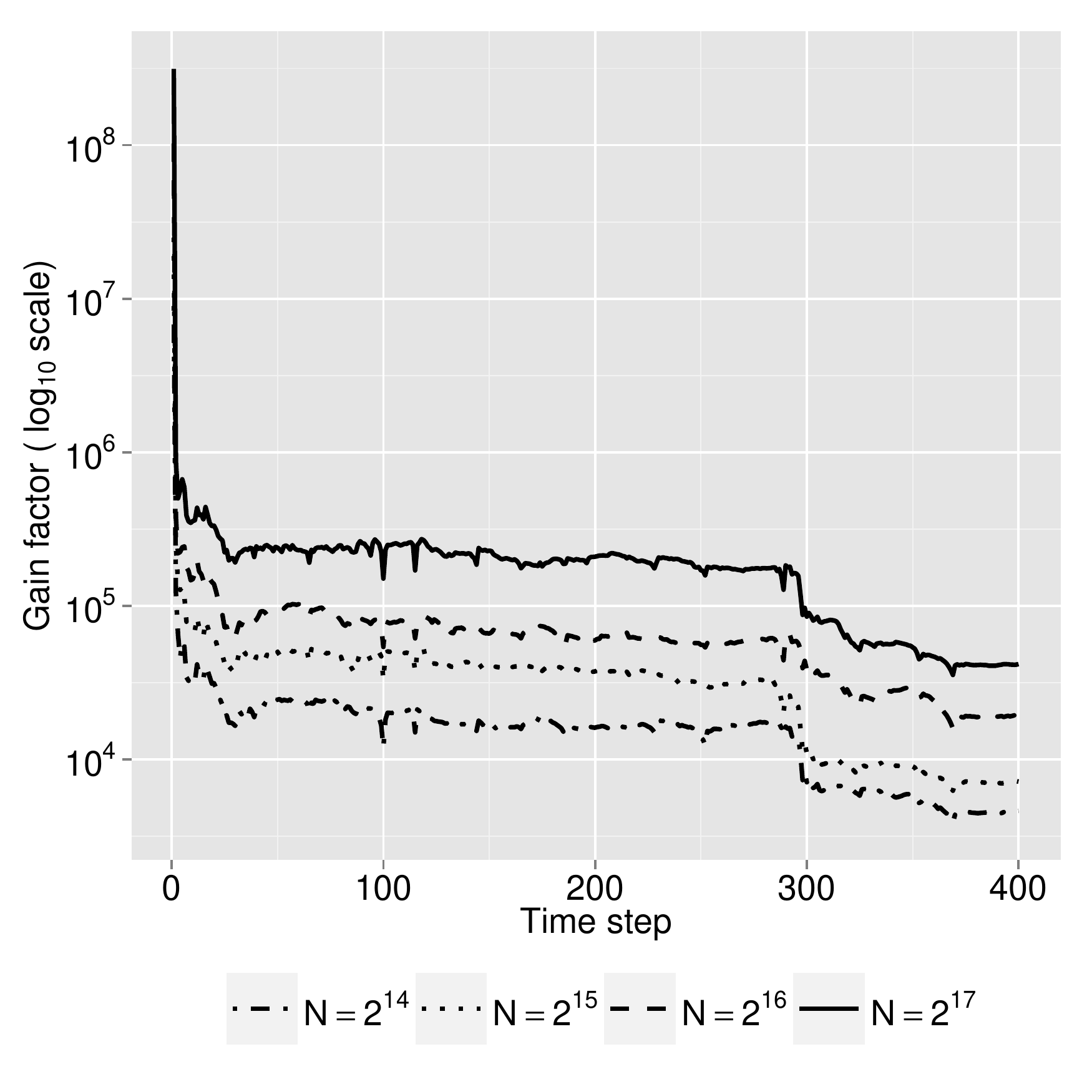}\\

\multicolumn{2}{c}{Bivariate SV model}\\

\includegraphics[scale=0.35]{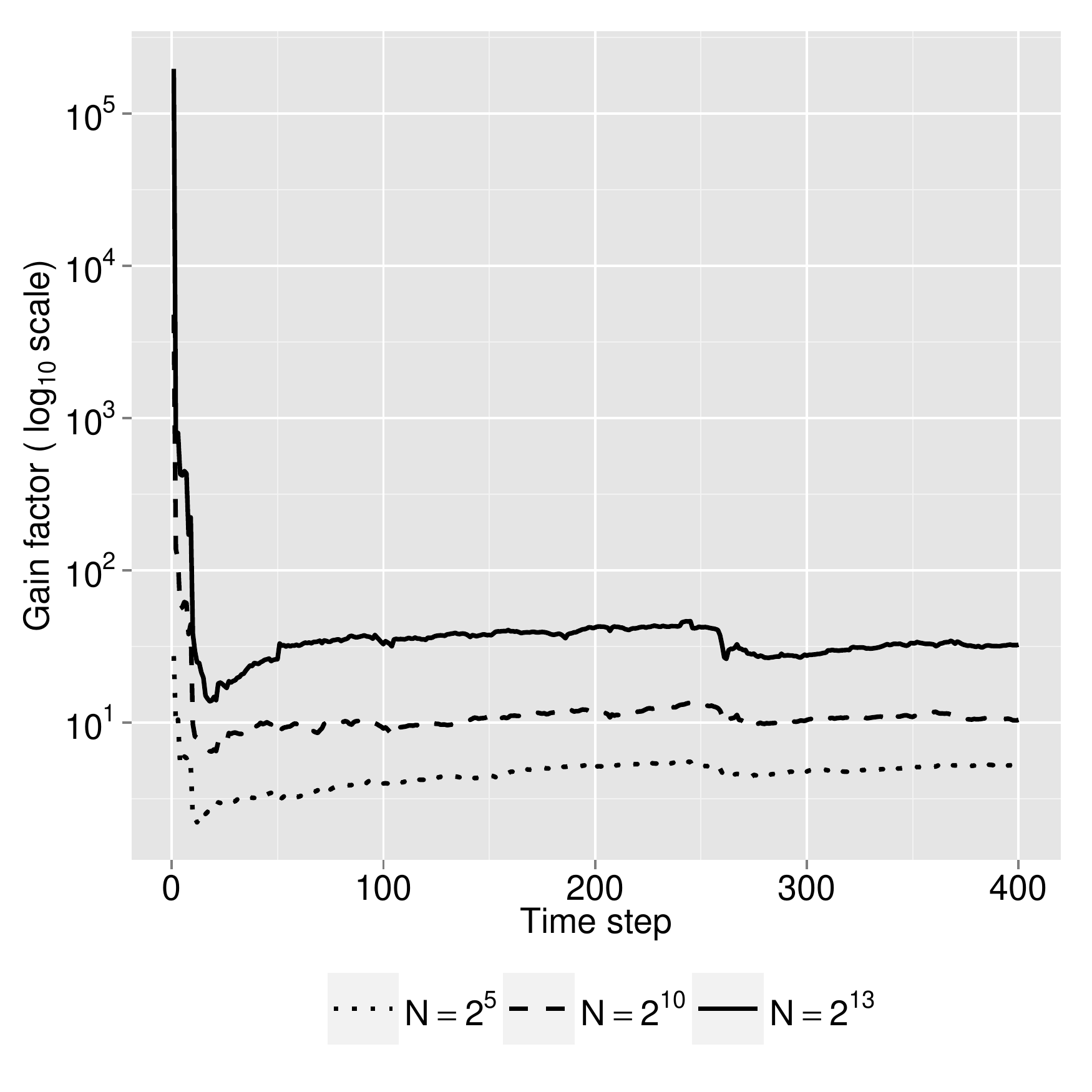}&
\includegraphics[scale=0.35]{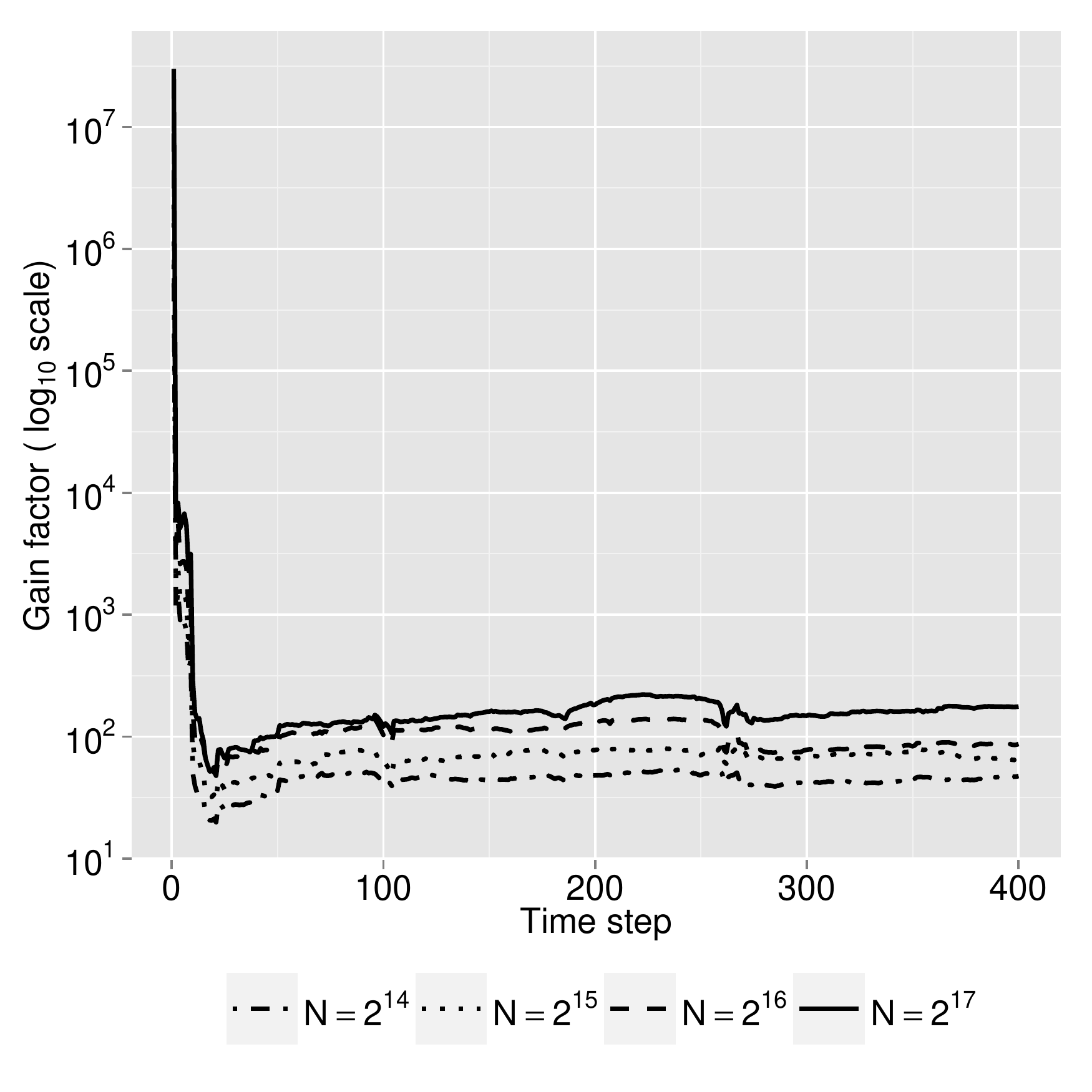}\\

\multicolumn{2}{c}{Four dimensional SV model}\\

\includegraphics[scale=0.35]{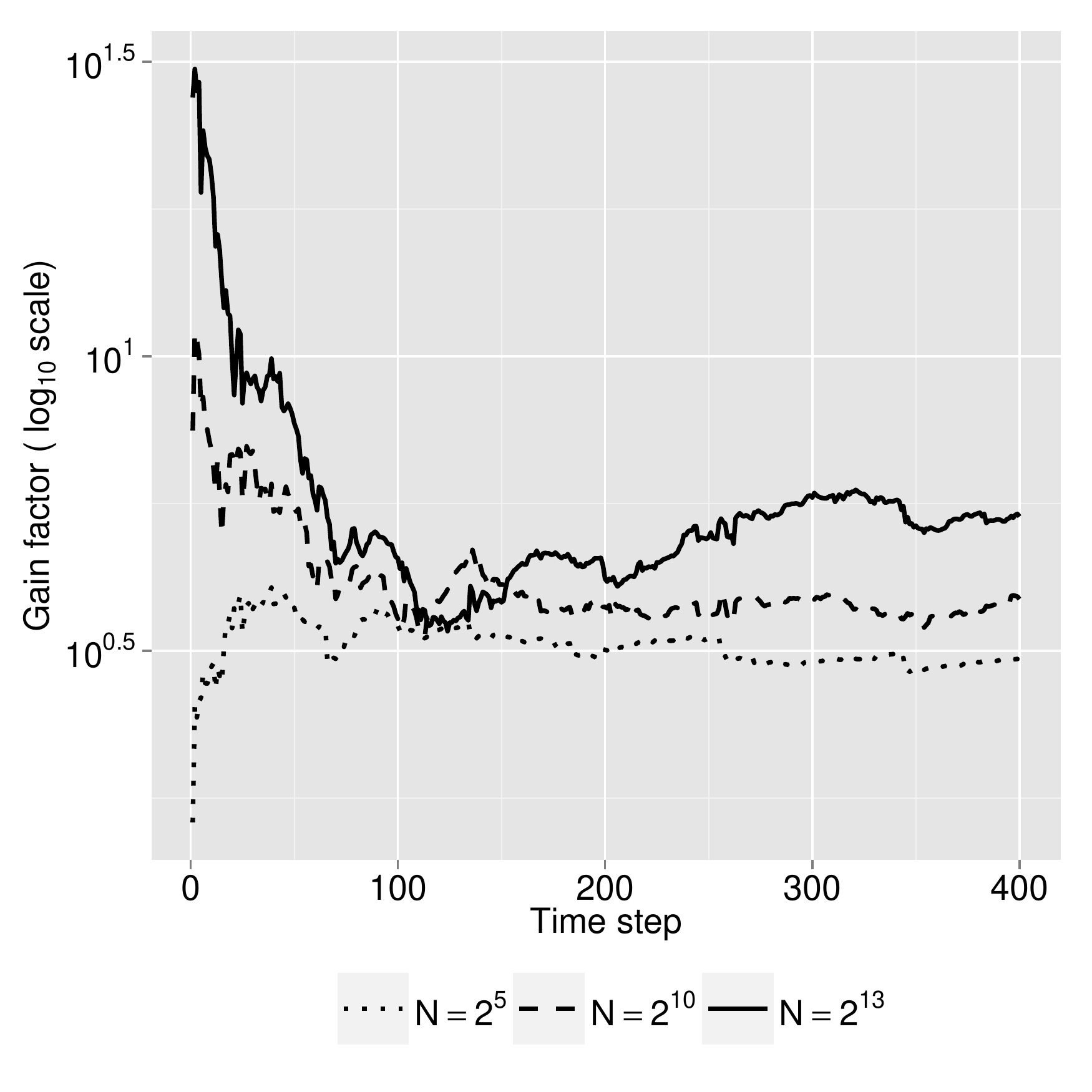}&
\includegraphics[scale=0.35]{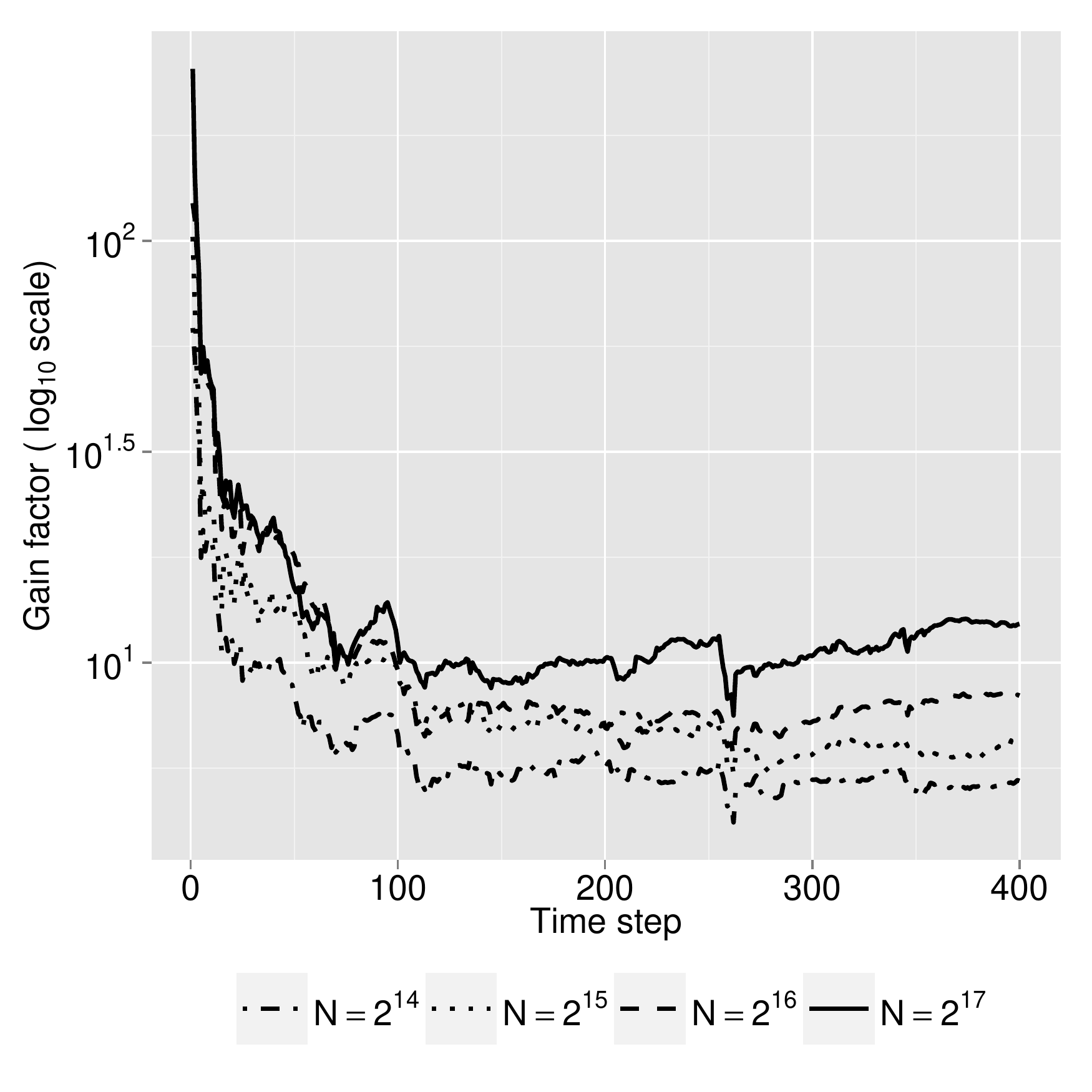}\\ 
\end{tabular}
\par\end{centering}

\caption{Log-likelihood estimation of the SV model \eqref{simu:eq:modelSV}:
gain factor as a function of $t$, obtained from  200 independent runs of SQMC and SMC.\label{fig:SVModel:LikT} }
\end{figure}

Figure \ref{fig:SVModel:LikT} represents the evolution with respect to $t$ 
of the MSE for the partial log-likelihood of data $y_{0:t}$ up to time $t$; gain factors are reported for different values of $N$. As we can see from these graphs, the performance of SQMC does not seem to  depreciate with $t$. 

\begin{figure}
\begin{centering}
\begin{tabular}{cc}
Bivariate SV model & Four dimensional  SV model\\
\includegraphics[scale=0.35]{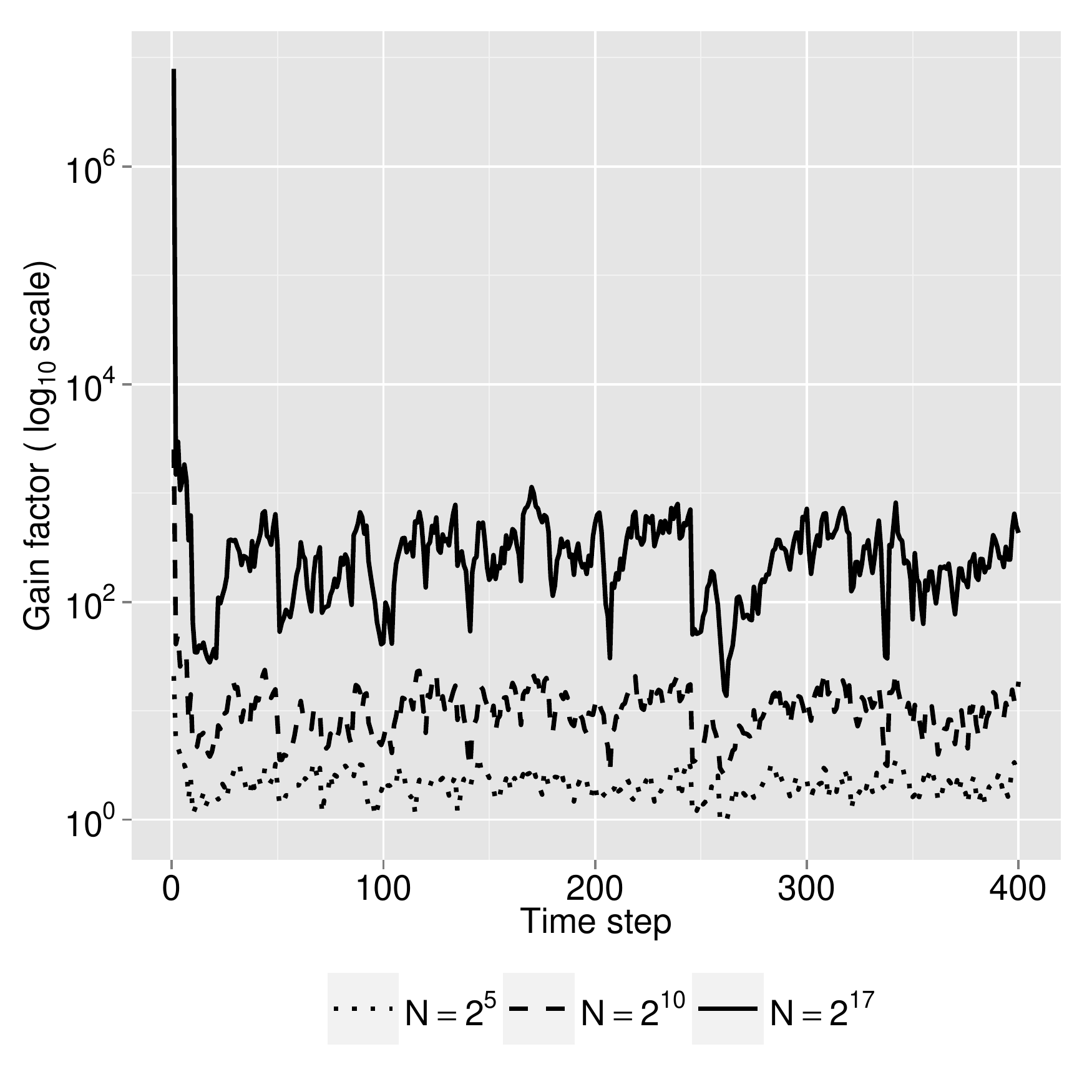}&\includegraphics[scale=0.35]{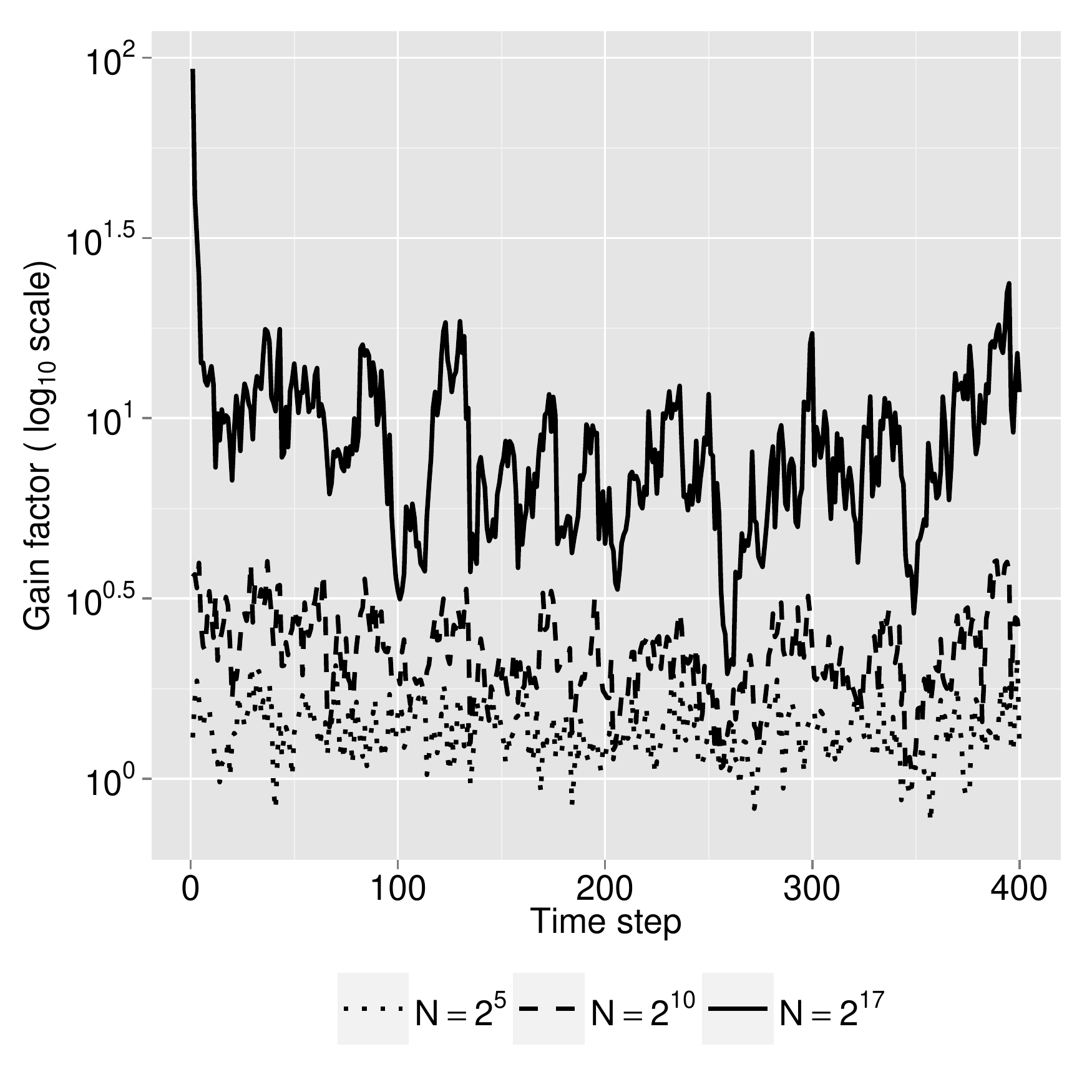}\\ 
\end{tabular}
\par\end{centering}
\caption{Filtering of the multivariate SV model \eqref{simu:eq:modelSV}: gain factor as a function of $t$ for the estimation of $\E[x_{1t}|\mathbf{y}_{0:t}]$, obtained from 200 independent runs of SQMC and SMC.\label{fig:SVModel:EX} }
\end{figure}

Finally, Figure \ref{fig:SVModel:EX} shows that SQMC also give impressive gain when $d>1$ concerning the estimation of the filtering expectation $\E[x_{1t}|\mathbf{y}_{0:t}]$ of the first component of $\bx_t$.

\subsection{Application: Bayesian estimation of MSV using PMMH on real data}\label{subsec:PMMH}

To compare SMC to SQMC when used as a  way to approximate the likelihood 
within a PMMH algorithm, as described in Section \ref{sub:evidence}, 
we turn our attention to the Bayesian estimation of the multivariate SV model (\ref{simu:eq:modelSV}), for $d=2$. As in \citet{Chan2006}, we take the following prior: 
$$
\phi_{ii}\sim \Unif((0,1)), \quad 1/\psi^2_{ii}\sim\mathrm{Gamma}(10\exp(-10),10\exp(-3))\quad i=1,\dots d,
$$
where  $\phi_{ii}$ and $\psi^2_{ii}$ denotes respectively the diagonal elements of $\Phi$ and $\Psi$, and a flat prior for $\bm{\mu}$. 
In addition, we assume that
$C$ is uniformly distributed on the space of  correlation matrices which are such that the errors terms $\boldsymbol{\epsilon_t}$ and $\boldsymbol{\nu_t}$ are independents (no leverage effects). To sample from the posterior distribution of the parameters we use a  Gaussian random walk Metropolis-Hastings algorithm with  covariance matrix $\Sigma$ calibrated
so that the acceptance probability of the algorithm becomes,  as $N\rightarrow +\infty$, close to   25\%.
The matrix $\Sigma$, as well as the starting point of the Markov chain, are calibrated using a pilot run of the algorithm with $\Sigma=0.011^2\mathcal{I}_8$ and starting at the value of the parameters we used above for the simulations.  To compare PMMH-SQMC with PMMH-SMC, we run the two algorithms during $10^5$ iterations and for values of $N$ ranging from 10 to 200, where $N$ increases from 10 to 100 by increment of 10 and then by increment of 50.

We consider the following  dataset: the two series are the mean-corrected daily return on the Nasdaq and S\&P 500 indices for the period ranging from the $3^{\text{rd}}$ January 2012 to the 21$^{\text{th}}$ October 2013 so that the data set contains $452$ observations.

\begin{figure}
\begin{centering}
\includegraphics[scale=0.35]{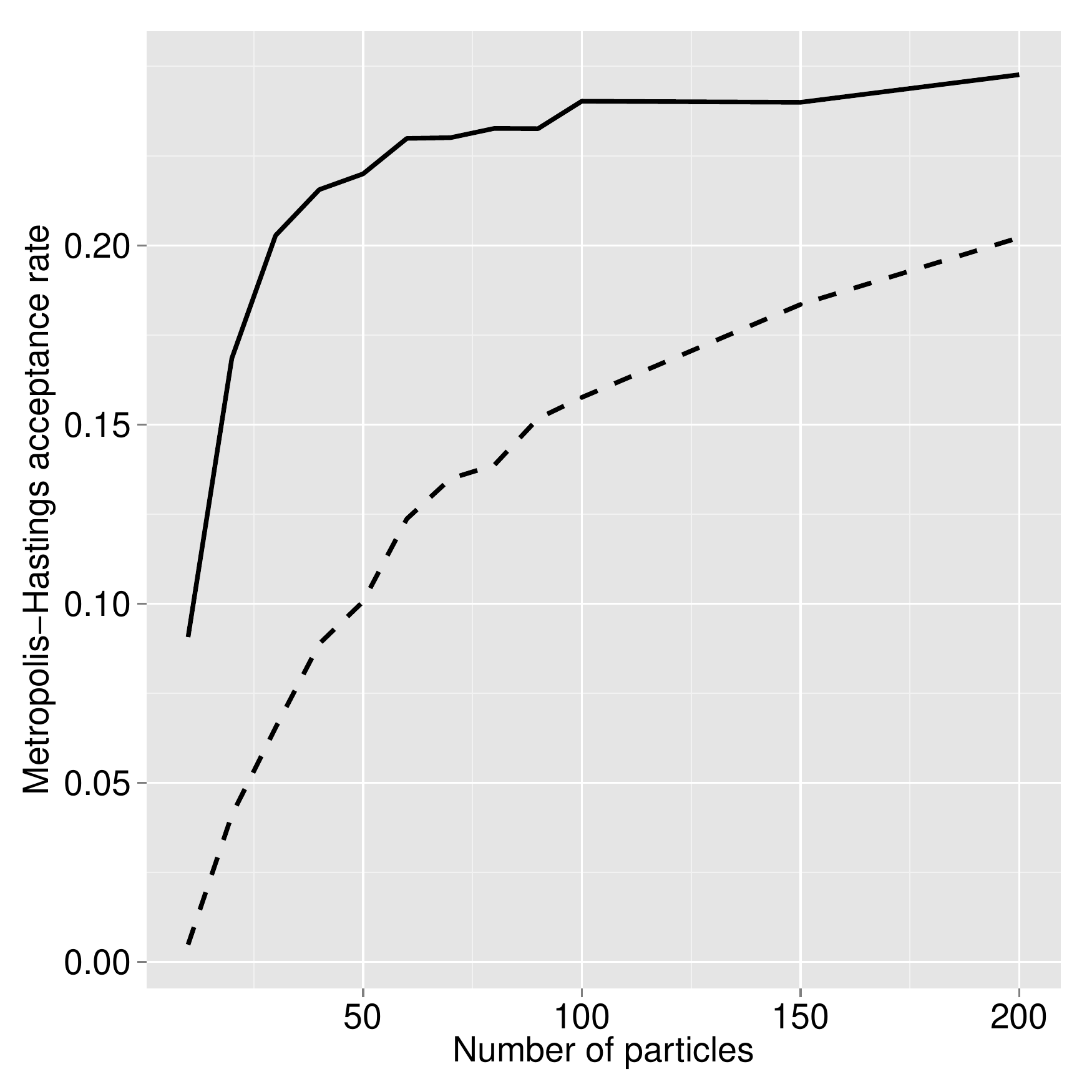}
\includegraphics[scale=0.35]{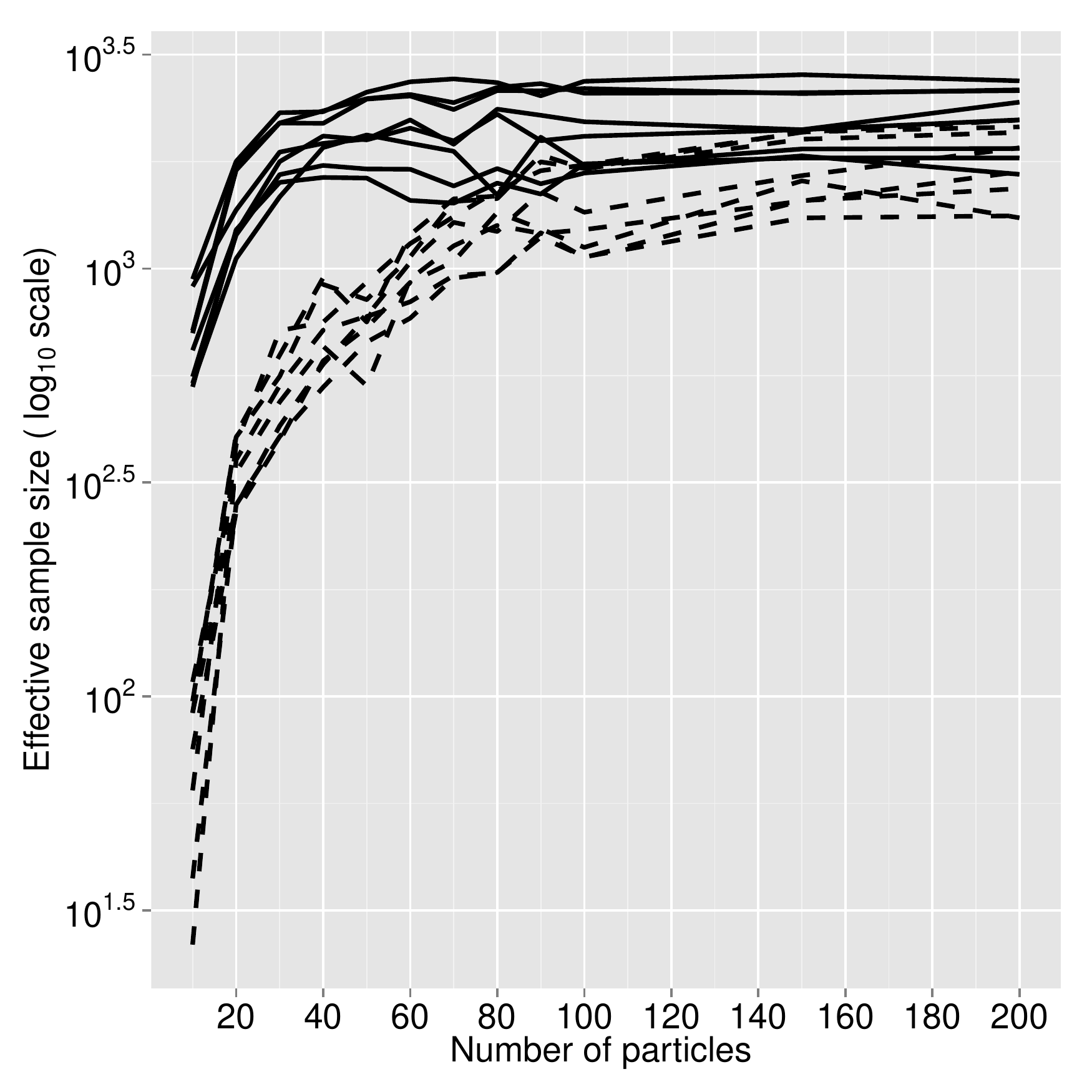}
\par\end{centering}
\caption{Metropolis-Hastings acceptance rate and effective sample sizes (one per parameter) for the multivariate SV model \eqref{simu:eq:modelSV}, $d=2$, and real data: the solid lines are for PMMH-SQMC while the dashed ones are for PMMH-SMC. The results are obtained from a Markov Chain of length $10^5$.\label{fig:SVModel:Proba} }
\end{figure}

Figure \ref{fig:SVModel:Proba} shows the Metropolis-Hastings acceptance rate and the effective sample sizes \citep[see][Section 12.3.5, for a definition]{RobCas}  for the PMMH-SQMC algorithm and for the standard PMMH algorithm. We first observe that the acceptance rate of PMMH-SQMC increases very quickly with $N$. Indeed, it is already of 20\% for only 30 particles   while for the same number of particles the acceptance rate for the standard PMMH is approximatively 6.5\%. As far as the acceptance rate is concerned, there is no significant gain to take $N>60$ for the PMMH-SQMC algorithm while for the plain Monte Carlo algorithm the acceptance rate is only about 20\% for $N=200$ and therefore much smaller than the target of 25\%. Looking at the results for the effective sample sizes (ESSs), we see that the same conclusions hold. More precisely, for the PMMH-SQMC algorithm, the ESSs increase  with $N$ much faster than for PMMH-SMC. Indeed, for $N\in 10:50$, the ESSs for the former is between 2.18 and 14.94 times larger than for PMMH-SMC. 

\subsection{Example 3: Neural decoding}

Neural decoding models are used for brain-machine interface in order to make inference about an organism's environment from its neural activity. More precisely, we consider the problem of decoding a set of environment variables $\mathbf{p}_t\in\mathbb{R}^{2}$, from the firing ensemble of $d_y$ neurons. The latent vector $\mathbf{p}_t$ may be interpreted as  two-dimensional hand kinetics for motor cortical decoding \citep[see][and references therein for more details about neural decoding models]{Koyama2010}. Noting $\dot{\mathbf{p}}_t$ the vector of velocities, the  neural decoding model we consider is given by \citep{Koyama2010}
\begin{equation}\label{sim:eq:neural2}
\begin{cases}
y_{ti}|\bx_{0:t}\sim\Poi
\left(\Delta\exp(\alpha_i+\beta^T_i\bx_t)\right), &i\in 1:d_y,\quad \hspace{1.47cm} t\geq 0\\
\bx_t= \Phi\bx_{t-1}+\Psi\boldsymbol{\epsilon}_t, &\boldsymbol{\epsilon}_t\sim\mathcal{N}_2(\bm{0}_2,\sigma^2\mathcal{I}_2),\quad t> 0
\end{cases}
\end{equation}
and $\bx_0\sim\mathcal{N}_4(\bm{0}_4,\mathcal{I}_4)$, where $\bx_t=(\mathbf{p}_t,\dot{\mathbf{p}}_t)$, the $y_{ti}$'s are conditionally independent,  $\Poi(\lambda)$ denotes the Poisson distribution with parameter $\lambda$,  $\Delta$ is the duration of the interval over which spikes are counted at each time step, and 
$$
\Phi=\begin{pmatrix}
\mathcal{I}_2&\Delta\mathcal{I}_2\\
\boldsymbol{0}_2&\mathcal{I}_2
\end{pmatrix},\quad \Psi^T=
\begin{pmatrix}
0&0&1&0\\
0&0&0&1
\end{pmatrix}.
$$

Realistic values for the parameters, see  \citet{Koyama2010}, that we will take in ours simulations, are $d_y=10$, $T=23$, $\Delta=0.03$, $\sigma^2=0.019$, $\alpha_i{\buildrel i.i.d \over \sim}\mathcal{N}_{1}(2.5,1)$,  $\beta_i\sim\Unif(\ui^d)$.

One important aspect of this model is that the dimension of the noise term $ \boldsymbol{\epsilon}_t$ is lower than the dimension of $\bx_t$. As a result, two components of $\bx_t$ are deterministic functions of $\bx_{t-1}$. Many tracking problems have a similar structure.

This requires us to slightly adapt SQMC as follows: one samples jointly 
the ancestor variables $a_{t-1}^{1:N}$ and the new velocities $\dot{\mathbf{p}}_t^n$
as in Steps (b) and (c) of Algorithm \ref{alg:SQMC}, 
then one obtains the new $\mathbf{p}_t^n$ as $\mathbf{p}_t^n=\mathbf{p}_{t-1}^n+\dot{\mathbf{p}}_t^n$, i.e. the deterministic linear transformation 
of $\mathbf{p}_{t-1}^{a_{t-1}^n}$ and $\dot{\mathbf{p}}_{t-1}^n$ defined by the model. 
Note that in this case the dimension of the point set $\bu_t^{1:N}$ is 3 for $t>0$;
we could say that $d=2$ in this case, even if the dimension of $\bx_t$ itself is $4$.

\begin{figure}
\begin{centering}
\includegraphics[scale=0.35]{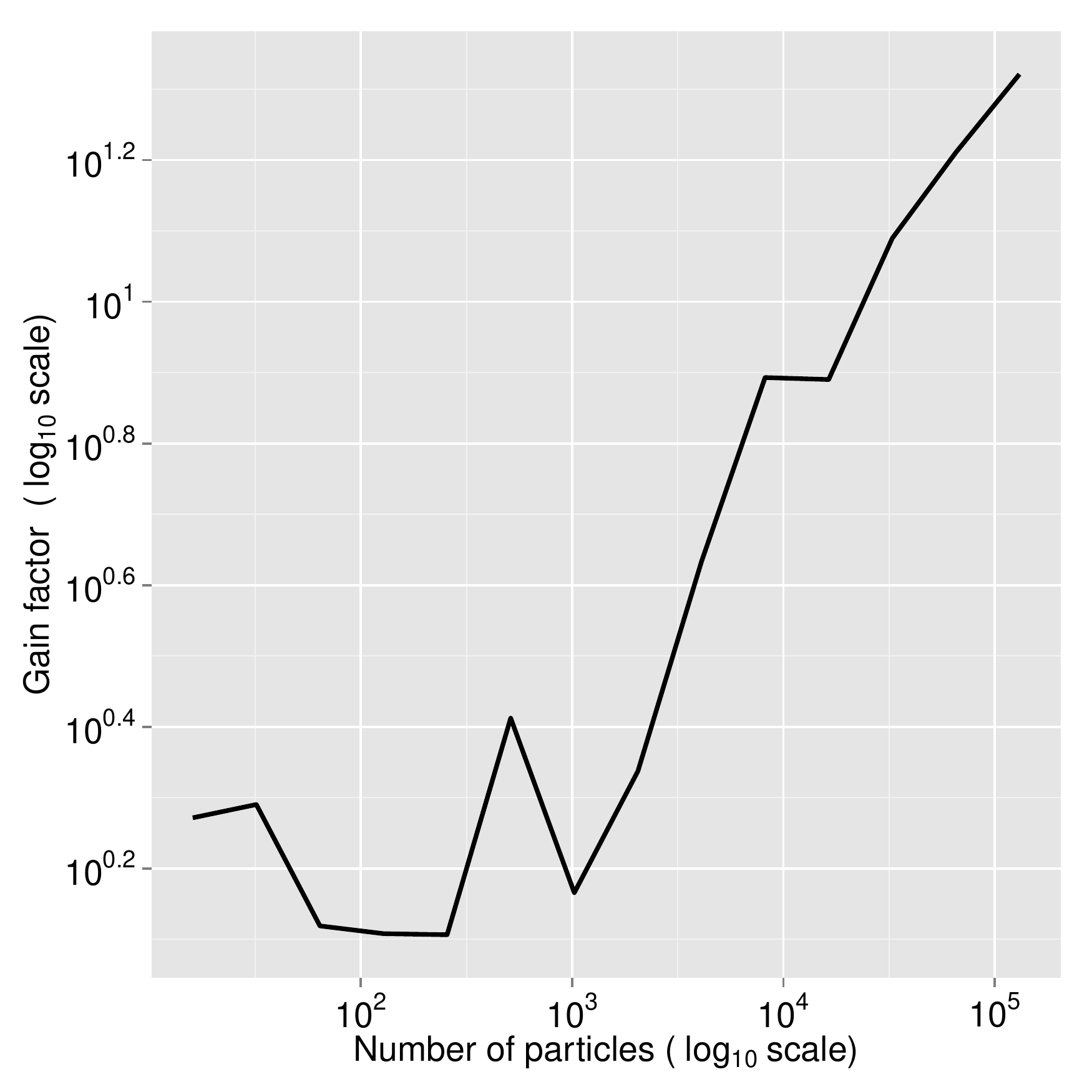}\includegraphics[scale=0.35]{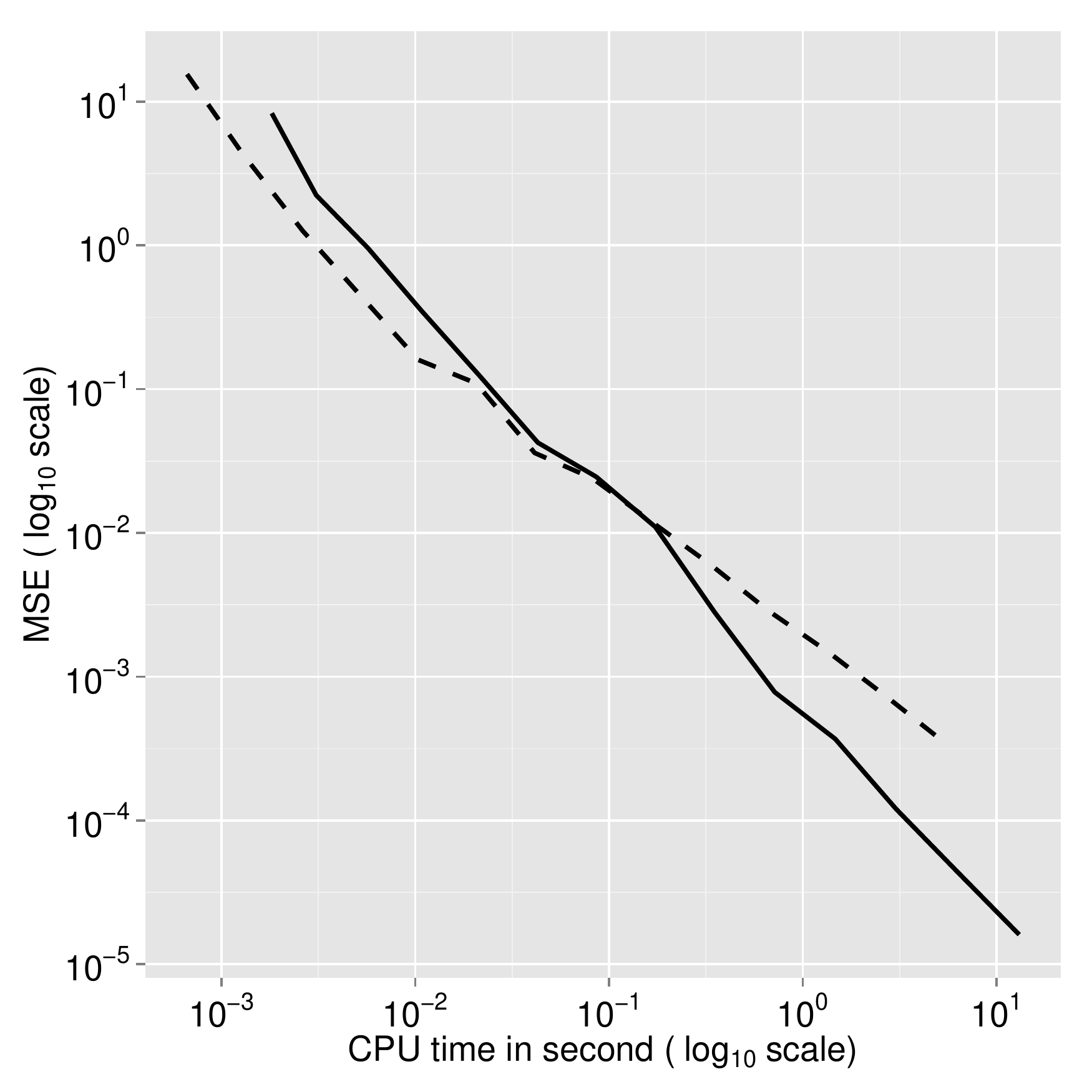}
\par\end{centering}

\caption{Log-likelihood estimation of the neural decoding model \eqref{sim:eq:neural2}.  The left graph gives the ratio of the SMC and the SQMC MSE. In the right graph, the solid line is for SQMC while the dashed line is for SMC. The graphs  are obtained from 200 independent runs of SQMC and SMC.}\label{fig:Neuro:Lik2}
\end{figure}

\begin{figure}
\begin{centering}
\includegraphics[scale=0.35]{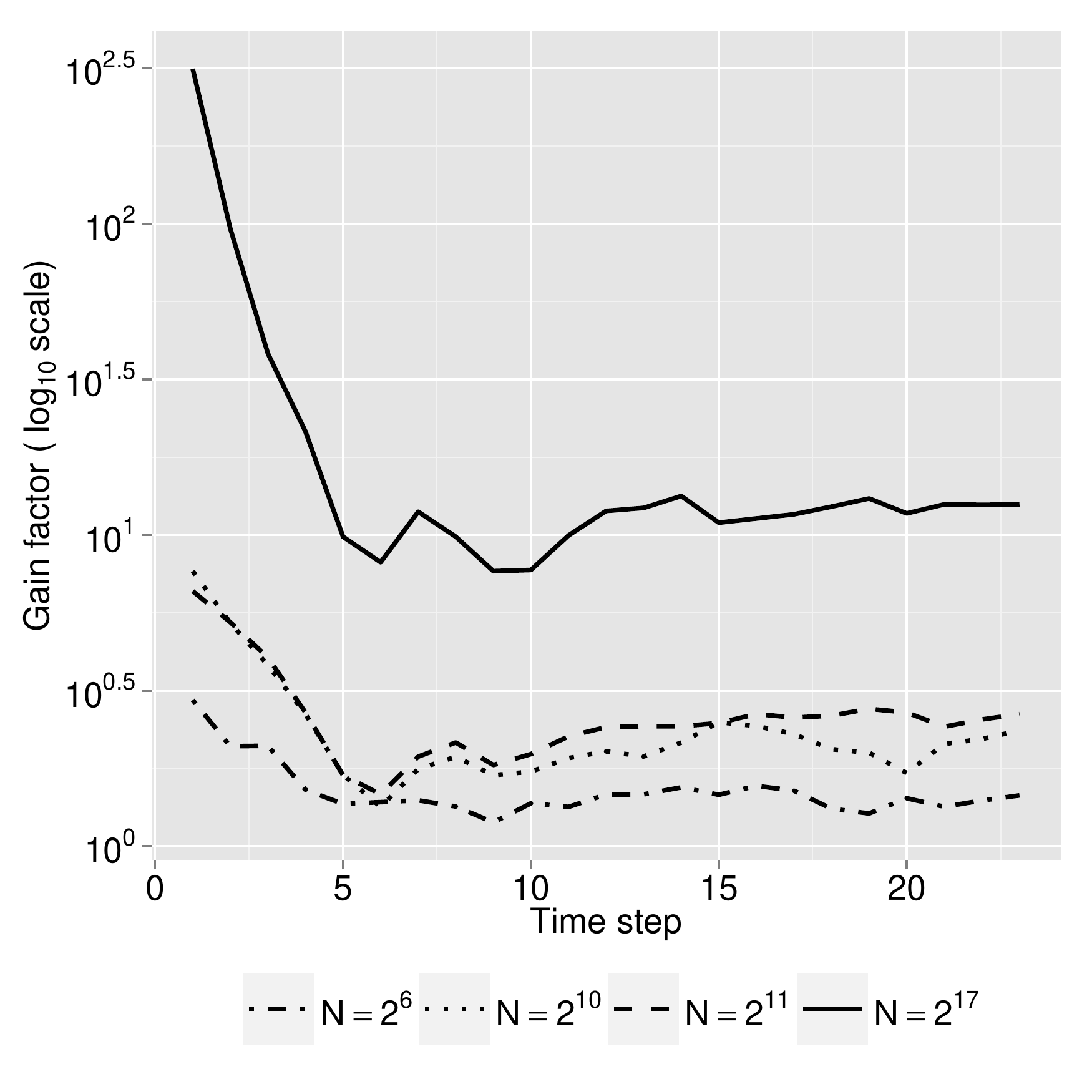}\includegraphics[scale=0.35]{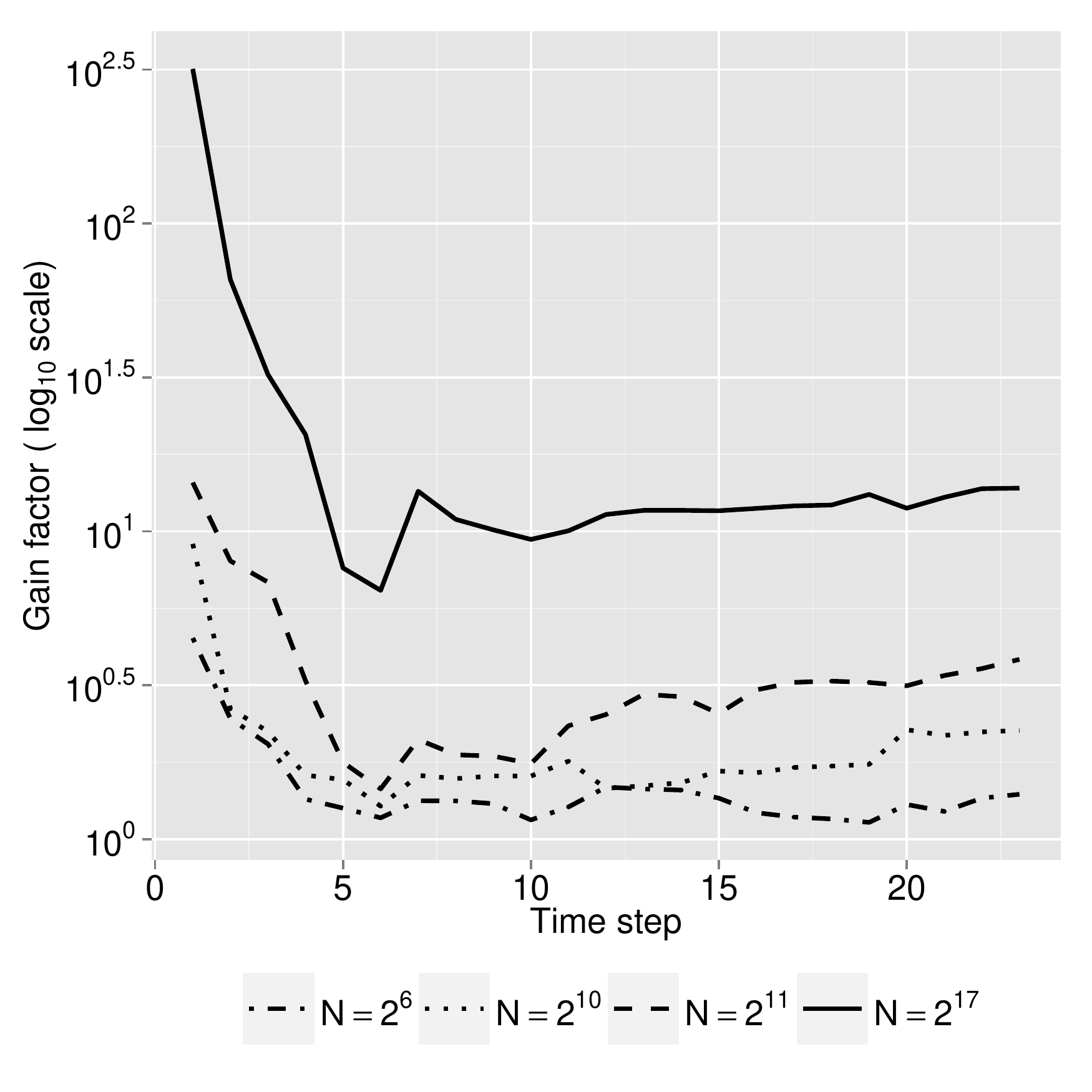} 

\includegraphics[scale=0.35]{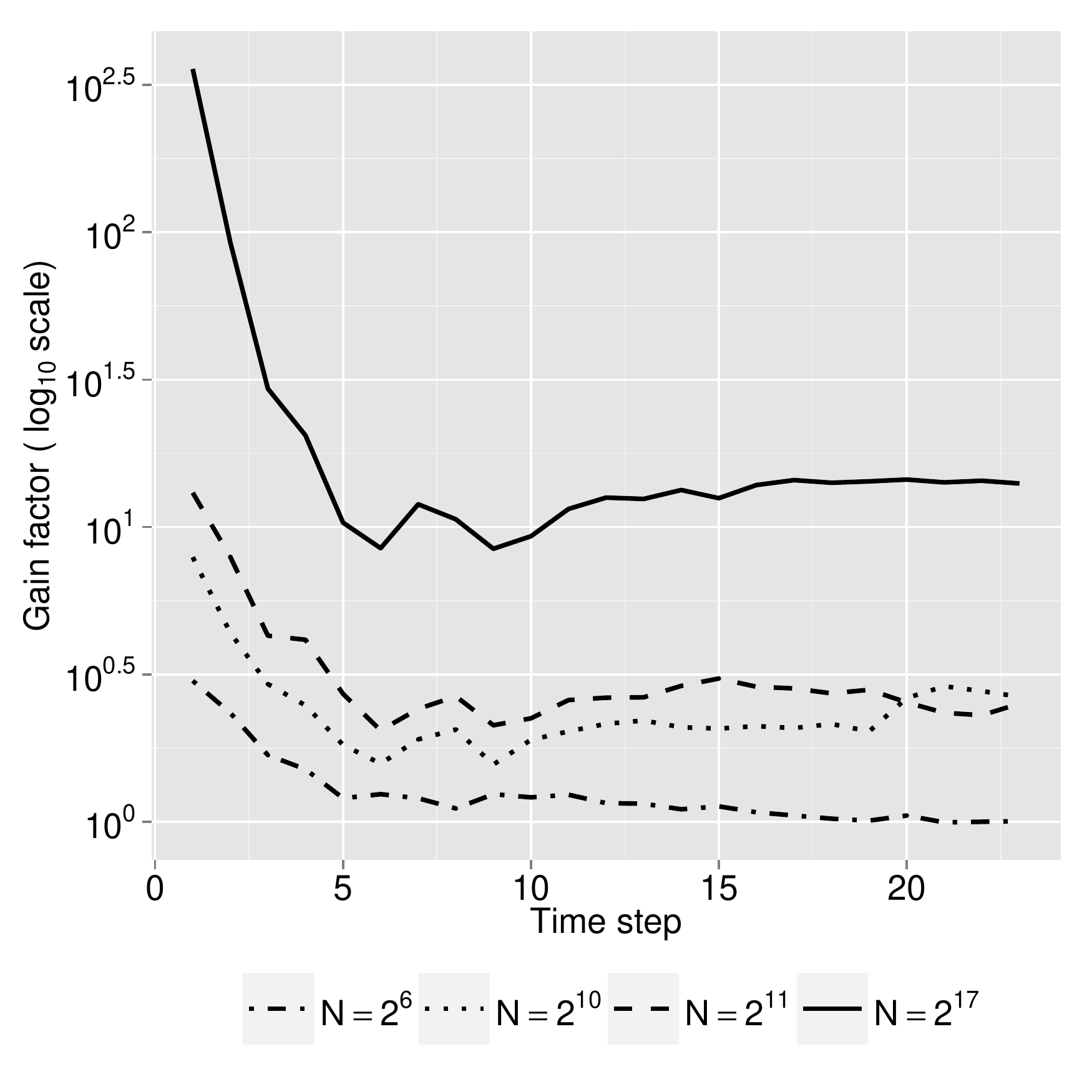}\includegraphics[scale=0.35]{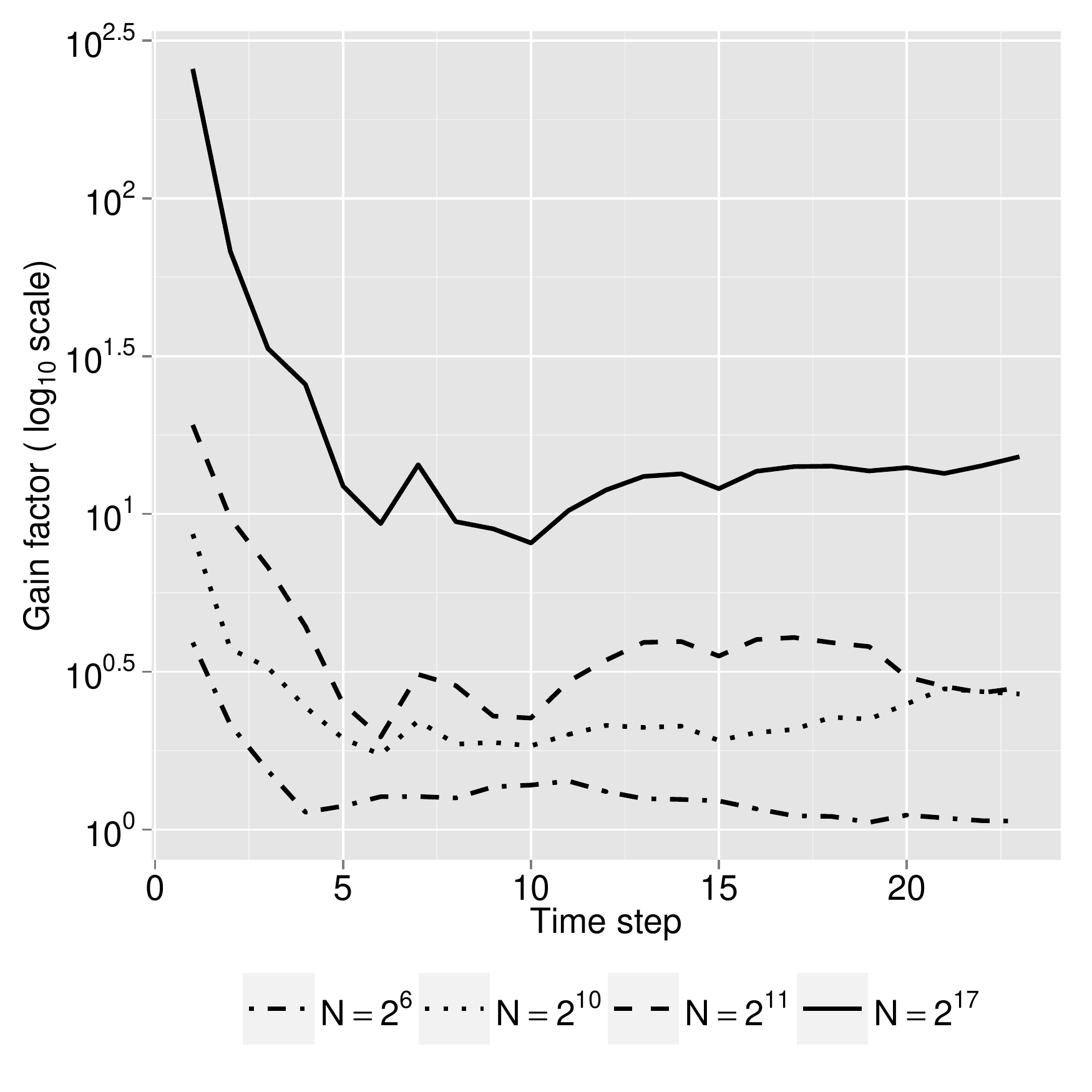} 
\par\end{centering}

\caption{Filtering of the Neuro decoding model \eqref{sim:eq:neural2}. From left to right and from top to bottom,  the graphs give the ratio of the SMC and SQMC MSE for the estimation of $\E[x_{kt}|\mathbf{y}_{0:t}]$ as a function of $t$, $k=1,\dots,4$,  and are obtained from 200 independent runs of SQMC and SMC.}\label{fig:Neuro:EX2}
\end{figure}

Figures \ref{fig:Neuro:Lik2} and \ref{fig:Neuro:EX2} present, respectively, results for the estimation of the log-likelihood (evaluated at the true value of the parameters) and for the estimation of the filtering expectation $\E[x_{ti}|\mathbf{y}_{0:t}]$ for $i\in 1:d$. Concerning the log-likelihood estimation we  observe fast increase of the gain factor after about $2^{11}$ particles with a maximum close to 21 when $N$ is very large. The gain of SQMC compensates its longer running time after only about $0.17$ seconds. Important and increasing (in $N$) gains  are also observed for the estimation of the filtering expectations.

\section{Conclusion and future work\label{sec:Conclusion}}

The main message of the paper is that SMC users should be strongly encouraged 
to switch to SQMC, as SQMC is ``typically'' much more accurate (produces estimates with
smaller errors) than SMC. We add the word ``typically'' to recall that our asymptotic
analysis, by construction, proves only that  the SQMC error is smaller than the SMC error
\emph{for $N$ large enough}. But our range of numerical examples, which are representative of real-world filtering problems, makes us optimistic than in most
practical  cases SQMC should outperform SMC even for moderate values of $N$. 

The main price to pay to switch to SQMC is that users should spend some time thinking
on how to write the simulation of $\bx_t^n$ given $\bx_{t-1}^n$ as $\bx_t^n=\Gamma_t(\bx_{t-1}^n,\bu_t^n)$, where $\bu_t^n\sim\Unif([0,1]^d)$ and $\Gamma_t$ is a deterministic function that is easy to evaluate. Fortunately, this is often straightforward. In fact, there are many models of interest where $\bx_t^n$ given $\bx_{t-1}^n$ is linear and Gaussian. Since this case is already implemented in
our program, adapting it to such a model should be just a matter of changing a few
lines of code (to evaluate the probability density of $\by_t$ given $\bx_t$). 

Regarding future work, the most pressing tasks seem (a)
to refine the convergence rate of the SQMC error; and (b) to establish that 
it does not degenerate over time 
\citep[in the spirit of time-uniform estimates for SMC, see p. 244 of][]{DelMoral:book}.
Regarding the former, \cite{He2014} make the interesting conjecture that 
the mean square error of SQMC converges at rate $\bigO(N^{-1-2/d})$. This would explain 
why the relative performance of SQMC decreases with the dimension. 
Fortunately, a majority of the state space models of interest in signal processing, finance, or other fields are such that $d\leq 6$. A notable exception
is geophysical data assimilation
(in e.g. meteorology or oceanography)
for which $d$ can be very large, but for such large-dimensional problems 
SMC seems to perform too poorly for practical use anyway
\citep{bocquet2010beyond}. 

Finally, it is also our hope that this paper will help QMC  garner wider  recognition 
in Bayesian computation and related fields. 
Granted, QMC is more technical than standard Monte Carlo, and there is perhaps something specific about particle filtering that makes the introduction of QMC
so effective. Yet we cannot help but think that the full potential of QMC in Statistics
remains under-explored.

\section*{Acknowledgements}


We thank the referees, 
Christophe Andrieu, 
Simon Barthelm\'e, 
Arnaud Doucet,
Paul Fearnhead, 
Simon Lacoste-Julien, 
and Art Owen
for excellent remarks that helped us to greatly improve the paper.  
The second author is partially supported by a grant from the French National
Research Agency (ANR) as part of the  ``Investissements d'Avenir'' program 
(ANR-11-LABEX-0047).

\bibliographystyle{apalike}
\bibliography{complete}

\appendix

\section{Proofs}

\subsection{Importance sampling: Theorems \ref{thm:IS_1} and \ref{thm:IS_2}}\label{sub:proofs_IS}

\subsubsection{Preliminary calculation}

Let $\hat{q}(\dx)=\mathcal{S}(\bx^{1:N})(\dx)=N^{-1}\sum_{n=1}^{N}\delta_{\bx^{n}}(\dx)$,
and, as a preliminary calculation, take $\varphi\in L_{2}\left(\ui^{d},\lambda_{d}\right)$
and
\begin{align}
\left|\pi^N(\varphi)-\pi(\varphi)\right| & =\left|\frac{N^{-1}\sum_{n=1}^{N}w(\bx^{n})\varphi(\bx^{n})}{N^{-1}\sum_{n=1}^{N}w(\bx^{n})}-\pi(\varphi)\right|\nonumber \\
 & \leq\left|\frac{N^{-1}\sum_{n=1}^{N}w(\bx^{n})\varphi(\bx^{n})}{N^{-1}\sum_{n=1}^{N}w(\bx^{n})}-N^{-1}\sum_{n=1}^{N}w(\bx^{n})\varphi(\bx^{n})\right|\nonumber \\
 & \quad+\left|N^{-1}\sum_{n=1}^{N}w(\bx^{n})\varphi(\bx^{n})-q(w\varphi)\right|\nonumber \\
 & \leq\frac{N^{-1}\sum_{n=1}^{N}w(\bx^{n})\left|\varphi(\bx^{n})\right|}{N^{-1}\sum_{n=1}^{N}w(\bx^{n})}\left|q(w)-\hat{q}(w)\right|\nonumber \\
 & \quad+\left|\hat{q}(w\varphi)-q(w\varphi)\right|.\label{eq:prelim_result_is}
\end{align}

We will use this inequality in the two following proofs. 

\subsubsection{Proof of Theorem \ref{thm:IS_1}}

\label{sub:Proof-IS1}

Take $\varphi=\ind_{B}$ 
for $B\in\mathcal{B}_{\uid}$ in \eqref{eq:prelim_result_is}. Consider the first term above. The
ratio is bounded by $1$, and (since $w$ is bounded) $\left|q(w)-\hat{q}(w)\right|\rightarrow0$
by portmanteau lemma \citep[Lemma 2.2]{VanderVaart2007}. Now consider the second
term.

We follow essentially the same steps as in \citet[Lemma 2.2]{VanderVaart2007}.
Without loss of generality we assume that $q(\dx)$
is a continuous probability measure \citep[the same argument as in][is used for the general case]{VanderVaart2007}.

Let $\epsilon>0$ and take $J\in\mathcal{B}_{\uid}$ such that $q(J^{c})\leq\epsilon$.
Since $J$ is compact, $w(\cdot{})$ is uniformly continuous on $J$.
Let $\eta>0$ be such that $\|\bx-\mathbf{y}\|\leq\eta\implies|w(\bx)-w(\mathbf{y})|\leq\epsilon$,
$\forall(\bx,\mathbf{y})\in J^{2}$. 
Let $\{J_{k}\}_{k=1}^{m}$ be a split of $J$ into a   finite collection of $m$ closed hyperrectangles with radius (at most) $\eta$. Let $g(\bx)=\sum_{k=1}^{m}w(\bx_{k})\mathbb{I}_{J_{k}}(\bx)$ and note  that $|w(\bx)-g(\bx)|\leq 2^d\epsilon$,
$\forall\bx\in J$. 
Thus 
\begin{align*}
\left|\int_{B}w(\bx)\left\{ \hat{q}(\dx)-q(\dx)\right\} \right| & \leq\left|\int_{B}\left\{ w(\bx)-g(\bx)\right\} \hat{q}(\dx)\right|+\left|\int_{B}g(\bx)\left\{ \hat{q}(\dx)-q(\dx)\right\} \right|\\
 & \quad+\left|\int_{B}\left\{ w(\bx)-g(\bx)\right\} q(\dx)\right|
\end{align*}
where for the first term we have 
\begin{align}
\left|\int_{B}(w(\bx)-g(\bx))\hat{q}(\dx)\right| & \leq\left|\int_{B\cap J}(w(\bx)-g(\bx))\hat{q}(\dx)\right|+\left|\int_{B\cap J^{c}}w(\bx)\hat{q}(\dx)\right|\nonumber \\
 & \leq 2^d\epsilon+\|w\|_{\infty}\hat{q}(J^{c})\nonumber \\
 & \leq\epsilon(2^d+2\|w\|_{\infty})\label{eq:thmIS_1}
\end{align}
as $\hat{q}(J^{c})$ converges to $q(J^{c})$, and thus $\hat{q}(J^{c})\leq2\epsilon$
for $N$ large enough; and for the second term 
\begin{equation}
\begin{split}\left|\int_{B}g(\bx)\left\{ \hat{q}(\dx)-q(\dx)\right\} \right| & \leq\sum_{k=1}^{m}w(\bx_{k})\left|\int_{\bar{J}_{k}\cap B}\left\{ \hat{q}(\dx)-q(\dx)\right\} \right|\\
 & \leq \|\hat{q}(\dx)-q(\dx)\stn\sum_{k=1}^{m}w(\bx_{k}).
\end{split}
\label{eq:thmIS_1:2}
\end{equation}
Finally, for the third term:
\begin{align}
\left|\int_{B}\left\{ w(\bx)-g(\bx)\right\} q(\dx)\right| & \leq\left|\int_{B\cap J}\left\{ w(\bx)-g(\bx)\right\} q(\dx)\right|+\left|\int_{B\cap J^{c}}w(\bx)q(\dx)\right|\nonumber \\
 & \leq\epsilon(2^d+\|w\|_{\infty}).\label{eq:thmIS_1:3}
\end{align}
Putting \eqref{eq:thmIS_1}-\eqref{eq:thmIS_1:3} together shows that,
for all $B\in\mathcal{B}_{\ui^d}$ 
\begin{align}
\left|\int_{B}w(\bx)\left\{ \hat{q}(\dx)-q(\dx)\right\} \right| & \leq\epsilon(2^{d+1}+3\|w\|_{\infty})+\|\hat{q}(\dx)-q(\dx)\stn\sum_{k=1}^{m}w(\bx_{k})\nonumber \\
 & \leq\epsilon(2^{d+2}+3\|w\|_{\infty})\label{eq:IS4}
\end{align}
for $N$ large enough (as $\|\hat{q}(\dx)-q(\dx)\stn\rightarrow0$)
which concludes the proof of Theorem \ref{thm:IS_1}.

\subsubsection{Proof of Theorem \ref{thm:IS_2}}\label{sub:Proof-IS2}

We prove first $L_1$ convergence (first part of Theorem \ref{thm:IS_2}).
We start again from \eqref{eq:prelim_result_is}, but for any $\varphi\in L_{2}\left(\ui^{d},\lambda_{d}\right)$.
For the second term, by Jensen's inequality 
\[
\E\left|\hat{q}(w\varphi)-q(w\varphi)\right|\leq\left[\mathrm{Var}\left\{ \hat{q}(w\varphi)\right\} \right]^{1/2}=\bigO(r(N)^{1/2})
\]
by assumption. For the first term, using Cauchy-Schwartz, $\E(\left|CD\right|)\leq\left\{ \E(C^{2})\E(D^{2})\right\} ^{1/2}$ with
\[
C=\frac{N^{-1}\sum_{n=1}^{N}w(\bx^{n})\left|\varphi(\bx^{n})\right|}{N^{-1}\sum_{n=1}^{N}w(\bx^{n})},\quad D=q(w)-\hat{q}(w),
\]
we have $\left\{ \E(D^{2})\right\} ^{1/2}=\bigO(r(N)^{1/2})$, and what
remains to prove is that $\E(C^{2})=\bigO(1)$.

From \eqref{eq:IS4}, and under Assumption \ref{H:thmIS_2:2}, one sees that there
exists $N_{\epsilon}$ such that with probability one $N^{-1}\sum_{n=1}^{N}w(\bx^{n})\geq1/2$
 as soon as $N\geq N_{\epsilon}$. Under Assumption \ref{H:thmIS_2:1},
a bound similar to \eqref{eq:IS4} is easily obtained by replacing
$\bx^{1:N}$ with $\bu^{1:N}$ and observing that $w\circ F_{q}^{-1}$
is continuous and bounded. Thus, for $N$ large enough 
$$
\E(C^{2}) 
\leq 4\E\left\{\left[N^{-1}\sum_{n=1}^{N}w(\bx^{n})\left|\varphi(\bx^{n})\right|\right]^{2}\right\} 
\leq\bigO(r(N))+\pi(\left|\varphi\right|)^{2}=\bigO(1).
$$

We now prove $L_{2}$ convergence (second part of Theorem \ref{thm:IS_2}):
\[
\var\left\{ \pi^N(\varphi)\right\} \leq\left[\text{Var}\left\{ \pi^N(\varphi)-\hat{q}(w\varphi)\right\} ^{1/2}+\text{Var}\left\{ \hat{q}(w\varphi)\right\} ^{1/2}\right]^{2},
\]
with $\text{Var}\left\{ \hat{q}(w\varphi)\right\} =\bigO(r(N))$ by
assumption, and for the first term: 
\begin{align*}
\E\left[\left\{ \pi^N(\varphi)-\hat{q}(w\varphi)\right\} ^{2}\right]= & \E\left[\left\{ \sum_{n=1}^{N}\left\{ W^{n}-N^{-1}w(\bx^{n})\right\} \varphi(\bx^{n})\right\} ^{2}\right]\\
= & \E\left[\left\{ 1-N^{-1}\sum_{n=1}^{N}w(\bx^{n})\right\} ^{2}\left\{ \sum_{n=1}^{N}W^{n}\varphi(\bx^{n})\right\} ^{2}\right]\\
= & \E\left[\frac{\left\{ 1-\hat{q}(w)\right\} ^{2}\hat{q}(w\varphi)^{2}}{\hat{q}(w)^{2}}\right]\\
\leq & 4\E\left[\left\{ 1-\hat{q}(w)\right\} ^{2}\hat{q}(w\varphi)^{2}\right]
\end{align*}
for $N$ large enough, using the same argument as above (as $\hat{q}(w)\rightarrow1$).
Then 
\begin{align*}
\E\left[\left\{ 1-\hat{q}(w)\right\}^{2}\hat{q}(w\varphi)^{2}\right]\leq & \E\left[\left\{ 1-\hat{q}(w)\right\} ^{2}\left\{ \hat{q}(w\varphi)-\pi(\varphi)\right\} ^{2}\right]-\pi(\varphi)^{2}\E\left[\left\{ 1-\hat{q}(w)\right\} ^{2}\right]\\
 & +2|\pi(\varphi)|\E\left[|\hat{q}(w\varphi)|\left\{ 1-\hat{q}(w)\right\} ^{2}\right]
\end{align*}
where for the second term, $\E\left[\left\{ 1-\hat{q}(w)\right\} ^{2}\right]=\var\left[\hat{q}(w)\right]=\bigO\left(r(N)\right)$,
for the first term 
\begin{align*}
\E\left[\left\{ 1-\hat{q}(w)\right\} ^{2}\left\{ \hat{q}(w\varphi)-\pi(\varphi)\right\} ^{2}\right] & =\E\left[\left\{ 1-\hat{q}(w)\right\} ^{2}\left\{ \hat{q}(w\varphi)-q(w\varphi)\right\} ^{2}\right]\\
 & \leq(1+\|w\|_{\infty})^{2}\var\left[\hat{q}(w\varphi)\right]\\
 & =\bigO\left(r(N)\right)
\end{align*}
and finally for the third term 
\begin{align*}
\E\left[|\hat{q}(w\varphi)|\left\{ 1-\hat{q}(w)\right\} ^{2}\right]\leq & \E\left[|\hat{q}(w\varphi)-q(w\varphi)|\left\{ 1-\hat{q}(w)\right\} ^{2}\right]+|q(w\varphi)|\var\left[\hat{q}(w)\right]
\end{align*}
with 
\begin{align*}
\E\left[|\hat{q}(w\varphi)-q(w\varphi)|\left\{ 1-\hat{q}(w)\right\} ^{2}\right] & \leq\left(1+\|w\|_{\infty}\right)\var\left[\hat{q}(w\varphi)\right]^{1/2}\var\left[\hat{q}(w)\right]^{1/2}\\
 & =\bigO\left(r(N)\right)
\end{align*}
which concludes the proof.

For subsequent uses (see the proof of Theorem \ref{thm:PF2}), we note that these computations imply, for $N$ large enough,
\begin{align}
&\var\{\pi^N(\varphi)\}\leq \left\{2(1+\|w\|_{\infty})\var[\hat{q}(w\varphi)]^{1/2}+(1-2|\pi(\varphi)|)\var[\hat{q}(w)]^{1/2} \right\}^2\label{eq:thmIS_BoundL2}\\
&|\pi^N(\varphi)-\pi(\varphi)|\leq [\var\{\hat{q}(w\varphi)\}]^{1/2}+2[\var\{\hat{q}(w)\}]^{1/2}\left[\var\{\hat{q}(w\varphi)\}+\pi(|\varphi|)^2\right]^{1/2}\label{eq:thmIS_BoundL1}.
\end{align}

\subsection{Hilbert curve and discrepancy: Theorems \ref{thm:Hilbert}
and \ref{thm:LD}\label{sub:proofs_hilbert}}

The proofs in this section rely on the properties of the Hilbert curved
laid out in Section \ref{sub:The-Hilbert-space-filling} and the corresponding
notations.

\subsubsection{Theorem \ref{thm:Hilbert}}\label{sub:Proof-hilbert1}

We first show that $\|\pi_h^N-\pi_h\stn=\sup_{0\leq a<b\leq1}|\pi_h^N([a,b))-\pi_h([a,b))|$. Because $\pi_h$ is a continuous probability measure on $\ui$, the result is obvious if $\pi_h^N$ is continuous as well. Let $0\leq a<b<1$ be such that $b$ is a discontinuity point of $F_{\pi_h^N}$ and let $\delta>0$ be small enough so that $\pi_h^N([a,b])=\pi_h^N([a,b+\delta))$ and $b+\delta\leq 1$. Then,
$$
\Big||\pi_h^N([a,b])-\pi_h([a,b])|-|\pi_h^N([a,b+\delta))-\pi_h([a,b+\delta))|\Big|\leq \pi_h([b,b+\delta]).
$$
By the bi-measure property of the Hilbert curve, the set $H([b,b+\delta])$ has Lebesgue measure $\delta$ in $\ui^d$ and therefore, $\pi_h([b,b+\delta])=\pi\big(H[b,b+\delta])\big)\leq \|\pi\|_{\infty}\delta$ where $\|\pi\|_{\infty}<+\infty$ by assumption. Hence, for all $\epsilon>0$ small enough, 
$$
\left|\|\pi_h^N-\pi_h\stn-\sup_{0\leq a<b\leq1}|\pi_h^N([a,b))-\pi_h([a,b))|\right|\leq \epsilon.
$$

To prove the theorem note that the above computations imply that
\begin{align*}
\|\pi_h^N-\pi_h\stn&\leq 2\sup_{b\in(0,1)}|\pi_h^N([0,b])-\pi_h([0,b])|.
\end{align*}
To bound the right-hand side, let $I=[0,b]$, $b\in(0,1)$, and  $m\in\mathbb{N}$ (which may depend on
$N$) and assume first that $b\geq2^{-dm}$, so that $I_{m}^{d}(0)\subseteq I$.
Take $\tilde{I}=[0,k^*2^{-dm}]$, where $k^*\leq(2^{dm}-1)$ is the largest
integer such that $k^*2^{-dm}\leq b$. Then 
\begin{align}
\left|\pi_h^N(I)-\pi_{h}(I)\right| & \leq\left|F_{\pi_h^N}\left(k^*2^{-dm}\right)-F_{\pi_{h}}\left(k^*2^{-dm}\right)\right|\nonumber \\
 & +\left|\pi_h^N(I)-F_{\pi_h^N}\left(k^*2^{-dm}\right)-\left\{ \pi_{\IHSFC}(I)-F_{\pi_{h}}\left(k^*2^{-dm}\right)\right\} \right|\nonumber \\
 & =\left|\pi^N(J)-\pi(J)\right|+\left|\pi_h^N\left((k^*2^{-dm},b]\right)-\pi_{\IHSFC}\left((k^*2^{-dm},b]\right)\right|\label{eq:proof_hilbert}
\end{align}
with $J=\HSFC(\tilde{I})$. Since $\tilde{I}$ is the union of $k^*$ intervals in $\mathcal{I}_m^d$, $J$ is the union of $k^*$ hypercubes in $\mathcal{S}_m^d$, and therefore  \citep[using a similar argument as above and][Proposition 2.4]{Niederreiter1992},
\begin{align*}
\left|\pi^N(J)-\pi(J)\right|&\leq c\|\pi^N-\pi\stn\leq 2^{dm}r(N)
\end{align*}
for a constant $c$ and where $r(N)=\|\pi^N-\pi\stn.$

For the second term of \eqref{eq:proof_hilbert}, by the properties
of the Hilbert curve, 
\begin{align*}
\left|\pi_h^N\left((k2^{-dm},b]\right)-\pi_{\IHSFC}\left((k2^{-dm},b]\right)\right| & \leq\pi_h^N\left(I_{m}^{d}(k)\right)+\pi_{\IHSFC}\left(I_{m}^{d}(k)\right)\\
 & =\pi^N\left(S_{m}^{d}(k)\right)+\pi\left(S_{m}^{d}(k)\right)\\
 & \leq2\pi\left(S_{m}^{d}(k)\right)+r(N)\\
 & =\bigO\left(2^{-dm}\vee r(N)\right)
\end{align*}
where the last inequality comes from the fact that $\pi(\bx)$ is
a bounded density.

In case $b<2^{-dm}$, similar computations show that 
\begin{align*}
 & \left|\pi_h^N(I)-\pi_{\IHSFC}(I)\right|\leq\pi_h^N(I_{m}^{d}(0))+\pi_{\IHSFC}(I_{m}^{d}(0))=\bigO\left(2^{-dm}\vee r(N)\right).
\end{align*}
To conclude, we choose $m$ so that $2^{-dm}=\bigO(r(N)^{1/2})$,
which gives 
\begin{align*}
\sup_{b\in (0,1)} \left|\pi_h^N([0,b])-\pi_{\IHSFC}([0,b])\right|=\bigO\left(r(N)^{1/2}\right).
\end{align*}
Finally, since replacing $[0,b]$ by $[0,a)$ changes nothing to the proof of the result above, one may conclude that $\sup_{I\in\mathcal{B}_{\ui}}|\pi_h^N(I)-\pi_h(I)|=\bigO(r(N)^{1/2})$.

\subsubsection{Proof of Theorem \ref{thm:LD}}\label{sub:Proof-hilbert3}

\paragraph{Preliminary computations}$ $

The proof of this result is based on \citet[][``Satz 2'']{Hlawka1972}. Compared to this latter, the main technical difficulty comes from the fact that the Rosenblatt transformation $F_{\pi_h^N\otimes K_h}$ is not continuous because $\pi_h^N$ is a weighted sum of Dirac measures. To control the ``jumps'' of the inverse Rosenblatt transformation $F^{-1}_{\pi_h^N\otimes K_h}$ introduced by the discontinuity of $\pi_h^N$, we first prove the following Lemma.

\begin{lem}\label{lemma:thmLD}
Consider the set-up of Theorem \ref{thm:LD}. For $n\in \onetoN$, let   $h_1^n=H(\bx^n_1)$  and assume that the points $h_1^{1:N}$ are labelled so that $n<m\implies h_1^n < h_1^m$. (Note that  the inequality is strict because, by Assumption \ref{H:thmLD:2} of Theorem \ref{thm:LD}, the points $\bx^{1:N}$ are distinct.) Without loss of generality, assume that $h_1^1>0$ and let $h_1^0=0$. Then,  as $N\rightarrow+\infty$,
$$
\max_{n\in 1:N}|h_1^{n}-h_1^{n-1}|\cvz.
$$
\end{lem}

To prove this Lemma, let $J_N=[h_1^{n^*-1},h_1^{n^*}]$ where $|h_1^{n^*}-h_1^{n^*-1}|=\max_{n\in 1:N}|h_1^{n}-h_1^{n-1}|$. Since $J_N$ contains at most two points, we have
$$
\pi_h(J_N)\leq \pi_h^N(J_N)+r_2(N)\leq 2r_1(N)+r_2(N)
$$
where $r_1(N)=\max_{n\in 1:N}W_N^n$ and $r_2(N)=\|\pi_h^N-\pi_h\stn$; note $r_1(N)\cvz$ by Assumption \ref{H:thmLD:2} of Theorem \ref{thm:LD} while $r_2(N)\cvz$ by Assumption \ref{H:thmLD:3} of Theorem \ref{thm:LD} and  by Theorem \ref{thm:Hilbert}. Therefore, $\pi_h(J_N)\cvz$ as $N\rightarrow+\infty$.

Assume now that $\max_{n\in 1:N}|h_1^{n}-h_1^{n-1}|\not\cvz$. Then, this means that there exists a $\epsilon\in (0,1)$ such that, for all $N>1$ there exists a $N^*\geq N$ for which $\lambda_1(J_{N^*})\geq \epsilon$. Assume  first that $J_{N^*}\subset [0,1-\frac{\epsilon}{2}]$. In that case,  we have $\pi_h(J_{N^*})\geq c_{\epsilon}$ for a constant $c_{\epsilon}>0$.  Indeed, by the continuity of the Hilbert curve, the set $H([0,1-\frac{\epsilon}{2}])$ is compact and therefore, $\forall \bx\in H([0,1-\frac{\epsilon}{2}])$,   $\pi(\bx)\geq \underline{\pi}^{(\epsilon)}$ for a constant $\underline{\pi}^{(\epsilon)}>0$ because the density $\pi(\bx)$ is continuous and strictly positive. Therefore, if $J_{N^*}\subset [0,1-\frac{\epsilon}{2}]$, we have 
$$
\pi_h(J_{N^*})=\pi(H(J_{N^*}))\geq  \underline{\pi}^{(\epsilon)}\lambda_d(H(J_{N^*}))
=\underline{\pi}^{(\epsilon)}\lambda_1(J_{N^*})
\geq \epsilon\underline{\pi}^{(\epsilon)}
$$  
where the second equality uses  the bi-measure property of the Hilbert curve.

Assume now that $J_{N^*}\not\subset [0,1-\frac{\epsilon}{2}]$. Write $J_{N^*}=[a_{N^*},b_{N^*}]$ and note that, since $\lambda_1(J_{N^*})\geq\epsilon$, we have $a_n^*<1-\epsilon$ and therefore
$$
\pi_h(J_{N^*})=\pi_h\left(\left[a_{N^*},1-\frac{\epsilon}{2}\right]\right)+ \pi_h\left(\left(1-\frac{\epsilon}{2},b_{N^*}\right]\right)\geq \left(1-\frac{\epsilon}{2}-a_{N^*}\right) \underline{\pi}^{(\epsilon)} \geq \frac{\epsilon}{2}\underline{\pi}^{(\epsilon)}.
$$
Thus, this shows that if $\max_{n\in 1:N}|h_1^{n}-h_1^{n-1}|\not\cvz$, then there exists a $\epsilon\in [0,1)$ such that  $\limsup_{N\rightarrow +\infty} \pi_h(J_N)\geq (\epsilon\underline{\pi}^{(\epsilon)})/2>0$. This  contradicts the fact that $\pi_h(J_N)\cvz$ as $N\rightarrow+\infty$ and the proof is complete.



\paragraph{Proof of Theorem \ref{thm:LD}} $ $

We use the shorthand $\opA(B)=\Sop(\bu^{1:N})(B)$ for any set $B\subset\ui^{1+d_2}$. One has 
\[
\|\Sop(P_h^{N})-\pi_h^N\otimes K_h\stn=\sup_{B\in\mathcal{B}^N_{[0,1)^{1+d_2}}}\left|\opA\left(E^N(B)\right)-\lambda_{1+d_2}\left(E^N(B)\right)\right|
\]
where 
$$
\mathcal{B}^N_{\ui^{1+d_2}}=\left\{B=[\bm{a},\bm{b}]\in \mathcal{B}_{\ui^{1+d_2}}:  \min_{n\in 1:N} h(\bx^n_1)\leq F_{\pi_h^N}(b_1)\leq \max_{n\in 1:N} h(\bx^n_1)\right\},
$$
and where, for an arbitrary set $\tilde{B}=[a_1,b_1]\times[\bm{a}',\bm{b}']$ with $0\leq a_1\leq b_1<1$ and  with $0\leq a_i'\leq b_i'<1$ for all $i\in 1:d_2$, we use the shorthand  $E^N(\tilde{B})$  for the set
$$
\left\{ (u_1,\bu_2)\in\ui^{1+d_2} : F_{\pi_h^N}(a_1)\leq  u_1\leq F_{\pi_h^N}(b_1), \bu_2 \in F_{K_h}\left(F^{-1}_{\pi_h^N}(u_1),[\bm{a}',\bm{b}']\right)\right\}.
$$ 

Let  $\mathcal{P}$ be a partition of $[0,1)^{1+d_{2}}$ in $L^{d_1+d_2}$ 
congruent hyperrectanges $W$ of size $L^{-d_{1}}\times L^{-1}\times...\times L^{-1}$
where $L\geq1$ is an arbitrary integer. Let $B=[a_1,b_1]\times [\bm{a}',\bm{b}']\in\mathcal{B}^N_{[0,1)^{1+d_{2}}}$,
$\mathcal{U}_{1}$ the set of the elements of $\mathcal{P}$ that
are strictly in $E^N(B)$, $\mathcal{U}_{2}$
the set of elements $W\in\mathcal{P}$ such that $W\cap\partial(E^N(B))\neq\emptyset$,
$U_{1}=\cup\text{ }\mathcal{U}_{1}$, $U_{2}=\cup\text{ }\mathcal{U}_{2}$,
and $U_{1}'=E^N(B)\setminus U_{1}$ so that
\[
\opA\left(E^N(B)\right)-\lambda_{1+d_{2}}\left(E^N(B)\right)=\opA(U_{1})-\lambda_{1+d_{2}}(U_{1})+\opA(U_{1}')-\lambda_{1+d_{2}}(U_{1}').
\]
To bound $\opA(U_{1}')-\lambda_{1+d_{2}}(U_{1}')$, note that we can
cover $U_{1}'$ with sets in $\mathcal{U}_{2}$, hence 
\[
\opA(U_{1}')-\lambda_{1+d_{2}}(U_{1}')\leq\opA(U_{2}),\quad\mbox{and }\opA(U_{1}')-\lambda_{1+d_{2}}(U_{1}')\geq-\lambda_{1+d_{2}}(U_{2})
\]
so that, by the definition of $D(\bu^{1:N})$, 
\begin{align*}
\left|\opA(U_{1}')-\lambda_{1+d_{2}}(U_{1}')\right| & \leq\left|\opA(U_{2})-\lambda_{1+d_{2}}(U_{2})\right|+\lambda_{1+d_{2}}(U_{2})\notag\\
&\leq\#\mathcal{U}_{2}\left\{ D(\bu^{1:N})+L^{-(d_1+d_2)}\right\} .
\end{align*}
We therefore have 
\begin{align*}
\left|\opA\left(E^N(B)\right)-\lambda_{1+d_{2}}\left(E^N(B)\right)\right|&\leq|\opA(U_{1})-\lambda_{1+d_{2}}(U_{1})|+\#\mathcal{U}_{2}\left\{ D(\bu^{1:N})+L^{-(d_1+d_2)}\right\}\\
&\leq L^{d_1+d_2}D(\bu^{1:N})+ \#\mathcal{U}_{2}\left\{ D(\bu^{1:N})+L^{-(d_1+d_2)}\right\}.
\end{align*}

The rest of the proof is dedicated to bounding $\#\mathcal{U}_{2}$,
the number of hyperrectangles in $\mathcal{P}$ required to cover
$\partial \left(E^N(B)\right)$. To that effect, first note that, using the continuity of $F_{K_h}$ and the fact that $B$  and $E^N(B)$ are  closed sets, we can easily show that $ E^N(\partial(B))\subset \partial(E^N(B))$. Let $\#\mathcal{U}^{(1)}_{2}$ and $\#\mathcal{U}_2^{(2)}$ be, respectively, the number of hyperrectangles in $\mathcal{P}$ we need to cover $E^N(\partial(B))$ and to cover $P(B):=\partial(E^N(B))\setminus E^N(\partial(B))$. Hence, $\#\mathcal{U}_{2} \leq \#\mathcal{U}^{(1)}_{2}+\#\mathcal{U}^{(2)}_{2}$ and we now bound $\#\mathcal{U}^{(i)}_{2}$, $i\in1:2$.

To bound $\# \mathcal{U}^{(1)}_{2}$  we first cover $\partial (B)$ with hyperrectangles belonging to a partition $\mathcal{P}'$ of the set $\ui^{1+d_2}$. We construct  $\mathcal{P}'$ as a partition of the
set $[0,1)^{1+d_{2}}$ into hyperrectangles $W'$ of size $L'{}^{-d_{1}}\times L'{}^{-1}\times...\times L'{}^{-1}$
such that, for all points $(h_1,\bx_2)$ and $(h_1',\bx_2')$ in
$W'$, we have 
\begin{equation}
\left\Vert F_{K_h}\left( h_1,\bx_2\right)-F_{K_h}\left(h_1',\bx_2'\right)\right\Vert_{\infty}=\left\Vert F_{K}\left(H(h_1),\bx_2\right)-F_{K}\left(H(h_1'),\bx_2'\right)\right\Vert _{\infty}\leq L^{-1}\label{eq:thmLD:cond1}
\end{equation}
and 
\begin{equation}
|F_{\pi_h^N}(h_1)-F_{\pi_h^N}(h_1')|\leq L^{-d_{1}}.\label{eq:thmLD:cond2}
\end{equation}

Let $L'=2^{m}$ for an integer $m\geq0$, so that $h_1$ and $h_1'$ are
in the same interval $I_{m}^{d_{1}}(k)\in\mathcal{I}_{m}^{d_{1}}$,
and $H(h_1)$ and $H(h_1')$ belong to the same hypercube in $\mathcal{S}_{m}^{d_{1}}$.
Let $C_{K}$ be the \Lip constant of $F_{K}$, then
\begin{align*}
\left\Vert F_{K}\left(H(h_1),\bx_2\right)-F_{K}\left(H(h_1'),\bx_2'\right)\right\Vert _{\infty} & \leq C_{K}\left\{ \|\bx_2-\bx_2'\|_{\infty}\vee\|H(h_1)-H(h_1')\|_{\infty}\right\}\\
& \leq C_{K}L'^{-1}
\end{align*}
and Condition \eqref{eq:thmLD:cond1} is verified as soon as $L'\geq C_{K}L$.
Let us now look at Condition \eqref{eq:thmLD:cond2}. We have: 
\begin{align*}
\left|F_{\pi_h^N}(h_1)-F_{\pi_h^N}(h_1')\right| & \leq 2\|F_{\pi_h^N}-F_{\pi_h}\|_{\infty}+|F_{\pi_h}(h_1)-F_{\pi_h}(h_1')|\\
 & \leq 2r_{2}(N)+\left|F_{\pi_h}(h_1)-F_{\pi_h}(h_1')\right|
\end{align*}
where, as in the proof of Lemma \ref{lemma:thmLD}, $r_2(N)=\|\pi_h^N-\pi_h\stn$. Since $h_1$ and $h_1'$ are in the same interval $I_{m}^{d_{1}}(k)\in\mathcal{I}_{m}^{d_{1}}$,
\[
\left|F_{\pi_h}(h_1)-F_{\pi_h}(h_1')\right|\leq\pi_h\left(I_{m}^{d_{1}}(k)\right)=\pi\left(S_{m}^{d_{1}}(k)\right)\leq\frac{\|\pi\|_{\infty}}{(L')^{d_{1}}}
\]
as $\pi$ is bounded. To obtain both \eqref{eq:thmLD:cond1} and \eqref{eq:thmLD:cond2},
we can take $L'=2^{m}$ to be the smallest power of 2 such that $L'\geq k_N L$ where
\[
k_{N}= C_K+\left(\frac{\|\pi\|_{\infty}}{(1-L^{d_{1}}2r_2(N))}\right)^{1/d_{1}}
\]
which implies that we assume from now on that $L^{-d_{1}}\geq 4r_2(N)$
for $N$ large enough.

Let $R\in\partial B$ be a $d_{2}$-dimensional face of $B$ and let
$\mathcal{R}$ be the set of hyperrectangles $W'\in\mathcal{P}'$
such that $R\cap W'\neq\emptyset$. Note that $\#\mathcal{R}\leq L'{}^{d_{1}+d_{2}-1}\leq (2k_{N}L)^{d_{1}+d_{2}-1}$. For each $W'\in\mathcal{R}$, take a point $\mathbf{r}^{W'}=(r_{1}^{W'},\mathbf{r}_{2}^{W'})\in R\cap W'$ and define 
\[
\tilde{\mathbf{r}}^{W'}=(\tilde{r}_{1}^{W'},\tilde{\mathbf{r}}_{2}^{W'})=F_{\pi_h^N\otimes K_h}(\mathbf{r}^{W'})\in E^N(R).
\]
Let $\tilde{\mathcal{R}}$ be the collection of hyperrectangles $\tilde{W}$
of size $4L^{-d_{1}}\times2L^{-1}\times...\times2L^{-1}$ and having
point $\tilde{\mathbf{r}}^{W'}$, $W'\in\mathcal{R}$, as middle point.

For an arbitrary $\bu=(u_1,\bu_{2})\in E^N(R)$,
let $h_1=a_1\vee F^{-1}_{\pi_h^N}(u_1)$ and $\bx_2=F^{-1}_{K_h}(h_1,\bu_2)$. Since $\bx=(h_1,\bx_2)\in R$,  $\bx$ is in one hyperrectangle $W'\in\mathcal{R}$. Hence,  using \eqref{eq:thmLD:cond1} and \eqref{eq:thmLD:cond2},
\begin{align*}
|u_1-\tilde{r}_{1}^{W'}|\leq |F_{{\pi}_h^N}(h_1)-F_{{\pi}_h^N}(r_{1}^{W'})|+|u_1-F_{{\pi}_h^N}(h_1)|\leq L^{-d_{1}}+r_1(N),
\end{align*}
where, as in the proof of Lemma \ref{lemma:thmLD}, $r_1(N)=\max_{n\in 1:N}W_N^n$, and 
\[
\|\bu_{2}-\tilde{\mathbf{r}}_{2}^{W'}\|_{\infty}=\|F_{K_h}\left(h_1,\bx_2\right)-F_{K_h}(r_{1}^{W'},\mathbf{r}_{2}^{W'})\|_{\infty}\leq L^{-1}.
\]
Assume from now on that $L^{-d_1}\geq  r_1(N)+4r_2(N)$. Then, this shows that $\bu$ belongs to the hyperrectangle $\tilde{W}\in\tilde{\mathcal{R}}$
with center $\tilde{\mathbf{r}}^{W'}$ so that $E^N(R)$
is covered by  at most  $\# \tilde{\mathcal{R}}=\# \mathcal{R}\leq (2k_{N}L)^{d_{1}+d_{2}-1}$
hyperrectangles $\tilde{W}\in\tilde{\mathcal{R}}$. To  go back to the initial partition of $[0,1)^{1+d_{2}}$
with hyperrectangles in $\mathcal{P}$, remark that every hyperrectangles in
$\tilde{\mathcal{R}}$ is covered by at most $c^*$ hyperrectangles in $\mathcal{P}$ for a constant $c^*$. Finally, since the set $\partial B$
is made of the union of $2(d_{2}+1)$ $d_{2}$-dimensional faces of $B$, we have 
\begin{align}\label{eq:thmLDBound}
\# \mathcal{U}_2^{(1)}\leq c_N L^{d_1+d_2-1}
\end{align}
where $c_N=c^*2(d_2+1)(2k_N)^{d_1+d_2-1}$.

We now consider the problem of bounding $\#\mathcal{U}_2^{(2)}$, the number of hyperrectangles in $\mathcal{P}$ we need to cover the set $P(B)=\partial(E^N(B))\setminus E^N(\partial(B))$.  Note that $P(B)$ contains the boundaries of the set $E^N(B)$ that are due to the discontinuities of $F_{\pi_h^N\otimes K_h}$.

To that effect, we show that there exists a finite collection $\{D^N_m\}_{m=1}^k$ of sets in $\mathcal{B}^N_{\ui^{1+d_2}}$ such that, for any  $\bu=(u_1,\bu_2)\in P(B)$, there exists a $m^*\in 1:k$ and a point $\tilde{\bu}=(\tilde{u}_1,\tilde{\bu}_2)\in E^N(\partial(D^N_{m^*}))$ which verifies $\tilde{u}_1=u_1$ and $\|\bu_2-\tilde{\bu}_2\|_{\infty}\leq Cr_3(N)^{1/d_1}$ for a constant $C$ and where $r_3(N)=\max_{n\in 1:N} |h_1^{n}-h_1^{n-1}|$; note that $r_3(N)\cvz$ as $N\rightarrow +\infty$ by Lemma \ref{lemma:thmLD}. Hence,   by taking $L$ small enough (i.e. such that $L^{-1}\geq 2 C r_3(N)^{1/d_1}$), we have $\#\mathcal{U}_2^{(2)}\leq \sum_{m=1}^k \#\mathcal{U}_2^{(D_m^N)}$ where $\#\mathcal{U}_2^{(D_m^N)}$  is the number of hyperrectangles in $\mathcal{P}$  we need to cover  $E^N(\partial(D^N_{m}))$. Then, because  the bound   we derived  above for the number of these hyperrectangles  required to cover $E^N(\partial(B)))$ is uniform in $B\in\mathcal{B}^N_{\ui^{1+d_2}}$, one can conclude using \eqref{eq:thmLDBound} that  $\mathcal{U}_2^{(2)}\leq k c_NL^{d_1+d_2-1}$.

To construct the collection $\{D^N_m\}_{m=1}^k$, let $\bu=(u_1,\bu_2)\in P(B)$, that is,  $u_1= F_{\pi_h^N}(h_1^{n^*})$ for a $n^*\in 1:N$  and  $\bu_2=F_{K_h}(h_1^{n^*},\bx^*)$ with  $\bx^*\in (\bm{a}',\bm{b}')$. By the definition of the boundary of a set,  for any $\epsilon>0$ there exists a $\bv=(v_1,\bv_2)\not\in E^N(B)$ such that $\|\bu-\bv\|_{\infty}\leq \epsilon$. Let  $\epsilon>0$  and assume that the point $\bv=(u_1-\epsilon,\bu_2)$ verifies this condition, that is, $\bu_2\not\in F_{K_h}(h_1^{n^*-1}, [\bm{a}',\bm{b}'])$, $n^*>1$. (The case  $v_1=(u_1+\epsilon,\bu_2)$ is treated in a similar way, just replace $n^*-1$ by $n^*+1$ in what follows.)

We now show that there exists a set $B^N\in \mathcal{B}^N_{\ui^{1+d_2}}$  and a point $\tilde{\bu}=(u_1,\tilde{\bu}_2)\in E^N(\partial(B^N))$ such that $\|\bu_2-\tilde{\bu}_2\|_{\infty}\leq C r_3(N)^{1/d_{1}}$ for a constant $C$. We consider the set $B^N=[a_1,b_1]\times [\bm{a}^N,\bm{b}^N]$ where $\bm{a}^N<\bm{b}^N\in \ui^{d_2}$.  In order to construct $[\bm{a}^N,\bm{b}^N]$,  we write $F_i(h_1, x_{1:i-1}, x_i)$ the $i$-th coordinate of $F_{K_h}(h_1,\bx)$ (with the natural convention $F_i(h_1, x_{1:i-1}, x_i)=F_1(h_1,x_1)$ when $i=1$).

Let  $i^*$ be smallest index $i\in 1:d_2$ such that $u_{2i}\neq  F_i(h_1^{n^*-1},x_{1:i-1},x_{i})$, $\forall \bx\in [\bm{a}',\bm{b}']$.
Then, for $i\in 1:(i^*-1)$,  set  $\tilde{u}_{2i}=u_{2i}$ and $\tilde{x}_{i}=x^*_{i}$, while, for $i\in 1:i^*$, we set $a_i^N=a_i'$ and $b_i^N=b'_i$. 

To choose $\tilde{u}_{2i^*}$ and $\tilde{x}_{i^*}$ we proceed as follows: if $F_{i^*}(h_1^{n^*-1},x^*_{1:i^*-1},b'_{i^*})<u_{2i^*}$, we take $\tilde{u}_{2i^*}=F_{i^*}(h_1^{n^*},x^*_{1:i^*-1},b'_{i^*})$ and $\tilde{x}_{i^*}=b'_{i^*}$ so that, noting $C_H$ the H\"{o}lder constant of $H$,
\begin{align*}
0\leq \tilde{u}_{2i^*}-u_{2i^*}&\leq F_{i^*}(h_1^{n^*},x^*_{1:i^*-1},b'_{i^*}) -F_{i^*}(h_1^{n^*-1},x^*_{1:i^*-1}, b'_{i^*})\leq C_K C_Hr_3(N)^{1/d_1}
\end{align*}
as required; if $F_{i^*}(h_1^{n^*-1},x^*_{1:i^*-1},a'_{i^*})>u_{2i^*}$, we take $\tilde{u}_{2i^*}=F_{i^*}(h_1^{n^*},x^*_{1:i^*-1},a'_{i^*})$ and $\tilde{x}_{i^*}=a'_{i^*}$ so that
\begin{align*}
0\leq u_{2i^*}-\tilde{u}_{2i^*}&\leq F_{i^*}(h_1^{n^*-1},x^*_{1:i^*-1},a'_{i^*}) -F_{i^*}(h_1^{n^*},x^*_{1:i^*-1},a'_{i^*})\leq C_K C_H r_3(N)^{1/d_1}
\end{align*}
as required.

Then, for $i\in(i^*+1): d_2$, take $\tilde{u}_{2i}=u_{2i}$ and $a_i^N=0$. Finally, to construct the right boundaries $b_i^N$, $i\in (i^*+1):d_2$, we define
$$
u_{2i}^*= \max_{n\in 1:N}\Big\{\sup \left\{ v\in F_i\left(h_1^n, [0,b'_{1:i}]\right)\right\}\Big\},\quad i=1,\dots d_2.
$$
Note that $u_{2i}^*\in (0,1)$ for all $i\in 1: d_2$. Indeed, the continuity of $F_i$ and the fact that $[0,b'_{1:i}]$ is compact imply that
$$
v_i^n:=\sup \left\{ v\in F_i\left(h_1^n, [0,b'_{1:i}]\right)\right\}\in F_i\left(h_1^n, [0,b'_{1:i}]\right).
$$
Then, since $b'_{i}\in (0,1)$ and $F_i$ is strictly increasing with respect to its $i$-th coordinate on $[0,1)$, we indeed have $v^n_i\in (0,1)$ for all $n\in 1:N$.

The right boundaries $b_i^N$, $i\in(i^*+1): d_2$ are then defined recursively  as follows:
\begin{align*}
&b^N_{i}=\inf \left\{c \in [0,1], g_i(c)\geq u^*_{2i} \right\}, \quad i=i^*+1,\dots, d_2
\end{align*}
where 
$$
g_i(c)= \min_{(h_1,x_{1:i-1})\in [a_1,b_1]\times [a^N_{1:i-1},b^N_{1:i-1}]} \tilde{F}_i(h_1,x_{1:i-1}, c),
$$
with  $\tilde{F}_i(\cdot{ })$  the continuous extension of $F_i(\cdot{ })$ on $[0,1]^{i+1}$. (Note that such an extension exists because $F_i$ is Lipschitz.) Because $\tilde{F}_i(h_1,x_{1:i-1},c)$ is continuous in $(h_1,x_{1:i-1},c)$ and  $ [a_1,b_1]\times [a^N_{1:i-1},b^N_{1:i-1}]\times [0,1]$ is compact, the function $g_i$ is continuous on $[0,1]$ with $g_i(0)=0$ and $g_i(1)=1$. Therefore, as $u^*_{2i}\in (0,1)$, we indeed have $b_i^N\in (0,1)$ for all $i\in (i^*+1):d_2$, as required.

To show that $\tilde{\bu}=(u_1,\tilde{\bu}_2)\in E^N(\partial(B^N))$, note that, by the construction of $\bm{b}^N$ we have, for all $i\in (i^*+1):d_2$,
$$
F_i(h_1,x_{1:i-1},b_i^N)\geq u^*_{2i}\geq u_{2i},\quad \forall (h_1,x_{1:i-1})\in   [a_1,b_1]\times [a^N_{1:i-1},b^N_{1:i-1}].
$$
Therefore, by the continuity of $F_i$, for any $(h_1,x_{1:i-1})\in  [a_1,b_1]\times [a^N_{1:i-1},b^N_{1:i-1}]$ there exists a $x_i\leq b_i^N$ such that $F_i(h_1,x_{1:i-1},x_i)=u_{2i}$. Hence, for $i\in (i^*+1):d_2$, $\tilde{x}_i$ is selected recursively as the unique solution of $F_i(h_1^{n^*},\tilde{x}_{1:i-1},\tilde{x}_i)= u_{2i}$. This concludes to show that there exists a $\tilde{\bx}\in B^N$ such that $\tilde{\bu}_2=F_{ K_h}(h_1^{n^*},\tilde{\bx})$ and $\|\bu_2-\tilde{\bu}_{2}\|_{\infty}\leq C_K C_Hr_3(N)^{1/d_1}$. Moreover, since $\tilde{x}_{i^*}=b'_{i^*}=b^N_{i^*}$,  we have $\tilde{\bx}\in\partial(B^N)$ and therefore $\tilde{\bu}\in \partial(E^N(B^N))$. 

Finally, note that the set $B^N$ depends only on   $i^*$, the smallest index $i\in 1:d_2$ such that $u_{2i}\neq  F_i(h_1^{n^*-1},x_{1:i-1},x_{i})$, $\forall \bx\in [\bm{a}',\bm{b}']$. Defining $D^N_{i^*}=B^N$, this shows that the collection $\{D^N_i\}_{i=1}^{d_2}$ of sets in $\mathcal{B}^N_{\ui^{d+1}}$ satisfies the desired properties.

Finally, we may conclude the proof as follows: 
\begin{align*}
\|\Sop(P_h^{N})-\pi_h^N\otimes K_h\stn
&\leq L^{d_1+d_2}D(\bu^{1:N})+(d_2+1)c_NL^{d_1+d_2-1}
\left(D(\bu^{1:N})+L^{-(d_1+d_2)}\right)
\end{align*}
where the optimal value of $L$ is such that $L=\bigO\left(D(\bu^{1:N})^{-\frac{1}{1+d_{1}+d_{2}}}\right)$. Let $r(N)=r_1(N)+2r_2(N)+(2C_KC_H)^{d_1} r_3(N)$. Then, 
if $r(N)D(\bu^{1:N})^{-\frac{d_{1}}{1+d_{1}+d_{2}}}=\bigO(1)$, $L$ verifies all the conditions above  and we have
$c_{N}=\bigO(1)$. Thus
\[
\|\Sop(P_h^{N})-\pi_h^N\otimes K_h\stn=\bigO\left(D(\bu^{1:N})^{\frac{1}{1+d_{1}+d_{2}}}\right).
\]
Otherwise, if $r(N)D(\bu^{1:N})^{-\frac{d_{1}}{1+d_{1}+d_{2}}}\rightarrow+\infty$,
let $L=\bigO(r(N)^{-\frac{1}{d_{1}}})$. Then $c_{N}=\bigO(1)$ and
\begin{align*}
 L^{d_{1}+d_{2}}D(\bu^{1:N}) 
& =\bigO(r(N))^{\frac{1}{d_1}-\frac{1+d_{1}+d_{2}}{d_{1}}}D(\bu^{1:N}) \\
& =\bigO(r(N)^{1/d_1}) \left(\bigO(r(N))^{-1}D(\bu^{1:N})^{\frac{d_{1}}{1+d_{1}+d_{2}}}\right)^{\frac{1+d_{1}+d_{2}}{d_{1}}}\\
 & =\smallo\left(r(N)^{1/d_{1}}\right).
\end{align*}
Therefore $\|\Sop(P_h^{N})-\pi_h^N\otimes K_h\stn=\smallo(1)$, which concludes
the proof. 

\subsection{Consistency: proof of Theorem \ref{thm:consistency}}

\label{sub:app_consistency}We first prove the following Lemma: 
\begin{lem}\label{lemma:Holder} 
Let $(\pi^N\otimes K)$ be a sequence of probability
measures on $\ui^{d_{1}+d_{2}}$. Assume that $\|\pi^N-\pi\stn=\smallo(1)$,
$\pi\in\mathcal{P}(\ui^{d_{1}})$ and that $F_{K}(\bx_{1},\bx_{2})$
is H\"older continuous with its $i$-th component  strictly increasing in $x_{2i}$, $i\in 1:d_2$. Then, as $N\rightarrow+\infty$, 
\[
\|\pi^N\otimes K-\pi\otimes K\stn\cvz.
\]

\end{lem}
To prove this result, let $B_{1}\times B_{2}\in\mathcal{B}_{\ui^{d_{1}+d_{2}}}$, $B_{2}=[\bm{a}_2,\bm{b}_{2}]$,
\begin{align*} 
\left|\int_{B_{1}\times B_{2}}(\pi^N\otimes K-\pi\otimes K)(\dx_{1},\dx_{2})\right|&=\left|\int_{B_{1}}K(\bx_{1},B_2)(\pi^N-\pi)(\dx_{1})\right|\\
&=\left|\int_{B_{1}}\lambda_{d_2}\left(F_K(\bx_1,B_2)\right)(\pi^N-\pi)(\dx_{1})\right|.
\end{align*}

The function $\bx_1\rightarrow \lambda_{d_2}\left(F_K(\bx_1,B_2)\right)$ is continuous and bounded
and therefore we proceed as in  the proof
of Theorem \ref{thm:IS_1}. But since $\lambda_{d_2}\left(F_K(\bx_1,B_2)\right)$
depends on $(\bm{a}_2,\bm{b}_{2})$ and we want to take the supremum over
$\bm{a}_2\,,\bm{b}_{2}\in(0,1)^{d_{2}}$, we need to make sure that, on a compact set $J$, for any
$\epsilon>0$ we can find $\eta>0$ which does not depend on $(\bm{a}_2,\bm{b}_{2})$
such that, for $\bx_1$, $\bx'_1\in J$,
\[
\|\bx_{1}-\bx_{1}'\|_{\infty}\leq\eta\implies|\lambda_{d_2}\left(F_K(\bx_1,B_2)\right)-
\lambda_{d_2}\left(F_K(\bx_1',B_2)\right)|\leq\epsilon.
\]
To see that this is true, note that $\partial \left(F_K(\bx_1,B_2)\right)=
F_K(\bx_1,\partial B_2)$. Hence, for any point $\mathbf{c}\in\partial F_K(\bx_1,B_2)$ there exists a $\bm{p}\in\partial B_2$ such that  $\mathbf{c}= F_K(\bx_1,\bm{p})$ and therefore, by the H\"older property of $F_K$, we have
$$
\|\bx_1-\bx_1'\|_{\infty}\leq \eta\implies \|\mathbf{c}-\mathbf{c}'\|_{\infty}\leq C_K\eta^{\kappa},\quad \mathbf{c}'= F_K(\bx_1',\bm{p})\in\partial F_K(\bx_1',B_2)
$$ 
where $C_K$ and $\kappa$ are respectively the H\"older constant and the H\"older exponent of $F_K$. Let   $\tilde{F}_K$ be the continuous extension of $F_K$ on $[0,1]^{d_1+d_2}$ (which exists because $F_K$ is  H\"{o}lder continuous on $\ui^{d_1+d_2}$). Let $w>0$, $\bx\in\ui^{d_1}$ and $\bm{a}\leq  \bm{b}$, $(\bm{a},\bm{b})\in[0,1]^{2d_2}$. Then, define
$$
A^+(w,\bx,\bm{a}, \bm{b})=
\left\{\bu\in\ui^{d_2}:\exists \bm{p}\in \partial[\bm{a},\bm{b}] \text{ such that }\|\bu-\tilde{F}_K(\bx,\bm{p})\|_{\infty}\leq C_Kw^{\kappa} \right\}
$$
and, noting $\tilde{F}_i(\bx_1,\bx_2)$ the $i$-th component of $\tilde{F}_K(\bx_1,\bx_2)$, $i\in 1:d_2$,
\begin{align*}
A^-(w,\bx,\bm{a},\bm{b})&=\left\{\bu\in \tilde{F}_K(\bx,[\bm{a},\bm{b}]):\text{ }\exists \bm{p}\in\partial [\bm{a},\bm{b}]\right.\\
&\left. \text{ such that }|u_i-\tilde{F}_i(\bx,\bm{p})|\geq C_Kw^{\kappa}, \text{ }\forall i\in 1:d_2 \right\}.
\end{align*}
Let $B^*=\{(\bm{a},\bm{b})\in[0,1]^{2d_2}:\, a_i\leq b_i\, i\in 1:d_2\}$ and  $f:\mathbb{R}^+\times \ui^{d_1}\times B^*\rightarrow [0,1]$ be the mapping
$$
(w,\bx,\bm{a},\bm{b})\in\mathbb{R}^+\times \ui^{d_1}\times B^*\mapsto f(w,\bx,\bm{a},\bm{b})=
\lambda_{d_2}(A^+(w,\bx,\bm{a},\bm{b}))-\lambda_{d_2}(A^-(w,\bx_,\bm{a},\bm{b})).
$$
Note that for a fix $w$ the function $f(w,\cdot{ })$ is continuous on  $\ui^d\times B^*$ (as  $\tilde{F}_K$ is continuous). Therefore, for all $\bx_1$ and $\bx_1'$ in $J$ such that $\|\bx_1-\bx_1'\|\leq \eta$, we have
\begin{align*}
|\lambda_{d_2}\left(F_K(\bx_1,B_2)\right)-
\lambda_{d_2}\left(F_K(\bx_1',B_2)\right)|&\leq f(\eta,\bx_1,\bm{a}_2,\bm{b}_2)\leq m(\eta)
\end{align*}
with
$$
m(\eta):=\max_{(\bx,\bm{a},\bm{b})\in J\times B^*}f(\eta, \bx,\bm{a},\bm{b}).
$$
Because  $f$ is continuous  and $J\times B^*$ is compact, $m(\eta)$ is continuous so that, for any $\epsilon>0$, there exists a $\eta>0$ (that depends only on $m(\cdot{})$ and therefore independent of $B_2$) such that $m(\eta)\leq \epsilon$. This concludes the proof of the Lemma.


We now prove Theorem \ref{thm:consistency}. By the result of \citet[``Satz 2'']{Hlawka1972} and Assumption \ref{H:thmPF1:3}, $\left(\bx_{0}^{1:N}\right)$
is such that $\|\Sop(\bx_{0}^{1:N})-m_{0}\stn=\smallo(1)$. In addition,
the importance weight function $\Q_{0}(\dx_{0})/m_{0}(\dx_{0})=G_{0}(\bx_{0})/m_{0}(G_{0})$
is continuous and bounded by Assumption \ref{H:thmPF1:2}. Therefore,
$\|\Qh_{0}^{N}-\Q_{0}\stn=\smallo(1)$ by Theorem \ref{thm:IS_1}. 

Assume that the result is true at time $t\geq 0$ and let $P_{t+1,h}^{N}=(h_{t}^{1:N}, \bx_{t+1}^{1:N})$
where $h_{t}^{n}=\IHSFC(\bx_{t}^{\sigma_t(a_{t}^{n})})$. Then, the
result is true at time $t+1$ if 
\begin{equation}
\|\Sop(P_{t+1,h}^{N})-\Q_{t,h}\otimes m_{t+1,h}\stn=\smallo(1).\label{eq:thmPF1_1}
\end{equation}
To see that, let $G_{t,h}(h_{t-1},\bx_t)=G_t(H(h_{t-1}),\bx_t)$ and $\Psi_{t+1}$ be the
Bolzmann-Gibbs transformation associated to $G_{t+1,h}$ \citep[see][Definition 2.3.3]{DelMoral:book}. Then, the importance weight function 
\[
\frac{\Psi_{t+1}(\Q_{t,h}\otimes m_{t+1})}{\Q_{t,h}\otimes m_{t+1}}(\dd(h_t,\bx_{t+1})=\frac{G_{t+1,h}(h_t,\bx_{t+1})}{\Q_{t}\otimes m_{t+1}(G_{t+1})}
\]
is continuous and bounded (by Assumption \ref{H:thmPF1:2} and the continuity of the Hilbert curve) and
therefore Theorem \ref{thm:IS_1} implies that $\|\Qh_{t+1}^{N}-\Q_{t+1}\stn=\smallo(1)$
if (\ref{eq:thmPF1_1}) is verified.

To show (\ref{eq:thmPF1_1}), note that 
\[
\|\Sop(P_{t+1,h}^{N})-\Q_{t,h}\otimes m_{t+1,h}\stn\leq\|\Sop(P_{t+1,h}^{N})-\Qb_{t+1,h}^{N}\stn+\|\Qb_{t+1,h}^{N}-\Q_{t,h}\otimes m_{t+1,h}\stn.
\]
By the inductive hypothesis, $\|\Qh_{t}^{N}-\Q_{t}\stn=\smallo(1)$
so that, by Theorem \ref{thm:Hilbert}, Assumption \ref{H:thmPF1:3}, the H\"older property of the Hilbert curve and Lemma \ref{lemma:Holder},
\[
\|\Qb_{t+1,h}^{N}-\Q_{t,h}\otimes m_{t+1,h}\stn=\|\Qh_{t,h}^{N}\otimes m_{t+1,h}-\Q_{t,h}\otimes m_{t+1,h}\stn=\smallo(1).
\]
Finally, note that 
$$
W_t^n\leq \frac{\|G_t\|_{\infty}}{\Sop(P_{t,h}^N)(G_{t,h})}=\smallo(1)
$$
because $\Sop(P_{t,h}^N)(G_{t,h})=\bigO(N^{-1})$ by the inductive hypothesis and the fact that $G_{t,h}$ is continuous and bounded  (by Assumption \ref{H:thmPF1:2} and the continuity of the Hilbert curve). Together with the inductive hypothesis and Assumptions \ref{H:thmPF1:1}, \ref{H:thmPF1:3}-\ref{H:thmPF1:4}, this
implies that all the assumptions of Theorem \ref{thm:LD} are verified
and therefore $\|\Sop(P_{t+1,h}^{N})-\Qb_{t+1,h}^{N}\stn=\smallo(1)$
as required.

\subsection{Stochastic bounds}
\label{sec:proof_stobounds}
\subsubsection{Setup of the proof of Theorem \ref{thm:PF2}}

The result is proved by induction. By Assumption \ref{H:thmPF1:2} of Theorem \ref{thm:consistency},
the weight function $\Q_{0}(\dx_{0})/m_{0}(\dx_{0})=G_{0}(\bx_{0})/m_{0}(G_{0})$
is continuous and bounded. Therefore, the continuity of $F^{-1}_{m_{0}}$, 
the assumptions on $(\bu_{0}^{1:N})$ (Assumptions \ref{H:thmPF2:1} and \ref{H:thmPF2:2}) and Theorem \ref{thm:IS_2} give
the result at time $t=0$. 

Assume that the result is true at time $t\geq 0$ and let 
$\hat{I}_{t+1}^N=\Qh_{t+1}^{N}(\varphi)$ 
where $\varphi:\ui^d\rightarrow\mathbb{R}$ verifies the conditions of the theorem. As mentioned previously,  iteration $t+1$ of SQMC is a QMC importance sampling step  from the proposal distribution $\Qb^N_{t+1,h}$ to the target $w^h_{t+1,h}(h_t,\bx_{t+1})\Qb^N_{t+1,h}(\dd(h_{t},\bx_{t+1}))$ where 
$$
w_{t+1,h}^{N}(h_{t},\mathbf{\bx}_{t+1}):=\frac{G_{t+1,h}\left(h_{t},\bx_{t+1}\right)}{C_{t+1}^{N}}
$$
with  $C^N_{t+1}=\Qb^N_{t+1,h}(G_{t+1,h})$ and $G_{t+1,h}$ as in the proof of Theorem \ref{thm:consistency}. To bound $\var\{\hat{I}^N_{t+1}\}$ and $\E|\hat{I}_{t+1}-\Q_{t+1}(\varphi)|$ we therefore naturally want to use expression \eqref{eq:thmIS_BoundL2} and \eqref{eq:thmIS_BoundL1} derived in the proof of Theorem \ref{thm:IS_2}. To that effect, we need to show that, for $N$ large enough  and almost surely, the assumptions given in Theorem \ref{thm:IS_2} on the weight function and on the point set at hand (Assumption \ref{H:thmIS_2:2} of Theorem \ref{thm:IS_2})  are satisfied.

To see that the conditions on the weight function are fulfilled, note first that $w_{t+1,h}^{N}$ is continuous  by Assumption
\ref{H:thmPF1:2} of Theorem \ref{thm:consistency} and by the continuity of the Hilbert curve. To show that $w_{t+1,h}^{N}$ is almost surely bounded for $N$ large enough, first note that, by Assumption \ref{H:thmPF2:1}, it is clear from the
proofs of Theorem \ref{thm:Hilbert} and of Theorem \ref{thm:consistency} that, for all $\epsilon>0$ and for
all $t\geq0$, there exists a $N^*_{\epsilon,t}$ such that, almost
surely, 
\[
\|\Qh_{t,h}^{N}-\Q_{t,h}\stn\leq\epsilon,\quad\forall N\geq N^*_{\epsilon,t}.
\]
In addition, under the assumptions of the theorem,  $(C_{t+1}^{N})^{-1}$ is almost surely bounded
above and below away from 0, for $N$ large enough. 
Indeed, by Lemma \ref{lemma:Holder} (and using the H\"older property of the Hilbert curve), $\|\Qb_{t+1,h}^{N}-\Q_{t,h}\otimes m_{t+1,h}\stn=\smallo(1)$
and, in particular, under the conditions of the theorem, for any $\delta>0$,
we have, almost surely, 
\begin{align}\label{eq:thm_PF2_Prel1}
\|\Qb_{t+1,h}^{N}-\Q_{t,h}\otimes m_{t+1,h}\|\leq\delta
\end{align}
for $N$ large enough (see the proof of Lemma \ref{lemma:Holder}
and the proof of Theorem \ref{thm:IS_1}). Writing $C_{t+1}=\Q_{t,h}\otimes m_{t+1,h}(G_{t+1})$, this observation, together
with the fact that 
\begin{align*}
 & |C_{t+1}^{N}-C_{t+1}|=|\Qb_{t+1,h}^{N}(G_{t+1,h})-\Q_{t,h}\otimes m_{t+1,h}(G_{t+1})|
\end{align*}
where $G_{t+1,h}$ is continuous and bounded (by Assumption \ref{H:thmPF1:2} of Theorem \ref{thm:consistency} and the continuity of the Hilbert curve), implies that, almost surely,
$C_{t+1}+\delta\geq C_{t+1}^{N}\geq C_{t+1}-\delta \eqdef c_\delta>0$ for $N$ large enough
(computations as in the proof of Theorem \ref{thm:IS_1}). Hence, almost surely, $\|w_{t+1,h}^{N}\|_{\infty}\leq c_\delta^{-1}\|G_{t+1}\|_{\infty}$,
for $N$ large enough.

Finally, to show that  the point set $P^N_{t+1,h}$ (defined as in the proof of Theorem \ref{thm:consistency}) verifies Assumption \ref{H:thmIS_2:2} of Theorem \ref{thm:IS_2}, note that, from Theorem \ref{thm:consistency} and under the assumptions of the theorem, for any $\epsilon>0$ there exists a $N_{\epsilon}$ such
that, almost surely, $\|\Sop(P_{t+1,h}^{N})-\Q_{t,h}\otimes m_{t+1,h}\stn\leq\epsilon$
for all $N\geq N_{\epsilon}$. Together with  \eqref{eq:thm_PF2_Prel1}, this shows that, as required,  for any $\epsilon>0$ we have, almost surely and for $N$ large enough, $\|\Sop(P_{t+1,h}^{N})-\Qb_{t+1,h}^N\stn\leq \epsilon$.

\subsubsection{Proof of Theorem \ref{thm:PF2}: $L_{2}$-convergence}

Using expression \eqref{eq:thmIS_BoundL2} given in the proof of Theorem \ref{thm:IS_2}, we have for $N$ large enough
\begin{equation}
\begin{split}\var\{\hat{I}_{t+1}^N\} & \leq\left[2(1+c^{-1}_\delta\|G_{t+1}\|_{\infty})\var\left\{\Sop(P_{t+1,h}^{N})(w_{t+1,h}^{N})\right\}^{1/2}+\right.\\
 & \left.\left\{ 1-2\E\left[\Sop(P_{t+1,h}^{N})(\varphi w_{t+1,h}^N)\right]\right\} \var\left\{\Sop(P_{t+1,h}^{N})(\varphi w_{t+1,h}^{N})\right\}^{1/2}\right]^{2}.
\end{split}
\label{eq:thmPF2:0}
\end{equation}

We first bound $\var\{\Sop(P_{t+1,h}^{N})(w_{t+1,h}^{N})\}$. Let $\mathcal{F}^N_{t}$ be the $\sigma$-algebra generated by the point set $(h_{1:t-1}^{1:N},\bx_{1:t}^{1:N})$. Then, by Assumption \ref{H:thmPF2:2}, 
\[
\var\left\{\Sop(P_{t+1,h}^{N})(w_{t+1,h}^{N})|\mathcal{F}^N_{t}\right\}\leq C^*r(N)\sigma_{N}^{2}
\]
with $C^*$ as in the statement of the theorem and $\sigma_{N}^{2}\leq\|w_{t+1,h}^{N}\|_{\infty}\leq c_\delta^{-1}\|G_{t+1}\|_{\infty}$
almost surely and for $N$ large enough. Therefore, since 
$\E\left[\Sop(P_{t+1,h}^{N})(w_{t+1,h}^{N})|\mathcal{F}^N_{t}\right]=1$,
we have 
\begin{equation}
\var\left\{\Sop(P_{t+1,h}^{N})(w_{t+1,h}^{N}))\right\}=\bigO(r(N)).\label{eq:thmPF2:1}
\end{equation}

Next, we need to bound $\var\{\Sop(P_{t+1,h}^{N})(\varphi w_{t+1,h}^{N})\}$.
Note that 
\begin{equation*}
\Qb_{t+1}^{N}\left(\left(C_{t+1}^{N}\right)^{-2}\varphi^{2}G_{t+1}^{2}\right) 
\leq\frac{1}{\left(C_{t+1}^{N}\right)^{2}}\|G_{t+1}\|_{\infty}^2\Qh_{t}^{N}\left(m_{t+1}(\varphi^{2})\right),
\end{equation*}
where the last factor is almost surely finite for all $N$. Indeed, since $\varphi\in L_{2}(\mathcal{X},\Q_{t+1})$,
$m_{t+1}(\varphi^{2})(\bx_{t})$ is finite for almost all $\bx_{t}\in\mathcal{X}$
and the integral with respect to $\Qh_{t}^{N}$ is a finite sum. Hence,
for all $N$, $\varphi \in L_{2}(\mathcal{X}^{2},\Qb_{t+1,h}^{N})$ almost surely
so that, by Assumption \ref{H:thmPF2:2}, we have almost surely 
\[
\var\left\{\Sop(P_{t+1,h}^{N})(\varphi w_{t+1,h}^{N})|\mathcal{F}^N_{t}\right\}\leq C^*r(N)\sigma_{N,\varphi}^{2}
\]
where, with probability one and for $N$ large enough, $\sigma_{N,\varphi}^{2}\leq c_{\delta}^{-2}\|G_{t+1}\|_{\infty}^2\Qb_{t+1}^{N}\left(\varphi^{2}\right)$.
We now need to show that $\E[\Qb_{t+1}^{N}\left(\varphi^{2}\right)]$
is bounded.

In order to establish this, we prove that for all $t\geq0$ and for 
$N$ large enough, we have, $\forall f\in L_{1}(\mathcal{X}^{2},\Q_{t}\otimes m_{t+1})$,
\begin{align}
\E[\Qb_{t+1}^{N}(f)]\leq c_{t+1}\Q_{t}\otimes m_{t+1}(|f|)\label{eq:thmPF2:2}
\end{align}
for constant $c_{t+1}$.

Equation \eqref{eq:thmPF2:2} is true for $t=0$. Indeed, let
$f\in L_{1}(\mathcal{X}^{2},\Q_{0}\otimes m_{1})$ and note that, under the conditions of the theorem, almost surely and for $N$ large enough, $\{\Sop(\bx_{0}^{1:N})(G_{0})\}^{-1}\leq \tilde{c}_0<\infty$ for a constant $\tilde{c}_0$. Hence, for 
$N$ large enough, we have 
\begin{align*}
\E[\Qb_{1}^{N}(f)] & =\E\left[\left\{\Sop(\bx_{0}^{n})(G_{0})\right\}^{-1}\frac{1}{N}\sum_{n=1}^{N}G_{0}(\bx_{0}^{n})\int_{\mathcal{X}}f(\bx_{0}^{n},\bx_{1})m_{1}(\bx_{0}^{n},\dx_{1})\right]\\
& \leq c_0\Q_{0}\otimes m_{1}(|f|)
\end{align*}
with $c_0=\tilde{c}_0 m_{0}(G_{0})$.
Assume that \eqref{eq:thmPF2:2}
is true for $t\geq 0$ and note that, under the conditions of the theorem, almost surely and for $N$ large enough, $\{\Sop(P_{t}^N)(G_{t})\}^{-1}\leq \tilde{c}_t<\infty$ for a constant $\tilde{c}_t$. Then, for $N$ large enough (with the convention $G_t(\bx_{t-1},\bx_t)=G_0(\bx_0)$ if $t=0$), 
\begin{align*}
\E[\Qb_{t+1}^{N}(f)] 
& =\E\left[\left\{\Sop(P^N_{t})(G_{t})\right\}^{-1}\frac{1}{N}\sum_{n=1}^{N}G_{t}(\bx_{t-1}^{\sigma_{t-1}(a_{t-1}^{n})},\bx_{t}^{n})\int_{\mathcal{X}}f(\bx_{t}^{n},\bx_{t+1})m_{t+1}(\bx_{t}^{n},\dx_{t+1})\right]\\
 & \leq \tilde{c}_{t}\frac{1}{N}\sum_{n=1}^{N}\E\left\{ \E\left[\int_{\mathcal{X}}G_{t}(\bx_{t-1}^{\sigma_{t-1}(a_{t-1}^{n})},\bx_{t}^{n})|f(\bx_{t}^{n},\bx_{t+1})|m_{t+1}(\bx_{t}^{n},\dx_{t+1})|\mathcal{F}^N_{t}\right]\right\} \\
 & =\tilde{c}_{t}\E\left[\int_{\mathcal{X}^{3}}G_{t}(\bx_{t-1},\bx_{t})|f(\bx_{t},\bx_{t+1})|\Qb_{t}^{N}\otimes m_{t+1}(\dx_{t-1:t+1})\right]\\
 &\leq \tilde{c}_{t}c_{t-1}\int_{\mathcal{X}^{3}}G_{t}(\bx_{t-1},\bx_{t})|f(\bx_{t},\bx_{t+1})|\Q_{t-1}\otimes m_t\otimes m_{t+1}(\dx_{t-1:t+1})\\
 & = c_t\int_{\mathcal{X}^{3}}|f(\bx_{t},\bx_{t+1})|\Psi_{t}\left(\Q_{t-1}\otimes m_{t}\right)\otimes m_{t+1}(\dx_{t-1:t+1}))\\
 & =c_t\int_{\mathcal{X}^{2}}|f(\bx_{t},\bx_{t+1})|\Q_{t}\otimes m_{t+1}(\dx_{t:t+1})\\
 & =c_t\Q_{t}\otimes m_{t+1}(|f|)
\end{align*}
with $c_t=c_{t-1}\tilde{c}_t \left[\Q_{t-1}\otimes m_t(G_t)\right]$, $\Psi_{t}$ be the
Bolzmann-Gibbs transformation associated to $G_{t}$ \citep[see][Definition 2.3.4]{DelMoral:book} and
 where the second inequality uses the inductive hypothesis and the fact
that the mapping 
\[
(\bx_{t-1},\bx_{t})\mapsto G_{t}(\bx_{t-1},\bx_{t})m_{t+1}(|f|)(\bx_{t})
\]
belongs to $L_{1}(\mathcal{X}^{2},\Q_{t-1}\otimes m_{t})$. This shows
\eqref{eq:thmPF2:2} and therefore, for $N$ large enough, $\E[\sigma_{N,\varphi}^{2}]\leq c$
for a constant $c$ so that 
$
\E [\var\{\Sop(P_{t+1,h}^{N})(\varphi w_{t+1,h}^{N})\big|\mathcal{F}^N_{t}\}]=\bigO(r(N))
$.
In addition 
\begin{align*}
  \E\left[\Sop(P_{t+1,h}^{N})(\varphi w_{t+1,h}^{N})|\mathcal{F}^N_{t}\right]&=\frac{\Qh_{t}^{N}\left(m_{t+1}(\varphi G_{t+1})\right)}{C_{t+1}^{N}}\\
 & =\frac{\Q_{t}\left(m_{t+1}(\varphi G_{t+1})\right)}{C_{t+1}^{N}}+\frac{(\Qh_{t}^{N}-\Q_{t})\left(m_{t+1}(\varphi G_{t+1})\right)}{C_{t+1}^{N}}
\end{align*}
where $\Q_{t}\left(m_{t+1}(\varphi G_{t+1})\right)<+\infty$ because $\varphi\in L_2(\setX,\Q_{t+1})$. Since
$$
\frac{\Q_{t}\left(m_{t+1}(\varphi G_{t+1})\right)}{C_{t+1}^{N}}  =C_{t+1}^{-1}\Q_{t}\left(m_{t+1}(\varphi G_{t+1})\right)+\frac{C_{t+1}-C_{t+1}^{N}}{C_{t+1}^{N}C_{t+1}}\Q_{t}\left(m_{t+1}(\varphi G_{t+1})\right),
$$
we therefore have, for $N$ large enough, 
\begin{align*}
\var\left\{\frac{\Q_{t}\left(m_{t+1}(\varphi G_{t+1})\right)}{C_{t+1}^{N}}\right\} & =[\Q_{t}\left(m_{t+1}(\varphi G_{t+1})\right)]^2\var\left\{\frac{C_{t+1}-C_{t+1}^{N}}{C_{t+1}^{N}C_{t+1}}\right\}\\
 & \leq\frac{[\Q_{t}\left(m_{t+1}(\varphi G_{t+1})\right)]^2}{(c_{\delta}C_{t+1})^{2}}\E\left[\left\{ (\Qh_{t}^{N}-\Q_{t})\left(m_{t+1}(G_{t+1})\right)\right\} ^{2}\right].
\end{align*}
Since $\|G_{t+1}\|_{\infty}<+\infty$, $m_{t+1}(G_{t+1})$ is bounded
and the inductive hypothesis implies that the term on the right of
the inequality sign is $\bigO(r(N))$. In addition, for all $N$ large
enough, 
\begin{equation*}
\var\left\{\frac{1}{C_{t+1}^{N}}\left(\Qh_{t}^{N}-\Q_{t}\right)\left(m_{t+1}(\varphi G_{t+1})\right)\right\}
 \leq c_{\delta}^{-2}\E\left[\left\{ (\Qh_{t}^{N}-\Q_{t})\left(m_{t+1}(\varphi G_{t+1})\right)\right\} ^{2}\right].
\end{equation*}
Since $\left[m_{t+1}(\varphi G_{t+1})(\bx_{t})\right]^{2}\leq\|G_{t+1}\|_{\infty}m_{t+1}(\varphi^{2}G_{t+1})(\bx_{t})$,
we have 
\[
\Q_{t}(\{m_{t+1}(\varphi G_{t+1})\}^2)\leq\|G_{t+1}\|_{\infty}C_{t+1}\Q_{t+1}(\varphi^{2})<+\infty
\]
by assumption. Therefore, $m_{t+1}(\varphi G_{t+1})\in L_{2}(\mathcal{X},\Q_{t})$
so that, by the inductive hypothesis, $\var\{\E [\Sop(P_{t+1,h}^{N})(\varphi w_{t+1,h}^{N})|\mathcal{F}^N_{t}]\}=\bigO(r(N))$.
Hence, 
\begin{equation}
\var\left\{\Sop(P_{t+1,h}^{N})(\varphi w_{t+1,h}^{N})\right\}=\bigO(r(N)).\label{eq:thmPF2:3}
\end{equation}

The last term of \eqref{eq:thmPF2:0} we need to control is  
\begin{align*}
\E\left[\Sop(P_{t+1,h}^{N})(\varphi w_{t+1,h}^{N})\right]=\E\left[\frac{1}{C_{t+1}^{N}}\Qh_{t}^{N}(m_{t+1}(\varphi G_{t+1}))\right].
\end{align*}
Since we saw that $m_{t+1}(\varphi G_{t+1})\in L_{2}(\mathcal{X},\Q_{t})$
, we have, for $N$ large enough, 
\begin{equation}
\begin{split}\E\left[\Sop(P_{t+1,h}^{N})(\varphi w_{t+1,h}^{N})\right] & =\E\left[\frac{1}{C_{t+1}^{N}}\right]\Q_{t}(m_{t+1}(\varphi G_{t+1}))\\
 & +\E\left[\frac{1}{C_{t+1}^{N}}(\Qh_{t}^{N}-\Q_{t})(m_{t+1}(\varphi G_{t+1}))\right]\\
 & \leq c_{\delta}^{-1}\left[\Q_{t}(m_{t+1}(\varphi G_{t+1}))+\bigO(r(N)^{1/2})\right]
\end{split}
\label{eq:thmPF2:4}
\end{equation}
using previous computations. 

Combining \eqref{eq:thmPF2:0}, \eqref{eq:thmPF2:1}, \eqref{eq:thmPF2:3}
and \eqref{eq:thmPF2:4}, one obtains  $\var\{\hat{I}_{t+1}^N\}=\bigO(r(N))$.

\subsubsection{Proof of Theorem  \ref{thm:PF2}: $L_{1}$-convergence}

Let $I_{t+1}=\Q_{t+1}(\varphi)$ and
$I_{t+1}^N=\Q_{t+1}^{N}(\varphi)$ so that 
$$
\E[|\hat{I}_{t+1}^N-I_{t+1}|]\leq\E[|\hat{I}_{t+1}^N-I_{t+1}^N|]+\E[|I_{t+1}^N-I_{t+1}|].
$$
Then, using expression \eqref{eq:thmIS_BoundL1}  in the proof of Theorem \ref{thm:IS_2}, we have, for $N$ large enough, 
\[
\begin{split}\E\left[|\hat{I}_{t+1}^N-I_{t+1}^N|\right] & \leq\var\left\{\Sop(P_{t+1,h}^{N})(\varphi w_{t+1,h}^{N})\right\}^{1/2}+2\left(\var\left\{\Sop(P_{t+1,h}^{N})(\varphi w_{t+1,h}^{N})\right\}\right.\\
 & +\left.\left\{\E\left[\Sop(P_{t+1,N})(\varphi w_{t+1,h}^{N})\right]\right\}^{2}\right) ^{1/2}\var\left\{\Sop(P_{t+1,h}^{N})(w_{t+1,h}^{N})\right\}^{1/2}\\
 & =\bigO(r(N)^{1/2})
\end{split}
\]
from above computations. In addition, 
\begin{align*}
\E\left[|I_{t+1}^N-I_{t+1}|\right] 
& =\E\left[\left|\left(\frac{\Qh_{t}^{N}}{C_{t+1}^{N}}-\frac{\Q_{t}}{C_{t+1}}\right)(m_{t+1}(\varphi G_{t+1}))\right|\right]\\
 & \leq\E\left[\frac{1}{C_{t+1}}\left|(\Qh_{t}^{N}-\Q_{t})(m_{t+1}(\varphi G_{t+1}))\right|\right]\\
 & +\E\left|\frac{|C_{t+1}-C_{t+1}^{N}|}{C_{t+1}^{N}C_{t+1}}\Qh_{t}^{N}(m_{t+1}(\varphi G_{t+1}))\right|.
\end{align*}
By the inductive hypothesis and the above computations, the first term after the inequality sign
is $\bigO(r(N)^{1/2})$. In addition,  for $N$ large enough, the second term after the inequality sign is bounded by 
\begin{align*}
 \E\left|\frac{|C_{t+1}-C_{t+1}^{N}|}{C_{t+1}^{N}C_{t+1}}\Qh_{t}^{N}(m_{t+1}(\varphi G_{t+1}))\right|
 & \leq\frac{\delta}{c_{\delta}C_{t+1}}\E\left|(\Qh_{t}^{N}-\Q_{t})(m_{t+1}(\varphi G_{t+1}))\right|\\
 & +\frac{\left|\Q_{t}(m_{t+1}(\varphi G_{t+1})\right|}{c_\delta C_{t+1}}
 \E\left|(\Qh_{t}^{N}-\Q_{t})(m_{t+1}(G_{t+1}))\right|.
\end{align*}
We saw above that the first term on the right-hand side is $\bigO(r(N)^{1/2})$.
In addition, $m_{t+1}(G_{t+1})$ belongs to $L_{2}(\mathcal{X},\Q_{t})$
because $\|G_{t+1}\|_{\infty}<+\infty$. Hence, by the inductive hypothesis,
the second term after the inequality sign is also $\bigO(r(N)^{1/2})$
and the proof is complete.

\subsubsection{Proof of Theorem \ref{thm:PF2Bis}} \label{sec:proof_smallo}

\label{p:thmPF2Bis}

To avoid confusion between  the $t$ of the time index and the $t$ of the $(t,s)$-sequence we replace the latter by $\tilde{t}$ in what follows.

The result is true at time $t=0$ by Theorem \ref{thm:IS_2}. To obtain
the result for $t\geq1$ we need to modify the steps in the proof
of Theorem \ref{thm:PF2} where we do not use the inductive hypothesis.
Inspection of this proof shows that we only need to establish that,
for any function $\varphi\in\mathcal{C}_{b}(\ui^{1+d})$, we have
\begin{align*}
\E\left[\var\left(\Sop(P_{t+1,h}^{N})(\varphi)|\mathcal{F}^N_{t}\right)\right]=\smallo(N^{-1}).
\end{align*}
Let $N=\lambda b^m$. Then, from the proof of \citet[][Theorem 1]{Owen1998}, and using
the same notations as in that paper (note in particular the new meaning for symbol $u$), we have 
\begin{align*}
\var\left(\Sop(P_{t+1,h}^{N})(\varphi)|\mathcal{F}^N_{t}\right)\leq\frac{c}{N}\sum_{|u|>0}\text{ }\sum_{|\kappa|>m-\tilde{t}-|u|}\sigma_{N,u,\kappa}^{2}
\end{align*}
for a constant $c$, where $|u|$ is the cardinal of $u\subseteq\{1,...,d+1\}$, $\kappa$ is a vector of $|u|$ nonnegative integers $k_j$, $j\in u$, and $|\kappa|=\sum_{j\in u} k_j$.  Note that $\kappa$ depends implicitly on $u$. The $\sigma_{N,u,\kappa}^2$'s are such that
\[
\sigma_{N}^{2}=\Qb_{t+1,h}^{N}(\varphi^{2})-\Qb_{t+1,h}^{N}(\varphi)^{2}=\sum_{|u|>0}\text{ }\sum_{\kappa}\sigma_{N,u,\kappa}^{2},
\]
with $\sigma_{N,u,\kappa}^{2}=\int_{\ui^{1+d}}\nu_{N,u,\kappa}(\bx)^{2}\dx$ and 
\begin{align*}
\nu_{N,u,\kappa}(\bx) & =\sum_{\tau(u,\kappa)}\sum_{\gamma(u)}<\varphi\circ F_{\Qb_{t+1,h}^{N}}^{-1},\psi_{u,\kappa,\tau,\gamma}>\psi_{u,\kappa,\tau,\gamma}(\bx)
\end{align*}
where $<f_{1},f_{2}>=\int f_{1}(\bx)f_{2}(\bx)\dx$, $\psi_{u,\kappa,\tau,\gamma}$
is bounded and all the sums in the definition of $\nu_{N,u,\kappa}(\bx)$ are
finite \citep[see][for more details]{Owen1997a}.

Similarly, let 
\[
\sigma^{2}=\Q_{t,h}\otimes m_{t+1,h}(\varphi^{2})-\Q_{t,h}\otimes m_{t+1,h}(\varphi)^{2}=\sum_{|u|>0}\text{ }\sum_{\kappa}\sigma_{u,\kappa}^{2},
\]
where $\sigma_{u,\kappa}^{2}=\int_{\ui^{1+d}}\nu_{u,\kappa}(\bx)^{2}\dx$ and with
\begin{align*}
\nu_{u,\kappa}(\bx) & =\sum_{\tau(u,\kappa)}\sum_{\gamma(u)}<\varphi\circ F_{\Q_{t,h}\otimes m_{t+1,h}}^{-1},\psi_{u,\kappa,\tau,\gamma}>\psi_{u,\kappa,\tau,\gamma}(\bx).
\end{align*}
We first want to establish that $|\sigma_{N,u,\kappa}^{2}-\sigma_{u,\kappa}^{2}|=\smallo(1)$
almost surely. Note that 
\[
\|\nu_{N,u,\kappa}-\nu_{u,\kappa}\|_{\infty}\leq c\sum_{\tau(u,\kappa)}\sum_{\gamma(u)}\left|<\varphi\circ F_{\Qb_{t+1,h}^N}^{-1},\psi_{u,\kappa,\tau,\gamma}>-<\varphi\circ F_{\Q_{t,h}\otimes m_{t+1,h}}^{-1},\psi_{u,\kappa,\tau,\gamma}>\right|
\]
for a constant $c>0$. To show that the term inside the absolute
value sign is almost surely $\smallo(1)$, assume that for all $\tilde{u}\in\ui$,
$|F_{\Qh_{t,h}^{N}}^{-1}(\tilde{u})-F_{\Q_{t,h}}^{-1}(\tilde{u})|=\smallo(1)$ almost
surely. Using the continuity of $\varphi$ and the continuity of the Hilbert curve $H$, and the fact that that
 $F_{m_{t+1}}^{-1}(\bx_t,\bx_{t+1})$ is a continuous function
of $\bx_{t}$ (Assumption \ref{H:thmPF3:3}), we have for any $(h_t,\bx_{t+1})\in\ui^{d+1}$
\[
\big|\varphi\circ F_{\Qb_{t+1,h}^N}^{-1}(h_t,\bx_{t+1})-\varphi\circ F_{\Q_{t,h}\otimes m_{t+1,h}}^{-1}(h_t,\bx_{t+1})\big|=\smallo(1),\quad \mbox{a.s.}
\]
and therefore, since $\varphi$ and $\psi_{u,\kappa,\tau,\gamma}$ are bounded, we have, by the dominated convergence
Theorem, 
\[
\left|<\varphi\circ F_{\Qb_{t+1,h}^N}^{-1},\psi_{u,\kappa,\tau,\gamma}>-<\varphi\circ F_{\Q_{t,h}\otimes m_{t+1,h}}^{-1},\psi_{u,\kappa,\tau,\gamma}>\right|\rightarrow0,\quad \mbox{a.s.}
\]
We now establish that, for all $\tilde{u}\in\ui$,  $|F_{\Qh_{t,h}^{N}}^{-1}(\tilde{u})-F_{\Q_{t,h}}^{-1}(\tilde{u})|\rightarrow 0$
almost surely. The proof of this result is inspired from \citet[Theorem 2]{Barvinek1991}.

First, note that because $p_t(\bx_t)>0$ for all $\bx_t\in \ui^d$ (Assumption \ref{H:thmPF1:4} of Theorem \ref{thm:consistency}) the function  $F_{\Q_{t,h}}$ is continuous and strictly increasing on $[0,1)$ (see the proof of Lemma \ref{lemma:thmLD}). Let $\epsilon>0$ and $\tilde{u}_{1}\in\ui$. Then, by the continuity of $F_{\Q_{t,h}}^{-1}$,
there exists a $\delta_{\tilde{u}_{1},\epsilon}>0$ such that, 
\begin{equation}
|\tilde{u}_{1}-\tilde{u}|\leq\delta_{\tilde{u}_{1},\epsilon},\text{ }\implies|F_{\Q_{t,h}}^{-1}(\tilde{u}_{1})-F_{\Q_{t,h}}^{-1}(\tilde{u})|\leq\epsilon.\label{eq:thmPF2Bis:1}
\end{equation}
In the proof of Theorem \ref{thm:PF2} we saw that, for any $\delta_{0}>0$, there exists a
$N_{\delta_{0}}$ such that, for all $N\geq N_{\delta_{0}}$, 
\begin{equation}
\|F_{\Qh_{t,h}^{N}}-F_{\Q_{t,h}}\|_{\infty}\leq\delta_{0},\quad\mbox{a.s.}\label{eq:thmPF2Bis:2}
\end{equation}
Let $x_{N}=F_{\Qh_{t,h}^{N}}^{-1}(\tilde{u}_{1})$ and $u_{N}=F_{\Q_{t,h}}(x_{N})$.
Then, by (\ref{eq:thmPF2Bis:2}), 
\[
|F_{\Qh_{t,h}^{N}}(x_{N})-F_{\Q_{t,h}}(x_{N})|\leq\delta_{0},\quad\forall N\geq N_{\delta_{0}},\quad\mbox{a.s.}
\]
Let $r_{N}(\tilde{u}_{1})=F_{\Qh_{t,h}^{N}}(F_{\Qh_{t,h}^{N}}^{-1}(\tilde{u}_{1}))-\tilde{u}_{1}$ so that
\begin{align*}
|F_{\Qh_{t,h}^{N}}(x_{N})-F_{\Q_{t,h}}(x_{N})|=|\tilde{u}_1+r_{N}(\tilde{u}_1)-u_{N}|\leq\delta_{0},\quad\forall N\geq N_{\delta_{0}},\quad \mbox{a.s.}
\end{align*}
Now note that $|r_{N}(\tilde{u}_{1})|\leq \frac{\|G_{t}\|_{\infty}}{N\Sop(P_{t}^{N})(G_{t})}$ a.s..
Then, it is easy to see that, for all $\delta'>0$, there exists a
$N_{\delta'}$ such that, a.s., $|r_{N}(\tilde{u}_1)|\leq\delta'$ for all
$N\geq N_{\delta'}$. Let $\delta=\delta_{0}+\delta'$ and set $N_{\delta}:=N_{\delta_{0}}\vee N_{\delta'}$.
Then, for $N\geq N_{\delta}$, we have almost surely $|\tilde{u}_1-u_{N}|\leq\delta$.
By taking $\delta_{0}$ and $\delta'$ such that $\delta=\delta_{\tilde{u}_1,\epsilon}$,
(\ref{eq:thmPF2Bis:1}) implies that 
\[
|F_{\Q_{t,h}}^{-1}(\tilde{u}_1)-F_{\Q_{t,h}}^{-1}(u_{N})|\leq\epsilon,\quad\forall N\geq N_{\delta_{\tilde{u}_1,\epsilon}},\quad\mbox{a.s.}
\]
In addition, 
$
F_{\Q_{t,h}}^{-1}(u_{N})=x_{N}=F_{\Qh_{t,h}^{N}}^{-1}(\tilde{u}_1)
$
and therefore , $\forall N\geq N_{\delta_{\tilde{u}_1},\epsilon}$, 
\[
|F_{\Q_{t,h}}^{-1}(\tilde{u}_1)-F_{\Qh_{t,h}^{N}}^{-1}(\tilde{u}_1)|\leq\epsilon,\quad\mbox{a.s.}
\]
Consequently, $\|\nu_{N,u,\kappa}-\nu_{u,\kappa}\|_{\infty}=\smallo(1)$
almost surely so that, by the dominated convergence Theorem, $\sigma_{N,u,\kappa}^{2}\rightarrow\sigma_{u,\kappa}^{2}$
almost surely. Also, because $\varphi$ is continuous and bounded,
$\sigma_{N}^{2}\rightarrow\sigma^{2}$ almost surely by Theorem \ref{thm:consistency} and portmanteau lemma \citep[][Lemma 2.2]{VanderVaart2007}.
To simplify the notations in what follows, let $
\tilde{\sigma}_{u,l}^{2}=\sum_{\kappa:|\kappa|=l}\sigma_{u,\kappa}^{2}$ and  $\tilde{\sigma}_{N,u,l}^{2}=
\sum_{\kappa:|\kappa|=l}\sigma_{N,u,\kappa}^{2}$ (remark that $l$ depends implicitly on $u$), and note that, for $m\geq \tilde{t}+ d+1$,
\[
\sum_{|u|>0}\text{ }\sum_{l>m-\tilde{t}-|u|}\tilde{\sigma}_{N,u,l}^{2}=\sigma_{N}^{2}-\sum_{|u|>0}\text{ }\sum_{l=0}^{\infty}\mathbb{I}(l\leq m-\tilde{t}-|u|)\tilde{\sigma}_{N,u,\kappa}^{2}.
\]
By Fubini's Theorem, 
\[
\E\left[\sum_{|u|>0}\text{ }\sum_{l>m-\tilde{t}-|u|}\tilde{\sigma}_{N,u,l}^{2}\right]=\E[\sigma_{N}^{2}]-\sum_{|u|>0}\text{ }\sum_{l=0}^{\infty}\mathbb{I}(l\leq m-\tilde{t}-|u|)\E[\tilde{\sigma}_{N,u,l}^{2}]
\]
where, by the dominated convergence Theorem, $\E[\sigma_{N}^{2}]\rightarrow\sigma^{2}$. In addition, since in the definition of $\tilde{\sigma}^2_{N,u,l}$ and $\tilde{\sigma}^2_{u,l}$ the notation $\sum_{\kappa:|\kappa|=l}$ denotes a finite sum, we have, for any $u$ and $l$, $\tilde{\sigma}^2_{N,u,l}\rightarrow \tilde{\sigma}^2_{u,l}$ almost surely  and therefore, by the dominated convergence Theorem,  $\E[\tilde{\sigma}_{N,u,\kappa}^{2}]\rightarrow\tilde{\sigma}_{u,\kappa}^{2}$
(because $\tilde{\sigma}_{N}^{2}$ is bounded by $\|\varphi\|_{\infty}^{2}$).
Hence, using Fatou's lemma, 
\begin{align*}
0 & \leq\limsup_{m\rightarrow +\infty}\E\left[\sigma_{N}^{2}-\sum_{|u|>0}\text{ }\sum_{l=0}^{\infty}\mathbb{I}(l\leq m-\tilde{t}-|u|)\tilde{\sigma}_{N,u,l}^{2}\right]\\
 & \leq\limsup_{m\rightarrow +\infty}\E[\sigma_{N}^{2}]+\limsup_{m\rightarrow +\infty}\left\{ -\sum_{|u|>0}\text{ }\sum_{l=0}^{\infty}\mathbb{I}(l\leq m-\tilde{t}-|u|)\E[\tilde{\sigma}_{N,u,l}^{2}]\right\} \\
 & \leq\sigma^{2}-\sum_{|u|>0}\text{ }\sum_{l=0}^{\infty}\liminf_{m\rightarrow +\infty}\mathbb{I}(l\leq m-\tilde{t}-|u|)\E[\tilde{\sigma}_{N,u,l}^{2}]\\
 & =\sigma^{2}-\sum_{|u|>0}\text{ }\sum_{l=0}^{\infty}\tilde{\sigma}_{u,l}^{2}=0,
\end{align*}
since the indicator functions converge to one.

\end{document}